\documentclass[11pt]{article}

\usepackage[margin=0.5in]{geometry}
\usepackage{setspace}
\usepackage{lineno}
\usepackage[authoryear,round]{natbib}
\usepackage{amssymb}
\usepackage{algcompatible}
\usepackage{amsmath}
\usepackage{amsthm}
\usepackage{epsfig}
\usepackage{booktabs}
\usepackage{graphicx}
\usepackage{graphics}
\usepackage{float}
\usepackage{multirow}
\usepackage{color}
\usepackage{caption}
\usepackage{subcaption}
\usepackage[normalem]{ulem} 
\usepackage{makeidx}
\usepackage{xspace}
\usepackage{wrapfig}
\usepackage{tikz}
\usepackage{hyperref}
\hypersetup{
colorlinks=true,
citecolor=blue,
linkcolor=blue,
filecolor=blue,      
urlcolor=blue,
}
\usepackage{setspace}
\usepackage{bm}
\usetikzlibrary{arrows.meta}

\usepackage{array}
\newcolumntype{C}[1]{>{\centering\arraybackslash}m{#1}}
\graphicspath{{./art/}}
\usepackage[plain,noend]{algorithm2e}
\usepackage{comment}

\newtheorem{theorem}{Theorem}
\newtheorem{proposition}{Proposition}
\newtheorem{assumption}{Assumption}
\newtheorem{definition}{Definition}
\newtheorem{corollary}{Corollary}

\newtheorem{lemma}{Lemma}

\newtheorem{example}{Example}
\newtheorem{remark}{Remark}
\newtheorem{condition}{Condition}

\usepackage{dsfont}

\usepackage[affil-it]{authblk}

\newenvironment{keywords}{
  \vspace{1em}
  \noindent\textbf{Keywords:}
  \begin{itshape}
}{
  \end{itshape}
}

\usepackage{comment}
\usepackage{pagecolor}
\usepackage{mdframed}

\usepackage{xcolor}

\begin{document}

\title{\LARGE Inward and Outward Spillover Effects of One Unit's Treatment on Network Neighbors under Partial Interference}
\author[1]{Fei Fang} \author[2]{Edoardo M. Airoldi} \author[1]{Laura Forastiere}
\affil[1]{Biostatistics, Yale School of Public Health, Yale University}
\affil[2]{Department of Statistics, Operations, and Data Science, Fox School of Business, Temple University}
\date{\today}
\maketitle
\vspace{-1cm}

\setcounter{page}{0}
\thispagestyle{empty}

\linespread{1.5}\selectfont

\begin{abstract}
In settings where interference is present, direct effects are commonly defined as the average effect of a unit's treatment on their own outcome while fixing the treatment status or probability among interfering units. Spillover effects measure the average effect of a change in the latter while the individual's treatment status is kept fixed. Here, we define the average causal effect of a unit's treatment status on the outcome of their network neighbors, while fixing the treatment probability in the remaining interference set. We propose two different weighting schemes defining two causal effects:
i) the outward spillover effect, which represents the average effect of a unit's treatment on their neighbors' potential outcomes, and ii) the inward spillover effect, which represents the impact of each neighbor's treatment on an individual's own potential outcome. 
We provide a necessary and sufficient condition for when outward and inward spillover effects differ--even in undirected networks--and show that they are equivalent under specific conditions.
We provide numerous examples illustrating the conditions for equivalence or discrepancy of the two spillover effects. We then compare their Horvitz-Thompson estimators, examining their relative variance under various graph structures and structural assumptions on potential outcomes. 
\end{abstract}

\begin{keywords}
causal inference; interference; outward and inward spillover effects; social networks.
\end{keywords}

\section{Introduction}
\subsection{Background}
\label{sec:background}
Causal inference typically relies on the Stable Unit Treatment Value Assumption (SUTVA) (\cite{rubin1980randomization}), which presumes the absence of interference among units. However, in numerous applications, interactions among units exist, either explicitly through social connections \citep[e.g.,][]{huang2023estimatingeffectslongtermtreatments, yuan2024twopartmachinelearningapproach}, or inherently within the same clusters \citep[e.g.,][]{papadogeorgou2019causal, tchetgen2012causal}. These interactions may potentially allow the treatments received by others to affect an individual's potential outcomes mainly through two major mechanisms. { The first mechanism is diffusion of treatments, where the $i$-th treatment affects the $j$-th treatment and then the $j$-th outcome. Examples include transmission of treatments such as flyers, and coupons \citep{an2018causal, tortú2021causaleffectshiddentreatment}. The second mechanism is diffusion of outcomes, i.e., $i$'s treatment affects $i$'s outcome, which in turn affects $j$'s outcome. Examples include vaccinated individuals protecting unvaccinated individuals' health through their own lower likelihood of infection  \citep{halloran2018estimating,vanderweele2011bounding}, and anti-conflict interventions on small groups of students influencing other students' behavior through the seed students' changed behavior \citep{paluck2016changing, aronow2017estimating}. }

When interference is assumed to be restricted to within clusters, a commonly adopted assumption is referred to as `partial interference' \citep{hudgens2008toward}. Under this assumption, direct and indirect effects are commonly defined by changing or fixing the individual's treatment status while the treatment assignment probability in the rest of the cluster is kept fixed or modified, respectively. For instance, \cite{hudgens2008toward} 
adopt this interference assumption and employ difference-in-mean estimators based on a two-stage design,
while \citet{basse2018analyzing} propose Horvitz-Thompson estimators with weights defined for both individual- or cluster-weighted estimands accommodating varying cluster sizes.

On the other hand, when social interactions are observed and interference is assumed to occur through the network of such interactions, it is common to assume that interference is restricted to network neighbors. Similar definitions are provided for the direct effect of a unit's treatment on its own outcome and for the spillover effect from neighbors' treatments. To reduce complexity, it is also common to assume an exposure mapping function {  that maps the treatment vector and graph structure to a low-dimensional quantity affecting potential outcomes \citep{aronow2017estimating}}
. Such assumptions enable unbiased or consistent estimators for spillover effects to be constructed using Horvitz-Thompson or H\'{a}jek estimators, with weights based on exposure probabilities \citep{eckles2017design}. Beyond direct estimation, several works have focused on detecting the presence of spillover effects through randomization-based testing \citep{aronow2012general, athey2018exact, basse2019randomization}.

Here, we consider a clustered network setting, where the sample is partitioned into clusters and networks are observed within clusters. Under partial interference, we define the causal effect of one unit's treatment on the outcomes of its neighbors. 
{  We consider neighborhood-level estimands for several reasons. First, interference effects often decay with network distance. Even when interference spans the full cluster, its strength typically varies by distance. Estimating neighborhood-specific effects lets researchers separate the impact of immediate neighbors from the broader cluster influence, offering a finer view of the interference mechanism \citep{xu2026higher}. Second, conditional spillover effects using neighborhood-type estimands help distinguish whether a unit influences others because of its network position (e.g., centrality) or because of intrinsic attributes (e.g., covariates). This distinction is critical for designing targeted interventions, such as deciding whether to prioritize high-centrality individuals or those with specific traits that strongly affect their neighbors. However, the causal estimand of interest should not be tied to the assumed interference structure. We adopt the partial interference assumption mainly as a tool for statistical identification and inference \citep{savje2024causal}. This assumption offers robustness to misspecification of the interference set (the set of units whose treatments affect an outcome), especially when practitioners are unsure whether interference extends beyond immediate neighbors. 

We then introduce two weighting schemes over neighbors, yielding outward and inward spillover effects.} The outward spillover effect is defined as the effect of an individual's treatment on the outcomes of their out-neighbors (Definition \ref{out_spillover}). The inward spillover effect is defined as the effect on an individual's outcome of the treatment of each of their in-neighbors (Definition \ref{in_spillover}). Both spillover effects are defined under partial interference and a hypothetical treatment assignment for the remaining units in the cluster that we keep fixed. 


The two estimands' weighting schemes provide two perspectives on evaluating spillover effects from one unit's treatment: the outward spillover effect is viewed from the perspective of the unit receiving the treatment, here referred to as the `sender';  the inward spillover effect corresponds to the perspective of the unit whose outcome is affected by the treatment of each of their neighbors, here referred to as the `receiver'. 
%
{ 
Most existing work defines spillovers from the receiver's perspective \citep{manski1993identification, cai2015social,leung2022causal}. 
%
Outward spillover effects, however, are also of substantive interest, for effect senders themselves and for policymakers seeking to target subpopulations that generate larger spillovers. In vaccination settings, for example, the inward estimand reflects the receiver's perspective, where the primary concern is the protection conferred by vaccinated neighbors \citep{barrios2022neighbors}. By contrast, policymakers aiming to increase population-level protection may prioritize individuals with the highest transmission potential to their contacts, thereby amplifying outward spillover effects \citep{benjamin2017spillover,carpenter2019direct}. 
A similar distinction arises in policing interventions. In hot-spots policing experiments, the inward perspective quantifies the extent to which policing intensity in surrounding areas reduces crime at a given location. By contrast, the outward perspective captures the reduction in crime at nearby locations induced by the policing intensity assigned to that location \citep{braga2012hot}.}
\subsection{Key results}
\label{sec:key_results}

When comparing outward and inward spillover effects, their difference primarily depends on the cardinalities of in- and out-neighborhoods, as well as the heterogeneity of unit-to-unit spillover effects. We establish a necessary and sufficient condition for the inequivalence of outward and inward spillover effects, detailed in Theorem \ref{difference_out_in}. { This condition is easily satisfied when the graph is non-regular and unit-to-unit spillover effects are heterogeneous.}

Conversely, we identify three sufficient conditions (Conditions \ref{ave_equivalence_cond_1}, \ref{ave_equivalence_cond_2}, and \ref{ave_equivalence_cond_3}) under which the two estimands coincide. These conditions restrict the heterogeneity of unit-to-unit spillover effects and the structure of out- and in-neighborhoods. Among these conditions, Condition \ref{ave_equivalence_cond_1} or \ref{ave_equivalence_cond_2} is often assumed when linear structural models are posited for outcomes \citep[e.g.,][]{cai2015social, manski1993identification}. 
{ Nevertheless, even when the two causal estimands coincide in such scenarios, their respective estimators may differ. When this is the case, we can leverage the estimators that have better statistical properties, such as 
higher efficiency, for estimating both effects.}
We therefore provide analytical comparisons of the design-based conservative and true variance of their Horvitz Thompson estimators (Propositions \ref{V_out_in_compare} and \ref{dif_var_out_var_in}). It turns out that the graph structure and the outcomes of out-neighbors influence the efficiencies of these estimators. 
{ We illustrate numerically how differences between the outward and inward weighting schemes affect the variances of the corresponding estimators in Section \ref{estimand_weight_efficiency}.}

\section{Causal estimands: outward and inward spillover effects}
\label{sec_estimand}
Let us consider a sample of $K$ clusters, with the k-th cluster containing $n_k$ units, indexed by $i=1, \dots, n_k$.  The set of all units in cluster $k$ is defined as $\mathcal{N}_k=\{ik: i=1, \dots, n_k\}$ and the set of all units across clusters is defined as $\mathcal{N}=\{ik: ik\in \mathcal{N}_k, k=1, \dots, K\}$. Within each cluster \( k \), the \( n_k \) units form a directed network denoted by \( \mathcal{G}_k = (\mathcal{N}_k, E_k) \), where \( E_k \) represents the set of directed edges among units in \( \mathcal{N}_k \). The overall network encompassing all clusters is denoted by \( \mathcal{G} \). For each unit \( i = 1, \dots, n_k \) in cluster \( k \), the sets of out-neighbors and in-neighbors are defined as \( \mathcal{N}^{out}_{ik} = \{jk \in \mathcal{N}_k : e_{ik,jk} \in E_k\} \) and \( \mathcal{N}^{in}_{ik} = \{jk \in \mathcal{N}_k : e_{jk,ik} \in E_k\} \), respectively, where \( e_{a,b} \) denotes the directed edge from unit \( a \) to unit \( b \). Their cardinalities are represented by \( |\mathcal{N}^{out}_{ik}| \) and \( |\mathcal{N}^{in}_{ik}| \). 
When the network is undirected, with $e_{a,b}=e_{b,a}$, \(\mathcal{N}^{out}_{ik} =\mathcal{N}^{in}_{ik}:= \mathcal{N}_{ik}\), where $\mathcal{N}_{ik}$ denotes the set of neighbors of unit $ik$, and thus \(|\mathcal{N}^{out}_{ik}| = |\mathcal{N}^{in}_{ik}|=|\mathcal{N}_{ik}| \). 

In the sample $\mathcal{N}$, the experimenter assigns a treatment vector \( \mathbf{Z} := (Z_{11}, \dots, Z_{n_K K}) \), with \( Z_{ik} \in \{0,1\} \), for each unit \( i = 1, \dots, n_k \) in cluster \( k = 1, \dots, K \). We consider the assignment mechanisms to be  based on a known parameter (or vector of parameters) \( \beta \). For instance, in a Bernoulli experiment where the treatment is assigned independently and with constant probability, $\beta$ simply represents this probability of treatment (type B parametrization in \cite{tchetgen2012causal}). Alternatively, in a two-stage randomization experiment \citep{hudgens2008toward}, $\beta$ consists of elements $(\nu, \phi, \psi)$, with $\nu$ and $1-\nu$ representing the probabilities of assigning a cluster to the individual treatment probabilities $\phi$ and $\psi$, respectively, to be used in the second stage. The hypothetical treatment assignment, i.e., the assignment of interest on the treatment vector \( \mathbf{Z} \), is similarly governed by a known parameter \( \alpha \). The hypothetical treatment assignment may or may not follow the same parametrization or take the same values as \( \beta \). 

The potential outcome for unit \( i \) in cluster \( k \) is denoted by \( Y_{ik}(\mathbf{Z} = \mathbf{z})\), or simply \( Y_{ik}(\mathbf{z}) \), where \( \mathbf{z}\) denotes a specific realization of the treatment vector. 
The observed outcome for the same unit is denoted by \( Y_{ik} \). 
Throughout, we assume partial interference, which restricts the dependence of potential outcomes to the treatment vector within the same cluster, as formalized below.

\begin{assumption}[Partial interference]
\label{part_intf} 
For any $\mathbf{z},\mathbf{z}^\prime \in \{0,1\}^{\sum_{h=1}^K n_h } $ such that $\mathbf{z}_k=\mathbf{z}^\prime _k$, the potential outcome satisfies $Y_{ik}(\mathbf{z})=Y_{ik}(\mathbf{z}^{\prime})$ for $i=1,\cdots, n_k$ and $k=1,\cdots,K$. 
\end{assumption}
Under this assumption, the potential outcome for unit $ik$ can be expressed as $Y_{ik}(\mathbf{z}_k)$.  
{  The assumption allows interference by any mechanism among units in the same cluster, as opposed to assuming that interference only travels through the network \citep{tchetgen2021auto, leung2022causal}. Consequently, under Assumption \ref{part_intf}, a unit's outcome can be affected by the treatment of any other cluster member. In particular, a unit's outcome can be affected by the treatment of isolated units in the same cluster. However, because isolated units have neither out- nor in-neighbors, they do not contribute as senders or receivers to the spillover estimands defined below in Definitions~\ref{out_spillover} and~\ref{in_spillover}.
}

Given a pair of units \( ik \) and \( jk \), let us write the potential outcome 
$Y_{ik}(\mathbf{z}_k)$ as $Y_{ik}({z}_{jk},\mathbf{z}_{-jk})$, where $\mathbf{z}_{-jk}$ is the realization of treatment vector \( \mathbf{Z}_{-jk} \) in cluster $k$, excluding unit $jk$.
 We now define 
$$\bar{Y}_{ik}(Z_{jk}=z,\alpha):= \mathbb{E}_{\mathbf{Z}_{-jk}|\alpha}\left[Y_{ik}(Z_{jk}=z,\mathbf{Z}_{-jk})\right]$$
{ which denotes the individual average potential outcome for unit \( ik \). Here, the treatment of unit \( jk \) is set at \( z \) and the rest of the cluster $k$, including unit $ik$ is randomized to treatment under a hypothetical treatment assignment parametrized by \(\alpha\).} 
Following this definition, we conceptualize the pairwise spillover effects, i.e., unit-to-unit spillover effects, as follows.
\begin{definition}[Pairwise spillover effects]
\label{pair_spillover}
The spillover effect from unit $jk$ to $ik$, with $j,i={1, \dots, n_k}$ and $k=1, \dots, K$, is defined as 
$$\tau_{ik,jk}(\alpha)=\bar{Y}_{ik}(Z_{jk}=1,\alpha)-\bar{Y}_{ik}(Z_{jk}=0,\alpha)$$
\end{definition}
$\tau_{ik,jk}(\alpha)$ can be interpreted as the spillover effect that a unit $jk$'s treatment status from $1$ to $0$ on the outcome of a unit $ik$ in the same cluster, while the rest of the cluster is randomly assigned to treatment with hypothetical parameter $\alpha$. We refer to $jk$ as the `sender' and to $ik$ as the `receiver'. { The quantity $\tau_{ik,jk}(\alpha)$ characterizes a unit-level spillover effect of the treatment assigned to one unit on the outcome of another. This pairwise spillover effect corresponds to a summand in the average indirect effect of \cite{hu2022average}.
} Under Assumption~\ref{part_intf}, \(\tau_{ik,jk}(\alpha)\) can be nonzero even when the two units are not directly connected, although we focus here on spillover effects among network neighbors. 
 To ensure that these spillover effects among network neighbors are well-defined, we impose the following assumption.  
{ 
\begin{assumption}[Intra-cluster edge existence]
\label{one_in_and_out_neigh}
For every cluster \(k=1,\ldots,K\), the set of intra-cluster edges is non-empty; that is, \(|E_k|\geq 1\).
\end{assumption}
Assumption~\ref{one_in_and_out_neigh} guarantees that each cluster in the sample contains at least one directed edge and, consequently, at least one unit with out-neighbors and at least one unit with in-neighbors. 
}

{ Depending on whether pairwise spillover effects are averaged from the sender's or the receiver's perspective, we define the outward and inward spillover effects as follows.}

\begin{definition}[Outward spillover effects]
\label{out_spillover}
For a unit $jk$ with $|\mathcal{N}^{out}_{jk}|>0$, let $\mu^{out}_{jk}(z, \alpha)$ be the average outcome among unit $jk$'s out-neighbors when the treatment of $jk$ is set to $z \in \{0,1\}$, while a random experiment with parameter $\alpha$ takes place in the rest of cluster $k$, i.e., 
$$\mu^{out}_{jk}(z_{jk}, \alpha): =\frac{1}{ |\mathcal{N}^{out}_{jk}|}  \sum_{ik \in \mathcal{N}^{out}_{jk}}\bar{Y}_{ik}(Z_{jk}=z, \alpha).$$
   The individual outward spillover effect of unit \(jk\) on the outcomes of its out-neighbors is defined as:
\begin{equation*}
    \begin{split}
        &  \tau^{out}_{jk}(\alpha)=\mu^{out}_{jk}(1, \alpha)-\mu^{out}_{jk}(0, \alpha).
    \end{split}
\end{equation*}
{ Under Assumption \ref{one_in_and_out_neigh}}, the average outward spillover effect is then given by
{ 
    \begin{equation*}
    \begin{split}
        \tau^{out}(\alpha)= \frac{1}{N^{out}} \sum_{k=1}^K  \sum_{jk \in \mathcal{N}_k^{out} }\tau^{out}_{jk}(\alpha)=\sum_{k=1}^K  \sum_{jk \in \mathcal{N}_k^{out} } \sum_{ik \in \mathcal{N}^{out}_{jk}}   \frac{1}{N^{out}|\mathcal{N}^{out}_{jk}|} \ \tau_{ik,jk}(\alpha)
    \end{split}
\end{equation*}}
where $\mathcal{N}^{out}_k= \left\lbrace jk\in \mathcal{N}: |\mathcal{N}^{out}_{jk}| >0  \right\rbrace $ and ${N}^{out}= \sum_{k=1}^{K} |\mathcal{N}^{out}_{k}|$. 
\end{definition}
{ $\tau^{out}_{jk}(\alpha)$ is the average influence that the treatment of unit \(jk\) has on each of its out-neighbors. This individual influence may differ across units because of (1) heterogeneity in the influence that each individual has on their out-neighbors (e.g., friends), or (2) heterogeneity in the characteristics of each individual's out-neighbors, which make them more or less susceptible to changing their outcomes in response to the treatments of their in-neighbors.} $\tau^{out}(\alpha)$ is then the average of $\tau^{out}_{jk}(\alpha)$ across all units in all clusters with at least one out-neighbor. We define the neighborhood spillover effect $\tau^{out}_{jk}(\alpha)$ as an average rather than a sum of neighbors' outcomes to capture the influence on neighbors' outcomes in average.

Our definition of the outward spillover effect can be considered as an analogue of the circle estimand defined in \cite{wang2024designbasedinferencespatialexperiments}. However, while the estimands in \cite{wang2024designbasedinferencespatialexperiments} are defined based on Euclidean distance to suit spatial settings, our estimands employ a measure of distance for social network, with a particular focus on network neighbors. Similarly, our definition aligns conceptually with the \( P \)-indexed direct effect proposed by \cite{zigler2021bipartite}. \( P \)-indexed direct effect captures the influence of a treated unit on all outcomes affected by this unit's treatment, which is consistent with the spirit of our outward spillover effect. 
However, a key distinction lies in our focus on the influence on network neighbors, rather than on all units affected by a given unit's treatment, even though in our setting, a unit's treatment could potentially influence outcomes across the entire cluster.
{  Another related concept is the average indirect effect of \cite{hu2022average}. By changing the order of summation, this estimand aggregates unit-level effects from one unit's treatment to all other units' average potential outcomes
. In contrast, our estimand focuses on average spillover effects of changing one unit's treatment on the average potential outcomes of its out-neighbors.} 

\begin{definition}[Inward spillover effect]
\label{in_spillover}
For a unit $ik$ with $|\mathcal{N}^{in}_{ik}|>0$, let $\mu^{in}_{ik}\left( z ,\alpha \right)$ be the average potential outcome of unit $ik$ when the treatment of one of its in-neighbors is set to   $z$, while a random experiment with parameter $\alpha$ takes place in the rest of cluster $k$, i.e., 
$$\mu^{in}_{ik}\left( z ,\alpha \right) :=  \frac{1}{|\mathcal{N}^{in}_{ik}|} \sum_{jk \in \mathcal{N}^{in}_{ik}} \bar{Y}_{ik} (Z_{jk}=z,\alpha).$$
The individual inward spillover effect from the treatments of $ik$'s in-neighbors on $ik$'s own outcome is defined as
\begin{equation*}
\begin{split}
    & \tau^{in}_{ik}\left(\alpha \right):=\mu^{in}_{ik}\left( 1 ,\alpha \right)- \mu^{in}_{ik}\left( 0 ,\alpha \right). 
\end{split}
\end{equation*}
{ Under Assumption \ref{one_in_and_out_neigh}}, the average inward spillover effect is then given by
{ 
\begin{equation*}
    \tau^{in}(\alpha)= \frac{1}{N^{in}} \sum_{k=1}^K \sum_{ik \in \mathcal{N}^{in}_k  }  \tau^{in}_{ik}(\alpha)= \sum_{k=1}^K \sum_{ik \in \mathcal{N}^{in}_k }\sum_{jk \in \mathcal{N}^{in}_{ik}} \frac{1}{N^{in }|\mathcal{N}^{in}_{ik}|} \tau_{ik,jk}(\alpha)
\end{equation*} }
where $\mathcal{N}^{in}_k= \{ik\in \mathcal{N}: |\mathcal{N}^{in}_{ik}| >0  \}$ and ${N}^{in}= \sum_{k=1}^{K} |\mathcal{N}^{in}_{k}|$.
\end{definition}
The individual inward spillover effect $\tau^{in}_{ik}\left(\alpha \right)$ quantifies, on average across in-neighbors, the influence exerted by treating one in-neighbor of unit $ik$. As with $\tau^{out}_{ik}\left(\alpha \right)$, the heterogeneity of $\tau^{in}_{ik}\left(\alpha \right)$ across units may be due to their own susceptibility to the treatment of their in-neighbors as well as to the characteristics of the latter. 
$\tau^{in}\left(\alpha \right)$ is then the average of $\tau^{in}_{ik}(\alpha)$ across all units in all clusters with at least one in-neighbor. 

{ A closely related concept of inward spillover effect is the average indirect effect in \cite{hu2022average}.} In its original order of summation, this estimand adopts the receiver's perspective by aggregating the effects of changes in all other units' treatment assignments on a given unit's potential outcome, regardless of whether those units are its in-neighbors. 
In addition, the \( M \)-indexed direct effect introduced by \cite{zigler2021bipartite} aligns conceptually with the inward spillover effect. In this estimand, the individual's average spillover effect is taken over all units in its interference set, whereas our individual inward spillover effect, $\tau^{in}_{ik}(\alpha)$ in Definition \ref{in_spillover}, captures the individual's average spillover effect from the treatments of its in-neighbors. 

{ 
Although the perspectives underlying the \(P\)- and \(M\)-indexed direct effects of \citet{zigler2021bipartite} are similar to our outward and inward spillover effects, the bipartite setting in \cite{zigler2021bipartite} differs substantively from the clustered network setting considered here. In bipartite interference, two distinct and non-overlapping types of units are considered, such as pollution sources and population locations, with treatments being assigned to interventional units and outcomes being measured on a distinct set of outcome units. In our setting, by contrast, the same units may both receive treatment and have their outcomes affected by others' treatments.} 

In both definitions of the average outward and inward spillover effects, each unit, i.e., the sender or the receiver, respectively, is equally weighted, regardless of their cluster size. { These definitions can be modified to yield cluster-weighted estimands, as in \citep{basse2018analyzing}.} 

\begin{remark}
\label{remark_direction_spillover}
The direction of the graph $\mathcal{G}$ we consider here aligns with the direction along which spillover effects are of interest.
The original graph may possess inherent directional information, such as the one provided by friendship nominations, where friends nominated by one individual are labeled as out-neighbors and vice versa.
However, the direction of spillover effects of interest may not align with the direction provided in network connections.
For instance, in a setting with behavioral outcomes, a training intervention and a friendship network,
individuals are often influenced by those they consider friends, implying that the spillover effect flows from the nominated friend to the nominator, thus in the opposite 
direction relative to the friendship nomination.
In this case, if we are interested in the spillover effect in this direction, we would define the graph $\mathcal{G}$ where the direction of the network edges, defining $\mathcal{N}^{in}_{ik}$ and $\mathcal{N}^{out}_{ik}$, is opposite to the direction of friendship nomination. However, a spillover effect in the opposite direction may still exist. If we were interested in estimating this spillover effect, we would keep the direction of the graph $\mathcal{G}$ as the original one in the friendship nomination network.

\end{remark}

\begin{remark}
Provided the direction of the graph discussed in Remark \ref{remark_direction_spillover}, individual spillover effects are always defined in such direction, that is from the treatment of in-neighbors on a receiver's outcome or, equivalently, from a sender's treatment on out-neighbors' outcomes.
Here, the terms "outward" and "inward" defining our individual and average spillover effects in Definitions \ref{out_spillover} and \ref{in_spillover} refer solely to the way we take averages over out-neighbors or in-neighbors, in line with the sender and the receiver  perspectives, respectively. Therefore, the two terms do not pertain to the actual directions of spillover effects incorporated into the causal estimands.
\end{remark}

We may further consider the outward and inward spillover effects, conditioning on  senders' covariates. These effects are conceptually similar to the average effects described in Definitions \ref{out_spillover} and \ref{in_spillover}. However, the conditional outward spillover effect specifically considers the influence of senders with particular covariate values on their neighbors. The conditional inward spillover effect focuses on the impact of treatments from in-neighbors with specific covariate values on a recipient's potential outcome. For instance, if we consider "female" as the covariate value of interest, the conditional outward spillover effect would examine the impact of treatments from females on the outcomes of their out-neighbors. Similarly, the conditional inward spillover effect assesses the influence of treatments received from female neighbors on an individual's potential outcome. The formal definitions are provided in Supplementary Material \ref{append_sec_estimand}.

\section{Comparison between outward and inward spillover effects}
\label{comp_out_in_spillover}
Despite the two different perspectives underlying the outward and inward spillover effects, both are defined by taking certain types of neighborhood averages, and they both capture the impact of one unit's treatment on their  neighbors' potential outcomes. In this section, we aim to explore whether and when these two causal estimands coincide or differ. { 
We provide a necessary and sufficient condition for the non-equivalence of outward and inward spillover effects (Theorem \ref{difference_out_in}), and show that they are equivalent under specific conditions }
(Conditions \ref{ave_equivalence_cond_1}, \ref{ave_equivalence_cond_2}, and \ref{ave_equivalence_cond_3}). 
{ 
\begin{theorem}
\label{difference_out_in}
Let the outward and inward estimand weights be  $S^{out}_{ik,jk}
=
(N^{out}|\mathcal{N}^{out}_{jk}|)^{-1}$, and  $S^{in}_{ik,jk}
=
(N^{in}|\mathcal{N}^{in}_{ik}|)^{-1}$, respectively. Define $\Delta_{ik,jk}
=
S^{out}_{ik,jk}-S^{in}_{ik,jk}$, for all pairs $(ik,jk)$ such that
$jk\in\mathcal{N}^{out}_{k}$ and $ik\in\mathcal{N}^{out}_{jk}$. Let $$L_1
=
\left\{
(ik,jk):
\Delta_{ik,jk}\tau_{ik,jk}(\alpha)\geq 0,\ 
jk\in\mathcal{N}^{out}_{k},\
ik\in\mathcal{N}^{out}_{jk},\
k=1,\ldots,K
\right\}, $$ and $$L_2
=
\left\{
(ik,jk):
\Delta_{ik,jk}\tau_{ik,jk}(\alpha)<0,\ 
jk\in\mathcal{N}^{out}_{k},\
ik\in\mathcal{N}^{out}_{jk},\
k=1,\ldots,K
\right\}.$$ Then $\tau^{out}(\alpha)\neq \tau^{in}(\alpha)$ if and only if
\begin{equation}
\label{theo_difference_out_in_int_1}
\sum_{(ik,jk)\in L_1}
\Delta_{ik,jk}\tau_{ik,jk}(\alpha)
+
\sum_{(ik,jk)\in L_2}
\Delta_{ik,jk}\tau_{ik,jk}(\alpha)
\neq 0.
\end{equation}
\end{theorem}
Theorem~\ref{difference_out_in} characterizes when the outward and inward average spillover effects differ. The difference is determined by the interaction between the graph-induced weight difference, $\Delta_{ik,jk}$, and the pairwise spillover effect, $\tau_{ik,jk}(\alpha)$. In particular, the two estimands differ whenever the positive contributions, $\Delta_{ik,jk}\tau_{ik,jk}(\alpha)\geq 0$, do not exactly offset the negative contributions, $\Delta_{ik,jk}\tau_{ik,jk}(\alpha)<0$. The relative magnitude of the two estimands is governed by how the weight difference $S^{out}_{ik,jk}-S^{in}_{ik,jk}$ co-varies with the pairwise spillover effect $\tau_{ik,jk}(\alpha)$ across receiver--sender pairs. When heterogeneity in the pairwise spillover effects is systematically related to network structure, e.g., $|\mathcal{N}^{out}_{jk}|$ and $|\mathcal{N}^{in}_{ik}|$, $\tau^{out}(\alpha)$ can differ substantially from $\tau^{in}(\alpha)$. We provide three simulation examples in Supplementary Material~\ref{Append:sim_dif_tau_out_tau_in} (Tables~\ref{setting_1:sim_tau_out_dif_tau_in}, \ref{setting_2:sim_tau_out_dif_tau_in}, and \ref{setting_3:sim_tau_out_dif_tau_in}) to illustrate this mechanism under heterogeneous pairwise spillover effects and general graph structures. When the data-generating process for $\tau_{ik,jk}(\alpha)$ is unrelated to $\Delta_{ik,jk}$, $\tau^{out}(\alpha)$ and $\tau^{in}(\alpha)$ can still differ in a finite population. We provide two simulation examples in Supplementary Material~\ref{Append:sim_dif_tau_out_tau_in} to illustrate this point. Further intuition for the non-equivalence characterized in Theorem~\ref{difference_out_in} is provided by Corollary~\ref{corol:suff_cond3_not_equal_relaxed} in Supplementary Material~\ref{append:tau_out_tau_in_unequal}, which gives an interpretable sufficient condition, and by the theoretical Example~\ref{examp_uncond_dif} in Supplementary Material~\ref{ave_spillover_example}.
}

Notably, Theorem \ref{difference_out_in} remains valid for undirected graphs. In such cases, \(N^{out} = N^{in}\), \( |\mathcal{N}^{out}_{jk}| = |\mathcal{N}^{in}_{jk}|= |\mathcal{N}_{jk}|\), and \( |\mathcal{N}^{out}_{ik}| = |\mathcal{N}^{in}_{ik}|=|\mathcal{N}_{ik}|\). From equation \eqref{theo_difference_out_in_int_1}, it follows that \(\tau^{out}(\alpha) \neq \tau^{in}(\alpha)\) if there exist some pairs of units \(jk\) and \(ik\), where \(ik\) is a neighbor of \(jk\), satisfying the following conditions: (1) the number of neighbors of \(jk\), $|\mathcal{N}_{jk}|$, differs from the number of neighbors of \(ik\), $|\mathcal{N}_{ik}|$, { so that
\(\Delta_{ik,jk}\neq 0\)}; (2) the pairwise spillover effect \(\tau_{ik,jk}(\alpha) \neq 0\); and (3) { the terms
\(\Delta_{ik,jk}\tau_{ik,jk}(\alpha)\) for pairs in \(L_1\) do not exactly
offset those for pairs in \(L_2\). We further provide an example in which
\(\tau^{out}(\alpha)\) differs substantially from \(\tau^{in}(\alpha)\) under
an undirected graph; see Example~\ref{examp:undirect_graph_tau_out_tau_in_dif}  in Supplementary Material~\ref{ave_spillover_example}.}


We then investigate sufficient conditions under which the two causal estimands 
$\tau^{out}(\alpha)$ and $\tau^{in}(\alpha)$ are equivalent. Each of the following conditions can lead to the equivalence. 
\begin{condition}
\label{ave_equivalence_cond_1}
    The graph $\mathcal{G}$ is undirected and $\tau_{ik,jk}(\alpha)=c_k$ for any pair $jk\in \mathcal{N}_{k}$ and $ik \in \mathcal{N}_{jk}$ in cluster $k=1, \dots,K$. 
\end{condition}
\begin{condition}
\label{ave_equivalence_cond_2}
    $\tau_{ik,jk}(\alpha)=c$ for any pair $jk\in \mathcal{N}^{out}_{k}$ and $ik \in \mathcal{N}^{out}_{jk}$ in cluster $k=1, \dots,K$.
\end{condition}
\begin{condition}
\label{ave_equivalence_cond_3}
    $ \frac{N^{out}}{N^{in}}= \frac{|\mathcal{N}^{in}_{ik}|}{|\mathcal{N}^{out}_{jk}|}$ 
    for any pair $jk \in \mathcal{N}^{out}_k$ and $ik \in \mathcal{N}^{out}_{jk}$ in cluster $k=1, \dots,K$. 
\end{condition}
Condition \ref{ave_equivalence_cond_1} demonstrates that if the pairwise spillover effects are homogeneous across different units within each cluster and $\mathcal{G}$ is undirected, then $\tau^{out}(\alpha)$ is equivalent to $\tau^{in}(\alpha)$. Condition \ref{ave_equivalence_cond_2} indicates that when the pairwise spillover effect is homogeneous across different units, regardless of the graph structure, $\tau^{out}(\alpha)$ is equal to $\tau^{in}(\alpha)$. { Condition \ref{ave_equivalence_cond_3} asserts that if the ratio between the out-degree of those with out-degree greater than one, and the in-degree of their out-neighbors is homogeneous and equal to the ratio of $N^{out}$ to $N^{in}$, then ${\tau}^{out}(\alpha) = {\tau}^{in}(\alpha)$. This condition is irrespective of the values of the pairwise spillover effects.} 
The proofs for Conditions \ref{ave_equivalence_cond_1}, \ref{ave_equivalence_cond_2}, and \ref{ave_equivalence_cond_3} leading to the equivalence of the two causal estimands are provided in Supplementary Material \ref{append_comp_out_in_spillover}. 
Examples satisfying Conditions \ref{ave_equivalence_cond_1} and \ref{ave_equivalence_cond_2} are presented in Supplementary Material \ref{ave_spillover_example}.

{ To illustrate when these conditions hold, consider a structural model for the potential outcomes that is linear and additive in a unit's own treatment and the treatments of its neighbors. If the coefficient on each neighbor's treatment is constant across units within each cluster, then Condition~\ref{ave_equivalence_cond_1} holds. If this coefficient is constant across all units and clusters, then Condition~\ref{ave_equivalence_cond_2} holds.}
Notably, Conditions \ref{ave_equivalence_cond_1} and \ref{ave_equivalence_cond_2} are proposed on average potential outcomes, allowing for more flexibility in the underlying structures of the potential outcomes. Condition \ref{ave_equivalence_cond_3}, on the other hand, holds when the graph is regular within clusters and all units have the same degree across clusters. { We introduce Example~\ref{exa_tau_out_equal_tau_in_2} as an illustration of this scenario. However, this condition can also hold for non-regular directed graphs, as shown in Example~\ref{examp2_ave_equivalence_cond_3}. Both examples are provided in Supplementary Material~\ref{ave_spillover_example}.}


For conditional out- and in-spillover effects, there are corresponding theorems and conditions that parallel those discussed in this section. The details are provided in Supplementary Material \ref{comp_cond_out_in_spillover}.

\section{Inference}
\label{sec:Inference}
In this section, we develop Horvitz-Thompson estimators for assessing both types of spillover effects. We derive design-based variance estimators and establish the consistency and asymptotic normality.
We assume the realized treatment assignment mechanism, governed by a vector of parameters $\beta$, is known, while the hypothetical treatment assignment, marginalizing our spillover effects in Definitions \ref{out_spillover} and \ref{in_spillover}, is governed by a vector of parameters $\alpha$. For both the realized and hypothetical treatment assignment mechanisms, the individual assignment probability may depend on covariates or may be constant, and the assignment  may be dependent. 
We develop the estimators according to our definitions of the  outward and inward spillover effects. 

\begin{definition}[Estimator for $\tau^{out}(\alpha)$]
\label{est_tau_out} For $k=1, \dots, K$ and $jk \in \mathcal{N}^{out}_{k}$, { define the average observed outcome among the out-neighbors of unit $jk$ as $\bar{Y}_{jk}^{out}
=
{|\mathcal{N}^{out}_{jk}|^{-1}}
\sum_{ik \in \mathcal{N}^{out}_{jk}} Y_{ik}$ and the estimator weight as $W_{jk}(\mathbf{Z}_k)
=
{\mathbb{P}_{\alpha}(\mathbf{Z}_{-jk})}/
{\mathbb{P}_{\beta}(\mathbf{Z}_k)}$ if $\mathbb{P}_{\beta}(\mathbf{Z}_k)>0$, and $W_{jk}(\mathbf{Z}_k)=0$ otherwise. Define the Horvitz--Thompson estimator of $\tau^{out}_{jk}(\alpha)$ as $\hat{\tau}^{out}_{jk}(\alpha)
:=
W_{jk}(\mathbf{Z}_k)
\left[Z_{jk}-(1-Z_{jk})\right]
\bar{Y}^{out}_{jk}$. The Horvitz--Thompson estimator for $\tau^{out}(\alpha)$ is 
\begin{small}
\begin{equation*}
    \begin{split}
     \hat{\tau}^{out}(\alpha) & =  \frac{1}{N^{out}}  \sum_{k=1}^K \sum_{jk \in  \mathcal{N}^{out}_k} \hat{\tau}^{out}_{jk}(\alpha)   =\sum_{k=1}^K \sum_{jk \in \mathcal{N}^{out}_{k}} \sum_{ik \in \mathcal{N}^{out}_{jk}} \frac{1}{N^{out} |\mathcal{N}^{out}_{jk}| } W_{jk}(\mathbf{Z}_k) \left[ Z_{jk}-(1-Z_{jk})  \right] Y_{ik}.\\ 
    \end{split}
\end{equation*}
\end{small}} 
\end{definition}
$\hat{\tau}^{out}(\alpha)$ is the difference between two weighted averages over the  $N^{out}$ units  of the average observed outcomes of out-neighbors. 
The weight \( W_{jk}(\mathbf{Z}_k) \) represents the ratio between the probability of observing the realized treatment vector $\mathbf{Z}_{-jk}$ in cluster $k$ excluding unit $jk$ under the hypothetical assignment mechanism $\alpha$ and the probability of the realized treatment in cluster \( k \) under the realized assignment mechanism $\beta$. This weight \( W_{jk}(\mathbf{Z}_k) \) is a random variable depending on the realization of \( \mathbf{Z}_k \). 
In the simplest case where the realized treatment assignment mechanism coincides with the hypothetical one, the weights of the Horvitz-Thompson estimators simplify \citep{wang2024designbasedinferencespatialexperiments}.
%
For any assignment mechanism, observed and hypothetical, the Horvitz-Thompson estimator for outward spillover effects in Definition \ref{est_tau_out} can also be implemented using a weighted regression with outcome $\bar{Y}^{out}_{jk}$ on the treatment $Z_{jk}$ with weight $W_{jk}(\mathbf{Z}_k)$ for $jk \in \mathcal{N}^{out}_{k}$, with $k=1, \dots,K$ \citep{wang2024designbasedinferencespatialexperiments}.
\begin{definition}[Estimator for $\tau^{in}(\alpha)$]
\label{est_tau_in}
For $k = 1, \dots, K$ and each $ik \in \mathcal{N}^{in}_k$, let $\tilde{W}^1_{ik}(\mathbf{Z}_k)= {|\mathcal{N}^{in}_{ik}|^{-1}} \sum_{jk \in \mathcal{N}^{in}_{ik}} W_{jk}(\mathbf{Z}_k) Z_{jk}$ and $
\tilde{W}^0_{ik}(\mathbf{Z}_k)= {|\mathcal{N}^{in}_{ik}|^{-1}} \sum_{jk \in \mathcal{N}^{in}_{ik}} W_{jk}(\mathbf{Z}_k) (1-Z_{jk})$, where $W_{jk}(\mathbf{Z}_k)$ is as in Definition \ref{est_tau_out}. {  Define $\hat{\tau}^{in}_{ik}(\alpha):= \tilde{W}^1_{ik}(\mathbf{Z}_k) Y_{ik}   -  \tilde{W}^0_{jk}(\mathbf{Z}_k) {Y}_{ik}$.  Then the Horvitz--Thompson Estimator for $\tau^{in}(\alpha)$ is
  \begin{equation*}
    \begin{split}
\hat{\tau}^{in}(\alpha) & = \frac{1}{N^{in}}  \sum_{k=1}^K \sum_{ik \in  \mathcal{N}^{in}_k} \hat{\tau}^{in}_{ik}(\alpha)= \sum_{k=1}^K \sum_{ik \in \mathcal{N}^{in}_k }\sum_{jk \in \mathcal{N}^{in}_{ik}} \frac{1}{N^{in }|\mathcal{N}^{in}_{ik}|} W_{jk}(\mathbf{Z}_k) \left[ Z_{jk}-(1-Z_{jk})  \right] Y_{ik}. \\
    \end{split}
\end{equation*} } 
\end{definition}
{  By Definition~\ref{in_spillover}, the individual inward spillover effect \(\tau_{ik}(\alpha)\) averages, over the in-neighbors of unit \(ik\), the effect of changing one in-neighbor's treatment on unit \(ik\)'s average potential outcome. Accordingly, the Horvitz--Thompson estimator \(\hat{\tau}^{in}(\alpha)\) in Definition~\ref{est_tau_in} is the difference between two weighted averages of the observed outcomes. The weight assigned to unit \(ik\) is obtained by averaging \(W_{jk}(\mathbf{Z}_k)\) over its treated or untreated in-neighbors, respectively.}

{  Analogously to ${\tau}^{out}(\alpha)$ and ${\tau}^{in}(\alpha)$, we compare $\hat{\tau}^{out}(\alpha)$ and $\hat{\tau}^{in}(\alpha)$. Only Condition \ref{ave_equivalence_cond_3} is sufficient for their equivalence (see Lemma \ref{est_out_in_equivalence} and its proof in Supplementary Material \ref{append_sec:Inference}); Conditions \ref{ave_equivalence_cond_1} and \ref{ave_equivalence_cond_2} are not, as they rely on pairwise spillover effects, which these estimators do not involve. Thus, $\hat{\tau}^{out}(\alpha)$ can differ from $\hat{\tau}^{in}(\alpha)$ even when ${\tau}^{out}(\alpha)={\tau}^{in}(\alpha)$. 
}

In the following propositions, we detail the properties of the two estimators $\hat{\tau}^{out}(\alpha)$ and $\hat{\tau}^{in}(\alpha)$. 
To establish these properties, we first introduce an overlap assumption, which ensures overlap between hypothetical and realized treatment assignments. This overlap is a necessary condition for proving the unbiasedness of the estimators.
\begin{assumption}[Overlap between hypothetical and realized treatment assignments]
    \label{unif_bound_weight}
For all \( \mathbf{z}_k \in \{0,1\}^{n_k} \) such that \( \mathbb{P}_{\alpha}(\mathbf{Z}_k = \mathbf{z}_k) > 0 \), it holds that \( \mathbb{P}_{\beta}(\mathbf{Z}_k = \mathbf{z}_k) > 0 \), for \( k = 1, \dots, K \).
\end{assumption}
Assumption \ref{unif_bound_weight} excludes scenarios where \( \mathbb{P}_{\alpha}(\mathbf{Z}_k = \mathbf{z}_k) > 0 \) but \( \mathbb{P}_{\beta}(\mathbf{Z}_k = \mathbf{z}_k) = 0 \), ensuring that any treatment vector with positive probability under the hypothetical assignment mechanism also has positive probability under the realized assignment mechanism. However, the assumption permits cases where \( \mathbb{P}_{\alpha}(\mathbf{Z}_k = \mathbf{z}_k) = 0 \) and \( \mathbb{P}_{\beta}(\mathbf{Z}_k = \mathbf{z}_k) > 0 \), allowing for more possible realized treatment vectors under the realized mechanism than under the hypothetical mechanism. Intuitively, this assumption ensures that the realized treatment assignment encompasses a broader range of possible treatment vectors compared to the hypothetical assignment. Combined with the definition of \( W_{jk}(\mathbf{Z}_k) \), this assumption guarantees the unbiasedness of the estimators for outward and inward spillover effects, as established below.

\begin{proposition}[Unbiasedness]
\label{unbiasedness}
   Under Assumptions \ref{part_intf} -- \ref{unif_bound_weight}, the proposed estimators satisfy \[
   \mathbb{E}_{\mathbf{Z}|\beta} \left( \hat{\tau}^{out}(\alpha) \right) = \tau^{out}(\alpha) \quad \text{and} \quad \mathbb{E}_{\mathbf{Z}|\beta} \left( \hat{\tau}^{in}(\alpha) \right) = \tau^{in}(\alpha).
   \] 
\end{proposition}

To conduct inference, we derive the Central Limit Theorem (CLT) for both estimators. This derivation relies on Lemma $1$ from \cite{ogburn2022causal} and Lemma A.5 from \cite{wang2024designbasedinferencespatialexperiments}, which is restated in Lemma \ref{Theorem_CLT_dependence} of Supplementary Material. Additionally, we introduce the following assumptions to ensure valid inference.

\begin{assumption}[Positivity of realized treatments]
\label{pos_realized_treatment}
For any treatment vector \( \mathbf{Z}_k \in \{0,1\}^{n_k} \) such that \( \mathbb{P}_{\beta}(\mathbf{Z}_k) > 0 \), there exists a constant \( c > 0 \) such that \( \mathbb{P}_{\beta}(\mathbf{Z}_k) \geq c \) for all \( k \in \{1,\dots, K\} \).    
\end{assumption}

Assumption \ref{pos_realized_treatment} ensures that the probability of observing any feasible treatment assignment is uniformly bounded below by a positive constant across clusters, as is typically assumed under partial interference \citep{papadogeorgou2019causal, park2022efficient}. Notably, this assumption imposes no restrictions on treatment assignments that are impossible under the realized assignment mechanism, i.e., those for which \( \mathbb{P}_{\beta}(\mathbf{Z}_k) = 0 \). This flexibility is particularly relevant in clustered treatment assignment settings, where certain treatment allocations may be infeasible. Moreover, the assumption does not preclude that $\mathbb{P}_\beta (\mathbf{Z}_k)$ may vary with $k$. The definition of \( W(\mathbf{Z}_k) \), together with Assumption \ref{pos_realized_treatment}, ensures that \( W(\mathbf{Z}_k) \) is bounded, satisfying \( 0 \leq W(\mathbf{Z}_k) \leq c^{-1} \) for all \( \mathbf{Z}_k \) and \( k \in \{1, \dots, K\} \).

\begin{assumption}[Constant cluster size]
\label{order1_cluster}
The cluster size $n_k=O(1)$ for $k=1, \dots, K$. 
\end{assumption}
Assumption \ref{order1_cluster} assumes that the sizes of the clusters are of the order of a constant. Therefore, the asymptotic behavior is governed by the number of clusters \(K\). We then postulate that the potential outcomes are bounded, as stated in the following assumption.
\begin{assumption}[Bounded potential outcomes]
    \label{unif_bound_pot_out} 
  For each unit $i = 1, \cdots, n_k$ and $k = 1, \cdots, K$, there exists a constant $C \in \mathbb{R}^{+}$ such that for any $\mathbf{z}_{k}\in \{0,1\}^{n_k}$, the potential outcome $|Y_{ik}(\mathbf{z}_{k})|\leq C$. 
\end{assumption}
Assumption \ref{unif_bound_pot_out} 
serves as a condition for establishing bounded variance for the estimators of both outward and inward spillover effects. We now state the asymptotic normality for $\hat{\tau}^{out}(\alpha)$ and $\hat{\tau}^{in}(\alpha)$. 
\begin{proposition}[Asymptotic normality of spillover effect estimators]
\label{clt_out_in_spillover}
Under Assumptions \ref{part_intf}- \ref{unif_bound_pot_out}, the following hold
{ \begin{equation*}
    \sqrt{N^{b}} \ (\Sigma_N^{b})^{-\frac{1}{2}} \left( \hat{\tau}^{b}(\alpha) - \tau^{b}(\alpha) \right) \overset{d}{\rightarrow} N(0, 1),
\end{equation*}
where $\Sigma_N^{b}=N^{b}V(\hat{\tau}^{b}(\alpha))$ for $b\in\{\mathrm{out},\mathrm{in}\}$. Here, $V(\hat{\tau}^{b}(\alpha))$ denotes the variance of $\hat{\tau}^{b}(\alpha)$, while $N^{out}$ and $N^{in}$ denote the numbers of units having at least one out-neighbor and at least one in-neighbor, respectively.
}
\end{proposition}

In our clustered setting, one unit in cluster $k$ can be at most correlated to all units within the same cluster, which is of constant order based on Assumptions \ref{part_intf} and \ref{order1_cluster}. Therefore, the key dependence condition in Lemma \ref{Theorem_CLT_dependence} is satisfied for both \(\hat{\tau}^{out}(\alpha)\) and \(\hat{\tau}^{in}(\alpha)\). The formulas for \( V(\hat{\tau}^{out}(\alpha)) \) and \( V(\hat{\tau}^{in}(\alpha)) \) are provided in Lemma \ref{formula_V_out_V_in} in Supplementary Material \ref{append_sec:Inference}.

{ Following from \cite{wang2024designbasedinferencespatialexperiments}, to construct confidence intervals for both estimators, we utilize conservative variances as follows.}

\begin{proposition}[Conservative variance]
\label{V_out_V_in}
For \( jk \in \mathcal{N}^{out}_k \) and \( k \in \{1, \ldots, K\} \), define  
\[
V_{zjk} = W_{jk}(\mathbf{Z}_k) \, 1\{Z_{jk} = z\} \, \bar{Y}^{out}_{jk}, \quad S_{zjk} = W_{jk}(\mathbf{Z}_k) \, 1\{Z_{jk} = z\} \, \widetilde{Y}^{in}_{jk}
\]  where $z \in \{0, 1\}$, $\bar{Y}^{out}_{jk}$ is defined in Theorem \ref{est_tau_out}, and $\widetilde{Y}^{in}_{jk}:=\sum_{ik \in \mathcal{N}^{out}_{jk}} |\mathcal{N}^{in}_{ik}|^{-1} Y_{ik}$. Then under Assumptions \ref{part_intf}, \ref{unif_bound_weight} and \ref{unif_bound_pot_out} , a conservative variance of $\hat{\tau}^{out}(\alpha)$ 
is 
\begin{equation*}
    \begin{split}
         V^c(\hat{\tau}^{out}(\alpha)) & = \frac{1}{(N^{out})^2} \sum_{k=1}^K \sum_{jk \in \mathcal{N}^{out}_k} |\mathcal{N}^{out}_{k}| \  \mathbb{E}_{\mathbf{Z}|\beta} \left(V_{1jk}^2+V_{0jk}^2 \right) \\
    \end{split}
\end{equation*}
and a conservative variance of $\hat{\tau}^{in}(\alpha)$ is 
\begin{equation*}
    \begin{split}
    {V}^c(\hat{\tau}^{in}(\alpha)) & = \frac{1}{ (N^{in})^2 } \sum_{k=1}^K \sum_{jk \in \mathcal{N}^{out}_k} |\mathcal{N}^{out}_k|  \cdot \mathbb{E}_{\mathbf{Z}|\beta} \left(S^2_{1jk}+S^2_{0jk}\right). 
    \end{split}
\end{equation*} 
\end{proposition}
The form of $V^c(\hat{\tau}^{in}(\alpha))$ is similar to that of $V^c(\hat{\tau}^{out}(\alpha))$ based on the equivalent form of $\hat{\tau}^{in}(\alpha)$ provided in equation \eqref{prof_difference_out_in_int_1} in the Supplementary Material. 
For both $V^c(\hat{\tau}^{out}(\alpha))$ and $V^c(\hat{\tau}^{in}(\alpha))$, the covariance among different units within the same cluster is accounted by $|\mathcal{N}^{out}_k|$ for $k=1, \dots, K$, { where $\mathcal{N}^{out}_k$ is defined in Definition~\ref{out_spillover}. Consequently, larger values of $|\mathcal{N}^{out}_k|$ lead to larger conservative variance.

We use the conservative variances rather than the exact variances because certain terms in \( V(\hat{\tau}^{out}(\alpha)) \) and \( V(\hat{\tau}^{in}(\alpha)) \) are not identifiable, a challenge similarly noted in \cite{aronow2017estimating}. Specifically,
\(V(\hat{\tau}^{out}(\alpha))\) contains
\(\mathbb{E}(V_{0jk})\mathbb{E}(V_{1jk})\), whereas
\(V(\hat{\tau}^{in}(\alpha))\) contains
\(\mathbb{E}(S_{0jk})\mathbb{E}(S_{1jk})\), for
\(jk\in\mathcal{N}^{out}_k\) and \(k=1,\dots,K\). Neither term is
identifiable.
}



We now consider the corresponding consistent estimators for \(V^c(\hat{\tau}^{out}(\alpha))\) and \(V^c(\hat{\tau}^{in}(\alpha))\). 
\begin{proposition}
\label{est_V_out_V_in}
Under Assumptions \ref{one_in_and_out_neigh}, \ref{pos_realized_treatment}, \ref{order1_cluster} and \ref{unif_bound_pot_out}, a consistent estimator of $ V^c({\hat{\tau}^{out}}(\alpha))$ is 
$\hat{V}^c(\hat{\tau}^{out}(\alpha) ) = {(N^{out})^{-2} } \sum_{k=1}^K \sum_{jk \in \mathcal{N}^{out}_k} |\mathcal{N}^{out}_{k}| \ ( V^2_{1jk}+V^2_{0jk} ) 
    $ and a consistent estimator of $V^c({\hat{\tau}^{in}}(\alpha))$ is 
$
    \hat{V}^c(\hat{\tau}^{in}(\alpha) )  = { (N^{in})^{-2} } \sum_{k=1}^K \sum_{jk \in \mathcal{N}^{out}_k} |\mathcal{N}^{out}_k| \ \left( S^2_{1jk}+ S^2_{0jk} \right)
$, where \( V_{zjk} \) and \( S_{zjk} \) for \( z \in \{0,1\} \), \( jk \in \mathcal{N}^{out}_{k} \), and \( k \in \{1, \dots, K\} \) are as defined in Proposition \ref{V_out_V_in}.
\end{proposition}

{ To clarify the factors driving the difference between
\(V^c(\hat{\tau}^{\cdot}(\alpha))\) and
\(V(\hat{\tau}^{\cdot}(\alpha))\), we derive a lower bound for the bias of
\(V^c(\hat{\tau}^{\cdot}(\alpha))\) relative to
\(V(\hat{\tau}^{\cdot}(\alpha))\); see
Proposition~\ref{dif_V_var} in the Supplementary Material.} 

For conditional outward and inward spillover effects, inference is similar to that for average effects, with two key differences: (1) the estimators are restricted to units with specific  covariate values or units whose in-neighbors have specific covariate values, respectively; (2) the asymptotic results rely on the assumption that the probability of units having specific covariate levels is bounded below by a positive constant. However, since the proofs for the average outward and inward spillover effects can be readily extended to the conditional effects by adjusting for these differences, the details on inference for conditional spillover effects are omitted.

\section{Comparison of variances}
\label{Comp_var}
When \( \tau^{out}(\alpha) \) equals \( \tau^{in}(\alpha) \), either \( \hat{\tau}^{out}(\alpha) \) or \( \hat{\tau}^{in}(\alpha) \) can be utilized to estimate both causal estimands. 
In such scenario,  both \(\hat{\tau}^{out}(\alpha)\) and \(\hat{\tau}^{in}(\alpha)\) are unbiased estimators, as established in Proposition \ref{unbiasedness}.
However, 
the efficiency of these estimators may differ substantially, depending on factors such as graph structures and pairwise spillover effects. We therefore compare \(V(\hat{\tau}^{out}(\alpha))\) with \(V(\hat{\tau}^{in}(\alpha))\), and \(V^c(\hat{\tau}^{out}(\alpha))\) with \(V^c(\hat{\tau}^{in}(\alpha))\).

Comparing the conservative variances \(V^c(\hat{\tau}^{out}(\alpha))\) and \(V^c(\hat{\tau}^{in}(\alpha))\) is important for two reasons: (1) they are the quantities we can actually estimate; (2) the scaled \(V^c(\hat{\tau}^{out}(\alpha))\) and \(V^c(\hat{\tau}^{in}(\alpha))\) are components of \(V(\hat{\tau}^{out}(\alpha))\) and \(V(\hat{\tau}^{in}(\alpha))\), respectively. When these components dominate, they effectively reflect the overall scales of \(V(\hat{\tau}^{out}(\alpha))\) and \(V(\hat{\tau}^{in}(\alpha))\).

{ This section is therefore organized as follows. We first characterize the difference between the conservative variances, which has a simpler structure and is estimable. We then give a proposition that provides an interpretable sufficient condition linking the conservative variance comparison to the graph structure, estimand weights, and variation in potential outcomes. Finally, we investigate the difference between the exact variances, identifying the additional sources contributing to the variance comparison and relating them to the comparison of conservative variances.}
{ 
\begin{proposition}[Difference between $V^c(\hat{\tau}^{out}(\alpha))$ and $V^c(\hat{\tau}^{in}(\alpha))$]
\label{V_out_in_compare}
    Let \( V_{jk} = V_{1jk} - V_{0jk} \) and \( S_{jk} = S_{1jk} - S_{0jk} \) where \( V_{zjk} \) and \( S_{zjk} \) for \( z \in \{0,1\} \), \( jk \in \mathcal{N}^{out}_{k} \), and \( k \in \{1, \dots, K\} \) are as defined in Proposition \ref{V_out_V_in}. The estimand weights $S^{out}_{ik,jk}$ and $S^{in}_{ik,jk}$ are as defined in Theorem \ref{difference_out_in}. Then the difference between $V^c(\hat{\tau}^{out}(\alpha))$ and $V^c(\hat{\tau}^{in}(\alpha))$ is as follows 
\begin{small}
\begin{equation}
\label{V_out_in_compare_eq}
    \begin{split}
        & V^c(\hat{\tau}^{out}(\alpha))-V^c(\hat{\tau}^{in}(\alpha)) =  \sum_{k=1}^K \sum_{jk \in \mathcal{N}^{out}_k } \sum_{z=0}^1 \frac{|\mathcal{N}^{out}_k|}{N^{out \ 2 }} \mathbb{E}_{\mathbf{Z}|\beta}(V^2_{zjk})-  \sum_{k=1}^K \sum_{jk \in \mathcal{N}^{out}_k } \sum_{z=0}^1 \frac{|\mathcal{N}^{out}_k|}{N^{in \ 2 }}  \mathbb{E}_{\mathbf{Z}|\beta}(S^2_{zjk})\\
        &= \sum_{k=1}^K \sum_{jk \in \mathcal{N}^{out}_k} \sum_{z=0}^1 |\mathcal{N}^{out}_k| \cdot \mathbb{E}_{\mathbf{Z}|\beta}\Big\lbrace W^2_{jk}(\mathbf{Z}_k)
1\{Z_{jk}=z\}  [
\sum_{ik\in\mathcal{N}^{out}_{jk}}
(S^{out}_{ik,jk}-S^{in}_{ik,jk})Y_{ik}
]\cdot[
\sum_{ik\in\mathcal{N}^{out}_{jk}}
(S^{out}_{ik,jk}+S^{in}_{ik,jk})
Y_{ik}]\Big\rbrace.\\
    \end{split}
\end{equation}
\end{small}
\end{proposition}
Proposition~\ref{V_out_in_compare} compares the conservative variances of
\(\hat{\tau}^{out}(\alpha)\) and \(\hat{\tau}^{in}(\alpha)\). The first line
of \eqref{V_out_in_compare_eq} compares the second moments of the aggregated
out-neighbor outcomes under the outward and inward estimand weights,
conditional on the sender having treatment status \(z\). The second line of
\eqref{V_out_in_compare_eq} provides a factorization of this difference. This
factorization shows that the conservative variance difference is governed by
two components: the difference and sum of the outward and inward estimand
weights, and their interaction with the outcomes within the sender's
out-neighborhood. 
We explore scenarios where one conservative variance is larger than the other by varying the graph structure. Detailed simulations are given in Section \ref{estimand_weight_efficiency} and in Supplementary Material~\ref{append:estimand_weight_efficiency} and \ref{Append:simulation_var_compare}.
} 

Moreover, Proposition \ref{V_out_in_compare} implies that Condition \ref{ave_equivalence_cond_3} is not only sufficient for ensuring \( \tau^{out}(\alpha) = \tau^{in}(\alpha) \) but also for equating \( V^c(\hat{\tau}^{out}(\alpha)) \) with \( V^c(\hat{\tau}^{in}(\alpha)) \). { This equivalence arises because \( S^{out}_{ik,jk}-S^{in}_{ik,jk} = 0 \) for all \(ik \in \mathcal{N}^{out}_{jk},\ jk \in \mathcal{N}^{out}_{k}\) and \(k = 1,\dots,K\)}, as stipulated by Condition \ref{ave_equivalence_cond_3}. For instance, if the graph is regular, as in Example \ref{exa_tau_out_equal_tau_in_2}, this results in \( V^{c}(\hat{\tau}^{out}(\alpha)) = V^{c}(\hat{\tau}^{in}(\alpha)) \). We also present an example of star graphs that satisfy Condition \ref{ave_equivalence_cond_3} and result in \( V^{c}(\hat{\tau}^{out}(\alpha)) = V^{c}(\hat{\tau}^{in}(\alpha)) \) in Supplementary Material \ref{ave_spillover_example}. 

{ 
Proposition~\ref{V_out_in_compare} identifies the main factors that determine the sign of $V^c(\hat{\tau}^{out}(\alpha))
-
V^c(\hat{\tau}^{in}(\alpha))$. However, how these factors influence the direction of the comparison is not readily apparent from Proposition~\ref{V_out_in_compare}. We therefore present the following proposition, which establishes an interpretable sufficient condition linking the conservative variance comparison to the graph structure, the estimand weights, and the variation in potential outcomes.

\begin{proposition}
\label{general_suff_cond_sender_type_3}
For each receiver--sender pair $(ik,jk)$, define the outward and inward estimand weights as in Theorem \ref{difference_out_in}. 
Then for each sender $jk\in\mathcal{N}^{out}_k$, partition its out-neighbors as $$H_{1,jk}
=
\bigl\{ik\in\mathcal{N}^{out}_{jk} : S^{in}_{ik,jk} \leq S^{out}_{ik,jk}\bigr\}, \quad H_{2,jk}
=
\bigl\{ik\in\mathcal{N}^{out}_{jk} : S^{in}_{ik,jk} > S^{out}_{ik,jk}\bigr\}.$$ Define $H_1
=
\bigcup_{k=1}^K
\bigcup_{jk\in\mathcal{N}^{out}_k}
\bigl\{(ik,jk) : ik\in H_{1,jk}\bigr\}$, $H_2
=
\bigcup_{k=1}^K
\bigcup_{jk\in\mathcal{N}^{out}_k}
\bigl\{(ik,jk) : ik\in H_{2,jk}\bigr\}$,
and assume $|H_2|>0$. Let 
\begin{small}$$\Delta_{H_1}
=
\frac{1}{|H_1|}
\sum_{k=1}^K
\sum_{jk\in\mathcal{N}^{out}_k}
\sum_{ik\in H_{1,jk}}
\bigl[
(S^{out}_{ik,jk})^2
-
(S^{in}_{ik,jk})^2
\bigr], \quad \Delta_{H_2}
=
\frac{1}{|H_2|}
\sum_{k=1}^K
\sum_{jk\in\mathcal{N}^{out}_k}
\sum_{ik\in H_{2,jk}}
\bigl[
(S^{in}_{ik,jk})^2
-
(S^{out}_{ik,jk})^2
\bigr],$$ \end{small} and define the ratio $R_s
=
\frac{|H_1|\,\Delta_{H_1}}{|H_2|\,\Delta_{H_2}}$. Suppose that the following conditions hold:
\begin{enumerate}
    \item[(i)] There exist constants $0 < a \leq b < \infty$ such that
    $a^2 \leq Y^2_{ik}(\mathbf{z}_k) \leq b^2$ for all $i = 1,\ldots,n_k$ and $k = 1,\ldots,K$.
    
    \item[(ii)] The hypothetical and realized treatment assignments are independent Bernoulli assignments with probability $\alpha = \beta$.
\end{enumerate}
If $R_s > b^2/a^2$, then $V^c(\hat{\tau}^{out}(\alpha)) > V^c(\hat{\tau}^{in}(\alpha))$. If $R_s < a^2/b^2$, then $V^c(\hat{\tau}^{out}(\alpha)) < V^c(\hat{\tau}^{in}(\alpha))$.
\end{proposition}
Proposition \ref{general_suff_cond_sender_type_3} gives sufficient conditions under which the conservative variance of one estimator dominates that of the other. The result separates the comparison into three interpretable components. The first is the relative size $|H_1|/|H_2|$ of the pairwise receiver--sender sets on which the outward weights dominate the inward weights, or vice versa. The second is the ratio $\Delta_{H_1}/\Delta_{H_2}$, which measures the relative average squared difference between the outward and inward weights over these two sets. The third is the outcome variation ratio $b^2/a^2$ and $a^2/b^2$, which controls how much heterogeneity in the potential outcomes. Together, these components determine $R_s={(|H_1|\Delta_{H_1})}/{(|H_2|\Delta_{H_2})}$. 

When \(R_s>b^2/a^2\), at least one of the following features are sufficiently satisfied: (i) the number of receiver--sender pairs for which the outward weight dominates the inward weight, is large, relative to the number of pairs for which the reverse holds, so that \(|H_1|/|H_2|\) is large; or (ii) the average squared weight difference over \(H_1\) is large relative to that over \(H_2\), so that \(\Delta_{H_1}/\Delta_{H_2}\) is large. Either mechanism, or both jointly, makes $R_s$ sufficiently large to imply $V^c(\hat{\tau}^{out}(\alpha))
>
V^c(\hat{\tau}^{in}(\alpha))$. 
Conversely, when $R_s<a^2/b^2$, the dominance is reversed, and $V^c(\hat{\tau}^{out}(\alpha))
<
V^c(\hat{\tau}^{in}(\alpha))$.

Proposition~\ref{general_suff_cond_sender_type_3} provides a testable condition for preferring one estimator over the other, since $R_s$, $a$, and $b$ are estimable. If a large number of receiver--sender pairs satisfy $S^{in}_{ik,jk} \leq S^{out}_{ik,jk}$, so that the set \(H_1\) is dominant, then Proposition~\ref{general_suff_cond_sender_type_3} implies $V^c(\hat\tau^{out}(\alpha))
>
V^c(\hat\tau^{in}(\alpha))$. In this case, the inward estimator has the smaller conservative variance. Conversely, if many pairs satisfy $S^{out}_{ik,jk}<S^{in}_{ik,jk}$, so that the set \(H_2\) is dominant, then $V^c(\hat\tau^{out}(\alpha))
<
V^c(\hat\tau^{in}(\alpha))$, and the outward estimator is preferable. When \(H_1\) and \(H_2\) are of comparable size, or when the pairwise discrepancies between \(S^{out}_{ik,jk}\) and \(S^{in}_{ik,jk}\) are small, the graph structure does not yield a clear ordering. In such cases, the relative efficiency of the two estimators, and thus the choice between them, is ambiguous.
} 

We now investigate the additional factors that characterize the difference between \( V(\hat{\tau}^{out}(\alpha) ) \) and \( V(\hat{\tau}^{in}(\alpha)) \) beyond those captured by \( V^c(\hat{\tau}^{out}(\alpha)) \) and \( V^c(\hat{\tau}^{in}(\alpha)) \). { This analysis also provides insight into the relation between the exact but non-identifiable variance difference, \(V(\hat{\tau}^{out}(\alpha))-V(\hat{\tau}^{in}(\alpha))\), and the estimable conservative variance difference, \(V^c(\hat{\tau}^{out}(\alpha))-V^c(\hat{\tau}^{in}(\alpha))\).} 

\begin{proposition}[Difference between $V(\hat{\tau}^{out})$ and $V(\hat{\tau}^{in})$] 
\label{dif_var_out_var_in}
{  Let \(V_{jk}\), \(S_{jk}\), \(V_{zjk}\), and \(S_{zjk}\), for \(z\in\{0,1\}\), be defined as in Proposition~\ref{V_out_in_compare}. Then the difference between \(V(\hat{\tau}^{out}(\alpha))\) and \(V(\hat{\tau}^{in}(\alpha))\) is  
\begin{equation*}
    \begin{split}
     & V\left( \hat{\tau}^{out}(\alpha) \right)- V(\hat{\tau}^{in}(\alpha) )= (a) + (b) - (c)\\ 
    \end{split}
\end{equation*}
where 
\begin{small}
\begin{equation*}
    \begin{split}
      & (a): =  \frac{1}{N^{out \ 2 }} \sum_{k=1}^K \sum_{jk \in \mathcal{N}^{out}_k }  \left( \mathbb{E}_{\mathbf{Z}|\beta}(V^2_{1jk})+ \mathbb{E}_{\mathbf{Z}|\beta}(V^2_{0jk}) \right) -  \frac{1}{N^{in \ 2 }} \sum_{k=1}^K \sum_{jk \in \mathcal{N}^{out}_k }  \left( \mathbb{E}_{\mathbf{Z}|\beta}(S^2_{1jk})+ \mathbb{E}_{\mathbf{Z}|\beta}(S^2_{0jk}) \right), \\  
    \end{split}
\end{equation*}
\begin{equation*}
    \begin{split}
     (b):= 2
    \sum_{k=1}^K
    \sum_{j_1k \in \mathcal{N}^{out}_k}
    \sum_{\substack{j_2k \in \mathcal{N}^{out}_k\\ j_1k\neq j_2k}}
    \left[
    \frac{1}{(N^{out})^2}
    \operatorname{cov}(V_{j_1k},V_{j_2k})
    -
    \frac{1}{(N^{in})^2}
    \operatorname{cov}(S_{j_1k},S_{j_2k})
    \right], \\  
    \end{split}
\end{equation*}
and 
\begin{equation*}
    \begin{split}
        (c) := \sum_{k=1}^K
    \sum_{jk \in \mathcal{N}^{out}_{k}}
    \left[
    \frac{1}{(N^{out})^2}\bar{\tau}^2_{jk}(\alpha)
    -
    \frac{1}{(N^{in})^2}\widetilde{\tau}^2_{jk}(\alpha) \right]
    \end{split}
\end{equation*}
\end{small}
with $\bar{\tau}_{jk}(\alpha)
=
\sum_{ik\in \mathcal{N}^{out}_{jk}}
\frac{1}{|\mathcal{N}^{out}_{jk}|}\tau_{ik,jk}(\alpha)$ and $\widetilde{\tau}_{jk}(\alpha)
=
\sum_{ik\in \mathcal{N}^{out}_{jk}}
\frac{1}{|\mathcal{N}^{in}_{ik}|}\tau_{ik,jk}(\alpha)$.}
\end{proposition}
{ Proposition~\ref{dif_var_out_var_in} concerns the comparison to the exact variances. Term $(a)$ is the difference between the second moments of the sender-level aggregated components in $\hat{\tau}^{out}(\alpha)$ and $\hat{\tau}^{in}(\alpha)$. When $|\mathcal{N}^{out}_k|$ is homogeneous across clusters, $(a)
={|\mathcal{N}^{out}_k|}^{-1}
\left[
V^c(\hat{\tau}^{out}(\alpha))
-
V^c(\hat{\tau}^{in}(\alpha))
\right]$. Thus, in this special case, term $(a)$ has the same sign as the difference between the two conservative variances. Term $(b)$ is the difference between the covariance contributions in $\hat{\tau}^{out}(\alpha)$ and $\hat{\tau}^{in}(\alpha)$. 
Term $(c)$ is the difference between the squared first moments of the sender-level aggregated pairwise spillover effects, where $\bar{\tau}_{jk}(\alpha)$ uses outward spillover estimand weights and $\widetilde{\tau}_{jk}(\alpha)$ uses inward spillover estimand weights. 

Because of the additional terms \((b)\) and \((c)\), the signs of \(V^c(\hat{\tau}^{out}(\alpha)) - V^c(\hat{\tau}^{in}(\alpha))\) and \(V(\hat{\tau}^{out}(\alpha)) - V(\hat{\tau}^{in}(\alpha))\) may reverse, even when \(|\mathcal{N}^{out}_k|\) is homogeneous across \(k \in \{1,\ldots,K\}\). The possible reversal in sign between them can be due to two opposing mechanisms operate. First, within each exact variance, the corresponding estimand weights may affect the variance and covariance components in opposite directions. In particular, within
\(V(\hat{\tau}^{out}(\alpha))\), the outward weights
\(S^{out}_{ik,jk}\) enter the within-sender variance component and the across-sender covariance component differently, so that a change in these weights may increase one component while decreasing the other. The same consideration applies to the inward weights
\(S^{in}_{ik,jk}\) within \(V(\hat{\tau}^{in}(\alpha))\). Second, the outward weights affect the first terms in components (a), (b), and (c), whereas the inward weights affect the corresponding second terms in these components 
need not be aligned. The resulting comparison therefore depends not only on how the weights influence the components within each exact variance, but also on how the outward and inward weighting schemes differ across the two exact variances. When these opposing effects are sufficiently strong, they provide an explanation for why $V^c(\hat{\tau}^{out}(\alpha))
-
V^c(\hat{\tau}^{in}(\alpha))$ and $V(\hat{\tau}^{out}(\alpha))
-
V(\hat{\tau}^{in}(\alpha))$ have different signs. 

When \( |\mathcal{N}^{out}_{k}| \) varies across \(k=1,\ldots,K\), the difference between the expression in \eqref{V_out_in_compare_eq} of Proposition~\ref{V_out_in_compare} and term \((a)\) lies in their cluster-specific scaling. Each summand in \eqref{V_out_in_compare_eq} is multiplied by \( |\mathcal{N}^{out}_{k}| \)-- which cannot be factored out due to its heterogeneity--
whereas the summands in term \((a)\) remain unscaled. Consequently, $V^c(\hat{\tau}^{out}(\alpha))
-
V^c(\hat{\tau}^{in}(\alpha))$ is not a constant multiple of term \((a)\). 
Therefore, the heterogeneous cluster-specific scaling, together with additional terms \((b)\) and \((c)\), can possibly lead the conservative and exact variance differences to have opposite signs.

It is also worth noting that, when \(|\mathcal{N}^{out}_k|\) is homogeneous across clusters, Condition~\ref{ave_equivalence_cond_3} is sufficient for \((a)=(b)=(c)=0\). Indeed, these terms involve differences of the form $S^{out}_{i_1k,j_1k}S^{out}_{i_2k,j_2k}
-
S^{in}_{i_1k,j_1k}S^{in}_{i_2k,j_2k}$, and Condition~\ref{ave_equivalence_cond_3} implies that $S^{out}_{i_1k,j_1k}=S^{in}_{i_1k,j_1k}$ and $S^{out}_{i_2k,j_2k}=S^{in}_{i_2k,j_2k}$. Consequently, under Condition~\ref{ave_equivalence_cond_3}, $V(\hat{\tau}^{out}(\alpha))
-
V(\hat{\tau}^{in}(\alpha))
=
0$. Equivalently, Lemma~\ref{est_out_in_equivalence} in the Supplementary Material shows that Condition~\ref{ave_equivalence_cond_3} is sufficient for $\hat{\tau}^{out}(\alpha)=\hat{\tau}^{in}(\alpha)$, which also implies equality of their variances.

} 

{ 
Finally, Propositions~\ref{V_out_in_compare} and~\ref{dif_var_out_var_in} show that, even when the two target estimands coincide, the corresponding estimators $\hat{\tau}^{out}(\alpha)$ and $\hat{\tau}^{in}(\alpha)$ may have different efficiencies. This motivates considering whether efficiency improves by using a weighted sum of the two estimators. We derive the weight that minimizes the conservative variance of this estimator and provide a sufficient condition under which the optimal weight lies strictly between $0$ and $1$
; see Proposition~\ref{prop:opt_lambda} and
Lemma~\ref{suf_cond_lambda_non_trivial} in
Supplementary Material~\ref{append:comb_estimators}.
}

\section{Simulation Studies of Variance Comparisons}
\label{estimand_weight_efficiency}
{ 
This section uses simulation examples to examine how differences between the
outward and inward estimand-weight distributions affect the conservative and
exact variances of the corresponding estimators through a mixture of inward- and outward-star graphs and other graph settings. 
}

{ 
We first consider a simulation design in which the outward and inward estimand-weight distributions can be varied through a single parameter. Specifically, we generate $K=2000$ independent clusters, each of size $n_k=10$. Each cluster is either an out-star graph or an in-star graph. A fraction $t$ of the clusters are in-stars and the remaining fraction $1-t$ are out-stars. In each star graph, the central node has degree $9$, with the direction of all edges determined by whether the graph is an in-star or an out-star. This design allows the distributions of the outward and inward estimand weights, $S^{out}_{ik,jk}$ and $S^{in}_{ik,jk}$, to be controlled through the single parameter $t$. Thus, the simulated population contains $tK$ in-star clusters and $(1-t)K$ out-star clusters. We vary the fraction of in-star clusters over $t\in \{0.1,0.2,\ldots,0.9\}$. Both the hypothetical and realized treatments are assigned independently via Bernoulli trials with probabilities \(\alpha =\beta=0.6\), respectively. The structural model for the potential outcomes is

\begin{equation}
\label{spec_outcome_model}
    \begin{split}
        Y_{ik}(\mathbf{z}_{k}) = \beta_0 + \beta_1 z_{ik} + \beta_2 \sum_{jk \in \mathcal{N}^{in}_{ik}} z_{jk}
    \end{split}
\end{equation}
where $\beta_0=2$, $\beta_1=0.5$ and $\beta_2=0.1$. Since $\beta_2$ is constant across units, the pairwise spillover effects $\tau_{ik,jk}(\alpha)$ are homogeneous. It follows from Condition~\ref{ave_equivalence_cond_2} that $\tau^{out}(\alpha)=\tau^{in}(\alpha)$. The model in \eqref{spec_outcome_model} excludes (fixed) noise terms to easily control the range of potential outcomes in this simulation study. We use $M=500$ Monte Carlo replications. Under the specified coefficient values in the structural model, the lower bound \(a\) and upper bound \(b\) for the potential outcomes defined in Proposition~\ref{general_suff_cond_sender_type_3} satisfy ${b^2}/{a^2}=2.890$, ${a^2}/{b^2}=0.346$. The distributions of $S^{out}_{ik,jk}$ and $S^{in}_{ik,jk}$ are displayed in Figure~\ref{fig:estimand-weight-density-all} in Supplementary Material~\ref{append:estimand_weight_efficiency}. The simulation results are reported in Table~\ref{tab:prop_in_star}.

\begin{table}[ht]
\centering
\small
\setlength{\tabcolsep}{2.5pt}
\renewcommand{\arraystretch}{1.15}
\caption{Comparison of the conservative and exact variances, together with a decomposition of the exact variance difference, across different proportions of in-star graphs. \protect\footnotemark}
\label{tab:prop_in_star}
\resizebox{\textwidth}{!}{
\begin{tabular}{c C{1.4cm} c c c C{1.9cm} c c C{1.7cm} c c c}
\hline
$t$
& ${[\mathbb{E}(S^{out}_{ik,jk})}-{\mathbb{E}(S^{in}_{ik,jk})}]\times 10^{5}$
& $R_s$
& $V^c(\hat{\tau}^{out}(\alpha))$
& $V^c(\hat{\tau}^{in}(\alpha))$
& $V^c(\hat{\tau}^{out}(\alpha))
   -V^c(\hat{\tau}^{in}(\alpha))$
& $V(\hat{\tau}^{out}(\alpha))$
& $V(\hat{\tau}^{in}(\alpha))$
& $V(\hat{\tau}^{out}(\alpha))
   -V(\hat{\tau}^{in}(\alpha))$
& $(a)\times 10^{3}$
& $(b)\times 10^{3}$
& $(c)\times 10^{6}$
\\
\hline
0.1 &  9.401 & 3.098 & 0.045 & 0.012 &  0.033 & 0.008 & 0.012 & -0.004 &  -4.565 & 0.037 &  -2.644 \\
0.2 &  7.377 & 2.247 & 0.041 & 0.014 &  0.028 & 0.006 & 0.013 & -0.008 &  -7.716 & 0.023 &  -3.996 \\
0.3 &  4.745 & 1.688 & 0.036 & 0.015 &  0.021 & 0.005 & 0.015 & -0.011 & -10.278 & 0.019 &  -5.042 \\
0.4 &  2.301 & 1.294 & 0.031 & 0.017 &  0.014 & 0.004 & 0.018 & -0.014 & -12.719 & 0.010 &  -6.040 \\
0.5 &  0.000 & 1.000 & 0.027 & 0.019 &  0.009 & 0.003 & 0.018 & -0.015 & -15.303 & 0.006 &  -7.111 \\
0.6 & -2.301 & 0.773 & 0.024 & 0.022 &  0.003 & 0.003 & 0.021 & -0.018 & -18.217 & 0.005 &  -8.341 \\
0.7 & -4.745 & 0.592 & 0.022 & 0.025 & -0.003 & 0.002 & 0.024 & -0.022 & -21.608 & 0.003 &  -9.786 \\
0.8 & -7.377 & 0.445 & 0.020 & 0.029 & -0.010 & 0.002 & 0.027 & -0.025 & -25.365 & 0.004 & -11.372 \\
0.9 & -9.401 & 0.323 & 0.018 & 0.033 & -0.015 & 0.002 & 0.031 & -0.029 & -27.060 & 0.003 & -12.045 \\
\hline
\end{tabular}
}
\end{table}
\footnotetext{The quantity $\mathbb{E}(S^{out}_{ik,jk})-\mathbb{E}(S^{in}_{ik,jk})$ is averaged over pairs $(ik,jk)$ satisfying $N^{out}|\mathcal{N}^{out}_{jk}|>0$ and $N^{in}|\mathcal{N}^{in}_{ik}|>0$. The quantity $R_s$ is defined in Proposition~\ref{general_suff_cond_sender_type_3}. The conservative variance difference, $V^c(\hat{\tau}^{out}(\alpha))-V^c(\hat{\tau}^{in}(\alpha))$, is defined in Proposition~\ref{V_out_in_compare}; the exact variance difference, $V(\hat{\tau}^{out}(\alpha))-V(\hat{\tau}^{in}(\alpha))$, and its components (a), (b) and (c) are defined in Proposition~\ref{dif_var_out_var_in}.}

The second column of Table~\ref{tab:prop_in_star}, together with
Figure~\ref{fig:estimand-weight-density-all}, shows that, as the proportion
of in-star graphs increases, the distribution of $S^{out}_{ik,jk}$ shifts
towards smaller values, whereas the distribution of $S^{in}_{ik,jk}$ shifts
towards larger values. By the definitions of $\Delta_{H_1}$ and
$\Delta_{H_2}$ in Proposition~\ref{general_suff_cond_sender_type_3}, this
implies that $\Delta_{H_1}$ decreases and $\Delta_{H_2}$ increases. The same change in the two weight distributions also affects the sizes of the sets $H_1$ and $H_2$. By the definitions of the receiver--sender
sets $H_1$ and $H_2$ in Proposition~\ref{general_suff_cond_sender_type_3}, a
concentration of $S^{in}_{ik,jk}$ below $S^{out}_{ik,jk}$ implies that many
receiver--sender pairs belong to $H_1$. Conversely, a concentration of
$S^{out}_{ik,jk}$ below $S^{in}_{ik,jk}$ implies that many such pairs belong
to $H_2$. Hence, as $t$ increases from $0$ to $1$, $|H_1|$ increases whereas
$|H_2|$ decreases. In this setting, $R_s$ can be shown to be decreasing in
$t$; see Lemma~\ref{suffi_cond3_example} in the Supplementary Material.


The numerical results also agree with the threshold comparison in
Proposition~\ref{general_suff_cond_sender_type_3}. Under the coefficient
values used in this simulation. When $R_s>b^2/a^2$, the table gives $V^c(\hat{\tau}^{out}(\alpha))
>
V^c(\hat{\tau}^{in}(\alpha))$, as in the first row of Table~\ref{tab:prop_in_star}. When $R_s<a^2/b^2$, the
table gives $V^c(\hat{\tau}^{out}(\alpha))
<
V^c(\hat{\tau}^{in}(\alpha))$, as in the last row of Table~\ref{tab:prop_in_star}. These findings are in
accordance with Proposition~\ref{general_suff_cond_sender_type_3}.

Moreover, the exact variances show a broadly similar trend to the conservative
variances. As \(t\) increases, both
\(V^c(\hat{\tau}^{out}(\alpha))\) and
\(V(\hat{\tau}^{out}(\alpha))\) decrease, whereas both
\(V^c(\hat{\tau}^{in}(\alpha))\) and
\(V(\hat{\tau}^{in}(\alpha))\) increase. Thus the conservative variances
capture the same qualitative movement in the relative variability of the
outward and inward estimators. The signs of $V(\hat{\tau}^{out}(\alpha))-V(\hat{\tau}^{in}(\alpha))$ and $V^c(\hat{\tau}^{out}(\alpha))-V^c(\hat{\tau}^{in}(\alpha))$ coincide for \(t\geq 0.7\), where both comparisons favor the outward
estimator.

For \(t\leq 0.6\), however, the two comparisons have opposite signs: the conservative variance is larger for the outward estimator, whereas the exact variance is smaller. To identify the source of this discrepancy, we examine components \((a)\), \((b)\), and \((c)\) in Proposition~\ref{dif_var_out_var_in}. The last three columns of Table~\ref{tab:prop_in_star} show that component \((a)\) is substantially larger in magnitude than components \((b)\) and \((c)\). The ordering of the exact variances is therefore driven primarily by component \((a)\).

The difference in signs arises because 
in this star-graph design, \(|\mathcal{N}^{out}_k|\) varies across clusters, so there is no common multiplicative factor relating $V^c(\hat{\tau}^{out}(\alpha))
-
V^c(\hat{\tau}^{in}(\alpha))$ to component \((a)\). As \(t\) decreases from \(0.9\) to \(0.1\), the number of units with out-neighbors decreases, and hence \(|\mathcal{N}^{out}_k|\) becomes smaller. At the same time, the estimand-weight comparison shifts from dominance of the inward weights to dominance of the outward weights; see the second column of Table~\ref{tab:prop_in_star} and Figure~\ref{fig:estimand-weight-density-all} in the Supplementary Material for $t$ from $0.9$ to $0.1$. Accordingly, \(S^{out}_{ik,jk}-S^{in}_{ik,jk}\) shifts from being predominantly negative to predominantly positive. As shown in the second line of
\eqref{V_out_in_compare_eq}, the conservative variance comparison contains contributions proportional to $\bigl(S^{out}_{ik,jk}-S^{in}_{ik,jk}\bigr)
|\mathcal{N}^{out}_k|Y_{ik}$. Because \(Y_{ik}\geq 0\) under the potential-outcome specification in \eqref{spec_outcome_model}, the sign of each contribution is determined by $\bigl(S^{out}_{ik,jk}-S^{in}_{ik,jk}\bigr)
|\mathcal{N}^{out}_k|$. The joint change in the estimand-weight difference and the cluster-specific factor \(|\mathcal{N}^{out}_k|\) causes the conservative variance difference to move from negative to positive, as shown in column~6 of Table~\ref{tab:prop_in_star} for $t$ from $0.9$ to $0.1$. By contrast, the exact variance difference assigns weight $1$ instead of $|\mathcal{N}^{out}_k|$ to these contributions, as discussed after Proposition~\ref{dif_var_out_var_in}. Consequently, as \(t\) decreases from \(0.9\) to \(0.1\), the negative contributions are not attenuated by the decreasing cluster-specific factor \(|\mathcal{N}^{out}_k|\), and the exact variance difference remains negative throughout. These explain why the conservative and exact variance comparisons can have opposite signs here.


We next consider three simulation settings to assess how different design and network parameters affect the estimand weights and the differences between the exact and conservative variances. In first simulation, we set $\alpha=0.5$ and $\beta=0.4$, while keeping the remaining simulation settings unchanged. The differences between the conservative variances and between the exact variances exhibit patterns similar to those reported above; see Table~\ref{tab:prop_in_star_alpha_not_equal_beta} in Supplementary Material~\ref{Append:simulation_var_compare}. In second simulation, we vary the number of clusters $K$, while keeping all other settings fixed. As $K$ increases, the magnitudes of $S^{out}_{ik,jk}$ and $S^{in}_{ik,jk}$ decrease, which in turn reduces the absolute differences between both the conservative variances and the exact variances; see Table~\ref{tab:increase_num_clusters} in Supplementary Material~\ref{Append:simulation_var_compare}. Finally, we consider a more realistic network model based on directed preferential attachment. A proportion $t$ of clusters is generated to favor in-degree attachment, while the remaining $1-t$ favors out-degree attachment. For all $t$, the signs of $V^c(\hat{\tau}^{out}(\alpha))-V^c(\hat{\tau}^{in}(\alpha))$ and $V(\hat{\tau}^{out}(\alpha))-V(\hat{\tau}^{in}(\alpha))$ coincide. The corresponding variance comparisons and decompositions are reported in Table~\ref{tab:prop_pref_attachment} in Supplementary Material~\ref{Append:simulation_var_compare}. This result indicates that the conservative and exact variance orderings can agree under a more heterogeneous and empirically plausible graph structure. Explanations for this agreement are discussed in Section~\ref{Append:simulation_var_compare} of the Supplementary Material.


}

\section{Discussion}
\label{Discussion}
In this paper, we introduce the outward and inward spillover effects, representing the spillover effect of one's treatment on the outcomes of their out-neighbors or the spillover effect on one's outcome from the treatment of one in-neighbor, under partial interference. While \citet{liu2014large}, 
and \cite {park2022efficient}  also define average spillover effects under the partial interference assumption, their focus lies on estimating the spillover effect by comparing two distinct treatment allocation strategies at the cluster level, while our estimands emphasize the impact of altering neighbors' treatment statuses under a certain treatment allocation strategy.
Moreover, unlike \citet{sussman2017elements} and \citet{aronow2017estimating}, 
we do not assume any exposure mapping function; instead, we marginalize over a treatment assignment strategy \( \alpha \) within each cluster. 

{   We derive a necessary and sufficient condition under which the two estimands differ, and we identify conditions under which they coincide. The proposed outward and inward spillover effects can differ because they generally assign different weights to the same set of pairwise spillover effects. We also derive Horvitz--Thompson estimators for the outward and inward spillover effects and compare their efficiencies in terms of both the exact and conservative variances.} The efficiency of these two estimators depends on the weights $W(\mathbf{Z}_k)$. As the cluster size increases, the estimators tend to exhibit instability.
To alleviate this issue, one can consider employing the H\'{a}jek estimators \citep{basu2011essay}, which, although potentially biased, is normalized to reduce variance (\cite{sarndal2003model}). 
These estimators for outward and inward spillover effects are presented in Supplementary Material \ref{hajek}.
Comparing the efficiencies of the H\'{a}jek estimators for outward and inward spillover effects remains challenging, even when the outward and inward spillover effects themselves are equal. 

Furthermore, our estimators provide valid inference when the cluster size is relatively small compared to the number of clusters, such as in small villages, households, and classrooms. 
If the cluster sizes are large relative to the number of clusters, a possible solution is to explore dimension reduction for the interference patterns \citep{hudgens2008toward}, distinguishing between interference from the neighborhood and from others. 

{  
We ultimately provide here recommendations for choosing an estimator for a given outward or inward estimand.
If the two estimands
differ, the estimator should be chosen to match the target estimand. If $\tau^{out}(\alpha)=\tau^{in}(\alpha)$, both estimators
are unbiased for the common estimand, and the choice can instead be guided by
efficiency. One may also consider a weighted combination of the two estimators. We develop the theory for this weighted estimator in the Supplementary Material, Section \ref{append:comb_estimators}, which 
gives the conservative-variance-minimizing
weights, and 
provides sufficient
conditions under which the optimal weight lies strictly between $0$ and $1$. In practice, Condition~\ref{ave_equivalence_cond_3} provides a testable sufficient condition for the equality of the two estimands, whereas Conditions~\ref{ave_equivalence_cond_1} and~\ref{ave_equivalence_cond_2} do not. Finally, even when the two estimands coincide, if preserving the alignment between the estimand weights and the weights used in the estimators is important for interpretation, one may still prefer the estimator corresponding to the estimand of interest. 
}

In addition to average spillover effects, we introduce conditional outward and inward spillover effects in Definitions \ref{cond_out_spillover} and \ref{cond_in_spillover}. These estimands are of great interest, particularly in contexts such as identifying influencers and developing targeting strategies. 
%
The conditions for determining the equivalence or inequivalence of these conditional spillover effects (Proposition \ref{difference_cond_out_in}, Conditions \ref{cond_equivalence_cond_1}, \ref{cond_equivalence_cond_2}, and \ref{cond_equivalence_cond_3}) are analogous to those for average effects. The difference is that Conditions \ref{cond_equivalence_cond_1} and \ref{cond_equivalence_cond_2} require homogeneity in pairwise spillover effects among units sharing a covariate value \( x \), and Condition \ref{cond_equivalence_cond_3} requires units who have neighbors with covariate \( x \) as well as those possessing the covariate \( x \) themselves. For inference, the CLT, conservative variances as well as conservative variance estimators also hold for the estimators of conditional spillover effects for a categorical covariate with the assumption that the probabilities of taking specific value $x$ for all possible $x$ are lower bounded by some constant. 

Three directions for future research are of particular interest. {   First, new outward and inward estimands can be defined using different subsets of units and alternative interference assumptions. Specifically, the characterization in Theorem \ref{difference_out_in} and Conditions \ref{ave_equivalence_cond_1}--\ref{ave_equivalence_cond_3} can be extended beyond partial interference, provided that the definitions of the average potential outcomes and pairwise spillover effects are modified accordingly. For example, if the clustered network structure is retained and treatment may affect units within graph distance $d \geq 1$, Theorem \ref{difference_out_in} and Conditions \ref{ave_equivalence_cond_1}--\ref{ave_equivalence_cond_3} continue to hold, with the average potential outcome $\bar{Y}_{ik}(z_{jk},\alpha)$ taken over the $d$-distance neighborhood rather than over the whole cluster. For the efficiency comparison between $\hat{\tau}^{out}(\alpha)$ and $\hat{\tau}^{in}(\alpha)$ under different interference assumptions, the proofs of Propositions \ref{V_out_in_compare} and \ref{dif_var_out_var_in} can be adapted with suitable modifications to accommodate the corresponding dependence structure. 
Similar characterizations can also be developed for estimands based on subsets other than a unit's neighborhood, with the conditions adapted to the relevant subsets of units. Finally, analogous results can be obtained under other network structures, following the same proof strategy as for Theorem \ref{difference_out_in} and Conditions \ref{ave_equivalence_cond_2}--\ref{ave_equivalence_cond_3}; Condition \ref{ave_equivalence_cond_1}, however, is specific to clustered network structures and would not apply when such structures are absent. The second research direction} concerns the development of estimands and inference methods for conditional spillover effects with continuous covariates or categorical covariates with few units observed for each category. 
A potential solution involves leveraging parametric assumptions.
The third research direction focuses on developing targeting strategies informed by these spillover effects. For instance, in vaccination programs (\cite{randolph2020herd}), policymakers are particularly interested in strategies that would encourage vaccination within specific communities to achieve protective effects for the whole population. Then either the outward or inward spillover effect--depending on specific policy goals--could be utilized for this purpose.

\section*{Acknowledgement}
Laura Forastiere and Fei Fang gratefully acknowledge support from the National Institutes of Health (NIH) Grant R01MH134715. 

\begin{center}
\noindent\textbf{\Large Supplementary Material for ``Inward and Outward Spillover Effects of One Unit's Treatment on Network Neighbors under Partial Interference''}
\end{center}

\bigskip

The Supplementary Material is organized into five sections: Section \ref{ave_spillover_example} presents various examples that illustrate scenarios where outward and inward spillover effects differ, conditions under which these two estimands are equivalent, and cases where the efficiencies of their estimators equate. { Section~\ref{append:tau_out_tau_in_unequal} provides an interpretable and sufficient condition under which \(\tau^{out}(\alpha)\) and \(\tau^{in}(\alpha)\) are not equivalent. Section~\ref{append:comb_estimators} derives the weight that minimizes the conservative variance of a weighted combination of the outward and inward estimators, and provides a sufficient condition under which the optimal weight is non-trivial. Section~\ref{append:lower_bound_V_c_V} examines the factors determining the gap between
\(V^c(\hat{\tau}^{\cdot}(\alpha))\) and
\(V(\hat{\tau}^{\cdot}(\alpha))\).} Section \ref{sec_cond_spill_over} provides detailed discussions on conditional spillover effects. 
Section \ref{Simulations} presents simulation results corresponding to different sections of the paper. Finally, Section \ref{sec_tech_details} includes all technical derivations and proofs related to both average and conditional spillover effects.

\appendix

\section{Examples}
\label{ave_spillover_example}
In this section, we present examples that agree with Theorem \ref{difference_out_in} or satisfy one of the conditions outlined in Section \ref{comp_out_in_spillover}.
\begin{example}
\label{examp_uncond_dif}
Consider the case where $K=1$ and the directed graph is as follows. \\
\begin{figure}[H]
\centering
\begin{tikzpicture}
\draw[{Latex[length=3mm]}-{Latex[length=3mm]}] (0.3,4) -- (2.7,4);
\draw (-0,4) circle (0.25 cm);
\node at (-0,4) {$1$};
\draw (3,4) circle (0.25 cm);
\node at (3,4) {$2$};
\draw[{Latex[length=3mm]}-] (1.2,1.3) -- (2.8,3.8);
\draw (1,1) circle (0.25 cm);
\node at (1,1) {$3$};
\draw[-{Latex[length=3mm]}] (4.8,1.2) -- (3.1,3.7);
\draw (5,1) circle (0.25 cm);
\node at (5,1) {$4$};
\end{tikzpicture}
\caption{Directed graph, with directed edges representing the presence and direction of links between nodes.}
\label{fig:exA1}
\end{figure}
\noindent The potential outcomes are $Y_{ik}(\mathbf{z}_k)= \beta_0+\beta_1 z_{ik}+  \sum_{jk \in \mathcal{N}^{in}_{ik}} \beta_{jk,ik} \cdot  z_{jk}+ \epsilon_{ik}$ for $i \in \{1,2,3,4\}$, with coefficients $\beta_{21,11}=\beta_{21,31}=3$ and $\beta_{11,21}=\beta_{41,21}=10$. Then we have 
\[
\begin{aligned}
    & \bar{Y}_{11}(z_{21} = 1, \alpha) - \bar{Y}_{11}(z_{21} = 0, \alpha) = 3, \quad \bar{Y}_{21}(z_{11} = 1, \alpha) - \bar{Y}_{21}(z_{11} = 0, \alpha) = 10, \\
    & \bar{Y}_{21}(z_{41} = 1, \alpha) - \bar{Y}_{21}(z_{41} = 0, \alpha) = 10, \quad \bar{Y}_{31}(z_{21} = 1, \alpha) - \bar{Y}_{31}(z_{21} = 0, \alpha) = 3.
\end{aligned}
\] 
Based on Definition \ref{out_spillover}, the outward spillover effect is 
\begin{equation*}
    \begin{split}
        & \frac{1}{3} \sum_{j \in \{1,2,4 \}} \frac{ 1  }{|\mathcal{N}^{out}_{j1}|} \sum_{i \in \mathcal{N}^{out}_{j1}}  \left( \bar{Y}_{i1}(z_{j1}=1, \alpha)- \bar{Y}_{i1}(z_{j1}=0, \alpha) \right) = \frac{1}{3} \left[ \frac{1}{1} \cdot 10 + \frac{1}{2} \cdot (3+3)+ \frac{1}{1} \cdot 10 \right] = \frac{23}{3}.
    \end{split}
\end{equation*}
The inward spillover effect based on the equivalent form \eqref{prof_difference_out_in_int_1} is 
\begin{equation*}
    \begin{split}
      &  \frac{1}{3} \sum_{j \in \{1,2,4\}}  \sum_{i \in \mathcal{N}^{out}_{j1}} \frac{1}{|\mathcal{N}^{in}_{i1}|} \left( \bar{Y}_{i1}(z_{j1}=1, \alpha)- \bar{Y}_{i1}(z_{j1}=0, \alpha) \right) = \frac{1}{3} \left[  \frac{1}{2} \cdot 10 + \left( \frac{1}{1} \cdot 3+ \frac{1}{1} \cdot  3  \right)    + \frac{1}{2}\cdot 10            \right]  = \frac{16}{3}
    \end{split}
\end{equation*} 
Therefore, the inward and outward spillover effects are different in this example. This result is common with a directed graph and heterogeneous pairwise spillover effects.
\end{example}

\begin{example}
\label{examp:undirect_graph_tau_out_tau_in_dif}
Consider $K=1$ and a graph structure as in Figure \ref{fig:ex1}. 
\begin{figure}[H]
\centering
\begin{tikzpicture}
\draw (1.5,6) circle (0.2 cm);
\node at (1.5,6) {$1$};

\draw[{Latex[length=2mm]}-] (1.6,6.2) -- (2.7,7.2);
\draw[-{Latex[length=2mm]}] (1.7,6) -- (2.8,7);
\draw (3,7.1) circle (0.2 cm);
\node at (3,7.1) {$2$};
\node[scale=0.8] at (1.9,6.8) {$-1$};
\node[scale=0.8] at (2.2,6.2) {$1$};

\draw (0,7.1) circle (0.2 cm);
\node at (0,7.1) {$5$};
\draw[-{Latex[length=2mm]}] (0.3,7.2) -- (1.4,6.2);
\draw[{Latex[length=2mm]}-] (0.2,7) -- (1.3,6.0);
\node[scale=0.8] at (0.75,6.2) {$1$};
\node[scale=0.8] at (1.1,6.8) {$-1$}; 

\draw (0,4.9) circle (0.2 cm);
\node at (0,4.9) {$4$};
\draw[-{Latex[length=2mm]}] (0.2,5.0) -- (1.3,6.0);
\draw[{Latex[length=2mm]}-] (0.3,4.8) -- (1.4,5.8);
\node[scale=0.8] at (0.7,5.8) {$-1$};
\node[scale=0.8] at (1,5.2) {$1$};

\draw (3,4.9) circle (0.2 cm);
\node at (3,4.9) {$3$};
\draw[{Latex[length=2mm]}-] (1.7,6.0) -- (2.8,5.0);
\draw[-{Latex[length=2mm]}] (1.6,5.8) -- (2.7,4.8);
\node[scale=0.8] at (2,5.2) {$1$};
\node[scale=0.8] at (2.3,5.8) {$-1$};
\end{tikzpicture}
\caption{The figure shows an undirected star graph, with edges representing the presence of links between units. Each pair of connected units is linked by two bi-directed edges, with the direction of each edge representing the direction of the spillover effect and the value assigned to each edge from $jk$ to $ik$ denoting the pairwise spillover effect $\tau_{ik, jk}(\alpha)$, for any $\alpha$, as defined in Definition \ref{pair_spillover}. For example, the value $-1$ from unit $5$ to unit $1$ denotes $\tau_{11,51}(\alpha)=\bar{Y}_{11}(Z_{51}=1,\alpha)- \bar{Y}_{11}(Z_{51}=0,\alpha)=-1$.}
\label{fig:ex1}
\end{figure}
\noindent 
The potential outcomes are defined as
\begin{equation*}
\begin{split}
  &  Y_{i1}(\mathbf{z}_1) 
  =  \begin{cases}
     \beta_0 + \beta_1 z_{i1}+  1 \cdot \sum_{j1 \in \mathcal{N}_{i1}} z_{j1} + \epsilon_{i1} & \text{if}\ i \in \{2,3,4,5\} \\
      \beta_0 + \beta_1 z_{i1}+ (-1) \cdot \sum_{j1 \in \mathcal{N}_{i1}} z_{j1} + \epsilon_{i1} & \text{if}\ i \in \{1\}.
    \end{cases}
\end{split}
  \end{equation*}
Then, the spillover effect from the central unit to a non-central unit is $1$, and the spillover effect from a non-central unit to the central unit is $-1$, i.e., 
  \begin{equation*}
\begin{split}
  & \tau_{i1,j1}(\alpha)={Y}_{i1}(Z_{j1}=1, \alpha)-{Y}_{i1}(Z_{j1}=0, \alpha) 
  = \begin{cases}
     1  & \text{if}\ j \in \{1\} 
     \\
     -1 & \text{if}\ j \in \{2,3,4,5\}. 
    \end{cases}
\end{split}
  \end{equation*}
The cardinality of the neighbor sets for each unit is:
\begin{equation*}
\begin{split}
  & |\mathcal{N}_{i1}|
  = \begin{cases}
     1  & \text{if}\ i \in \{2,3,4,5\} 
     \\
     4 & \text{if}\ i \in \{1\}. 
    \end{cases}
\end{split}
  \end{equation*}
By Definitions \ref{out_spillover} and \ref{in_spillover}, $\tau^{out}(\alpha)=-\frac{3}{5}$ and $\tau^{in}(\alpha)=\frac{3}{5}$.
\end{example}

\begin{example}[Example for Condition \ref{ave_equivalence_cond_1}]
\label{examp_ave_equivalence_cond_1}
Consider $K=2$ clusters with a graph structure represented in Figure \ref{fig:exA2}. The value assigned to each edge is the pairwise spillover effect defined in Definition \ref{pair_spillover}. 
\begin{figure}[H]
\centering
\begin{tikzpicture}
\draw (1.5,6) circle (0.2 cm);
\node at (1.5,6) {$1$};

\draw[{Latex[length=2mm]}-] (1.6,6.2) -- (2.7,7.2);
\draw[-{Latex[length=2mm]}] (1.7,6) -- (2.8,7);
\draw (3,7.1) circle (0.2 cm);
\node at (3,7.1) {$2$};
\node[scale=0.8] at (1.9,6.8) {$1$};
\node[scale=0.8] at (2.2,6.2) {$1$};

\draw (0,7.1) circle (0.2 cm);
\node at (0,7.1) {$5$};
\draw[-{Latex[length=2mm]}] (0.3,7.2) -- (1.4,6.2);
\draw[{Latex[length=2mm]}-] (0.2,7) -- (1.3,6.0);
\node[scale=0.8] at (1.1,6.8) {$1$};
\node[scale=0.8] at (0.75,6.2) {$1$};

\draw (0,4.9) circle (0.2 cm);
\node at (0,4.9) {$4$};
\draw[-{Latex[length=2mm]}] (0.2,5.0) -- (1.3,6.0);
\draw[{Latex[length=2mm]}-] (0.3,4.8) -- (1.4,5.8);
\node[scale=0.8] at (0.75,5.8) {$1$};
\node[scale=0.8] at (1,5.2) {$1$};

\draw (3,4.9) circle (0.2 cm);
\node at (3,4.9) {$3$};
\draw[{Latex[length=2mm]}-] (1.7,6.0) -- (2.8,5.0);
\draw[-{Latex[length=2mm]}] (1.6,5.8) -- (2.7,4.8);
\node[scale=0.8] at (2.2,5.8) {$1$};
\node[scale=0.8] at (2,5.2) {$1$};

\node[scale=1] at (1.5,4.2) {$k=1$};

\draw (6.5,6) circle (0.2 cm);
\node at (6.5,6) {$1$};

\draw[{Latex[length=2mm]}-] (6.6,6.2) -- (7.7,7.2);
\draw[-{Latex[length=2mm]}] (6.7,6) -- (7.8,7);
\draw (8,7.1) circle (0.2 cm);
\node at (8,7.1) {$2$};
\node[scale=0.8] at (6.9,6.8) {$2$};
\node[scale=0.8] at (7.2,6.2) {$2$};

\draw (5,7.1) circle (0.2 cm);
\node at (5,7.1) {$5$};
\draw[-{Latex[length=2mm]}] (5.3,7.2) -- (6.4,6.2);
\draw[{Latex[length=2mm]}-] (5.2,7) -- (6.3,6.0);
\node[scale=0.8] at (6.1,6.8) {$2$};
\node[scale=0.8] at (5.75,6.2) {$2$};

\draw (5,4.9) circle (0.2 cm);
\node at (5,4.9) {$4$};
\draw[-{Latex[length=2mm]}] (5.2,5.0) -- (6.3,6.0);
\draw[{Latex[length=2mm]}-] (5.3,4.8) -- (6.4,5.8);
\node[scale=0.8] at (5.75,5.8) {$2$};
\node[scale=0.8] at (6,5.2) {$2$};

\draw (8,4.9) circle (0.2 cm);
\node at (8,4.9) {$3$};
\draw[{Latex[length=2mm]}-] (6.7,6.0) -- (7.8,5.0);
\draw[-{Latex[length=2mm]}] (6.6,5.8) -- (7.7,4.8);
\node[scale=0.8] at (7,5.2) {$2$};
\node[scale=0.8] at (7.2,5.8) {$2$};

\node[scale=1] at (6.5,4.2) {$k=2$};
\end{tikzpicture}
\caption{Undirected graph for $K=2$ clusters, with edges representing the presence of links between nodes.  The value assigned to each edge from $jk$ to $ik$ is the pairwise spillover effect $\tau_{ik, jk}(\alpha)$, for any $\alpha$.}
\label{fig:exA2}
\end{figure}
\noindent Then we have 
$$\tau^{out}(\alpha)=\frac{1}{5}\cdot \left[ \frac{1}{4}\cdot (1\cdot 4) + (\frac{1}{1} \cdot 1) \cdot 4   \right] + \frac{1}{5}\cdot \left[ \frac{1}{4}\cdot (2\cdot 4) + (\frac{1}{1} \cdot 2) \cdot 4   \right]=3 $$
$$\tau^{in}(\alpha)=\frac{1}{5}\cdot \left[ \frac{1}{4}\cdot (1\cdot 4) + (\frac{1}{1} \cdot 1) \cdot 4   \right] + \frac{1}{5}\cdot \left[ \frac{1}{4}\cdot (2\cdot 4) + (\frac{1}{1} \cdot 2) \cdot 4   \right]=3$$
\end{example}

\begin{example}[Example for Condition \ref{ave_equivalence_cond_2}]
   \label{examp_ave_equivalence_cond_2}
Consider  $K=2$ clusters with a graph structure and pariwise spillover effects represented in Figure \ref{fig:exA3}.
\begin{figure}[H]
\centering
\begin{tikzpicture}
\draw (1.5,6) circle (0.2 cm);
\node at (1.5,6) {$1$};

\draw[-{Latex[length=2mm]}] (1.7,6) -- (2.8,7);
\draw (3,7.1) circle (0.2 cm);
\node at (3,7.1) {$2$};
\node[scale=0.8] at (2.2,6.2) {$1$};

\draw (0,7.1) circle (0.2 cm);
\node at (0,7.1) {$5$};
\draw[-{Latex[length=2mm]}] (0.3,7.2) -- (1.4,6.2);
\draw[{Latex[length=2mm]}-] (0.2,7) -- (1.3,6.0);
\node[scale=0.8] at (1.1,6.8) {$1$};
\node[scale=0.8] at (0.75,6.2) {$1$};

\draw (0,4.9) circle (0.2 cm);
\node at (0,4.9) {$4$};
\draw[-{Latex[length=2mm]}] (0.2,5.0) -- (1.3,6.0);
\draw[{Latex[length=2mm]}-] (0.3,4.8) -- (1.4,5.8);
\node[scale=0.8] at (0.75,5.8) {$1$};
\node[scale=0.8] at (1,5.2) {$1$};

\draw (3,4.9) circle (0.2 cm);
\node at (3,4.9) {$3$};
\draw[{Latex[length=2mm]}-] (1.7,6.0) -- (2.8,5.0);
\node[scale=0.8] at (2.2,5.8) {$1$};

\node[scale=1] at (1.5,4.2) {$k=1$};

\draw (6.5,6) circle (0.2 cm);
\node at (6.5,6) {$1$};

\draw[-{Latex[length=2mm]}] (6.7,6) -- (7.8,7);
\draw (8,7.1) circle (0.2 cm);
\node at (8,7.1) {$2$};
\node[scale=0.8] at (7.2,6.2) {$1$};

\draw (5,7.1) circle (0.2 cm);
\node at (5,7.1) {$5$};
\draw[{Latex[length=2mm]}-] (5.2,7) -- (6.3,6.0);
\node[scale=0.8] at (5.75,6.2) {$1$};

\draw (5,4.9) circle (0.2 cm);
\node at (5,4.9) {$4$};
\draw[{Latex[length=2mm]}-] (5.3,4.8) -- (6.4,5.8);
\node[scale=0.8] at (6,5.2) {$1$};

\draw (8,4.9) circle (0.2 cm);
\node at (8,4.9) {$3$};
\draw[-{Latex[length=2mm]}] (6.6,5.8) -- (7.7,4.8);
\node[scale=0.8] at (7,5.2) {$1$};

\node[scale=1] at (6.5,4.2) {$k=2$};

\end{tikzpicture}
\caption{Directed graph for $K=2$ clusters, with directed edges representing the presence and direction of links between nodes.  The value assigned to each edge from $jk$ to $ik$ is the pairwise spillover effect $\tau_{ik, jk}(\alpha)$, for any $\alpha$.}
\label{fig:exA3}
\end{figure}
\noindent Then we have 
\begin{equation*}
    \begin{split}
        & \tau^{out}(\alpha) 
         = \frac{1}{4+1} \cdot \left[ \left( \frac{1}{3} \cdot (1\cdot 3)+ (\frac{1}{1}\cdot 1) \cdot 3 \right) + \left(  \frac{1}{4} (1\cdot 4) \right)   \right]   = 1
    \end{split}
\end{equation*}
\begin{equation*}
    \begin{split}
        & \tau^{in}(\alpha) 
         = \frac{1}{4+4} \cdot \left[ \left( \frac{1}{3} \cdot (1\cdot 3)+ (\frac{1}{1}\cdot 1) \cdot 3 \right) + \left( ( \frac{1}{1} \cdot 1 )\cdot 4 \right)   \right]   = 1
    \end{split}
\end{equation*}
\end{example}

\begin{example}[Example for Condition \ref{ave_equivalence_cond_3}]
\label{exa_tau_out_equal_tau_in_2}
Consider a regular graph shown below in Figure \ref{fig:ex2}, with $K=1$ and degree $|N_{i1}|=2$ for any $i=1, \dots, 4$. 
\begin{figure}[H]
\centering
\begin{tikzpicture}

\draw (-0.5,7) circle (0.2 cm);
\node at (-0.5,7) {$1$};
\draw[-{Latex[length=2mm]}] (-0.3,7.1) -- (3,7.1);
\draw[{Latex[length=2mm]}-] (-0.3,6.9) -- (3,6.9);
\node[scale=0.8] at (1.35,7.3) {$1$};
\node[scale=0.8] at (1.35,6.7) {$2$};

\draw (3.2,7) circle (0.2 cm);
\node at (3.2,7) {$2$};
\draw[{Latex[length=2mm]}-] (3.1,6.8) -- (3.1,5.2);
\draw[-{Latex[length=2mm]}] (3.3,6.8) -- (3.3,5.2);
\node[scale=0.8] at (2.9,6) {$3$};
\node[scale=0.8] at (3.5,6) {$4$};

\draw (3.2,5) circle (0.2 cm);
\node at (3.2,5) {$3$};
\draw[-{Latex[length=2mm]}] (-0.3,5.1) -- (3,5.1);
\draw[{Latex[length=2mm]}-] (-0.3,4.9) -- (3,4.9);
\node[scale=0.8] at (1.35,5.3) {$5$};
\node[scale=0.8] at (1.35,4.7) {$6$};

\draw (-0.5,5) circle (0.2 cm);
\node at (-0.5,5) {$4$};
\draw[{Latex[length=2mm]}-] (-0.6,6.8) -- (-0.6,5.2);
\draw[-{Latex[length=2mm]}] (-0.4,6.8) -- (-0.4,5.2);
\node[scale=0.8] at (-0.2,6) {$7$};
\node[scale=0.8] at (-0.8,6) {$8$};
\end{tikzpicture}
\caption{The figure shows a regular graph, with edges representing the presence of links between units. Each pair of connected units is linked by two bi-directed edges, with the direction of each edge representing the direction of the spillover effect, and the value assigned to each edge from $jk$ to $ik$ denoting the pairwise spillover effect $\tau_{ik, jk}(\alpha)$, for any $\alpha$, as defined in Definition \ref{pair_spillover}.}
\label{fig:ex2}
\end{figure}
\noindent Then based on Definitions \ref{out_spillover} and \ref{in_spillover}, 
\begin{equation*}
    \begin{split}
        & \tau^{out}(\alpha) 
         = \frac{1}{4} \cdot \left[  \frac{1}{2} \cdot (1+7)+ \frac{1}{2} \cdot (2+4)+\frac{1}{2} \cdot (3+6) + \frac{1}{2} \cdot (5+8) \right]   = \frac{9}{2}
    \end{split}
\end{equation*}
\begin{equation*}
    \begin{split}
        & \tau^{in}(\alpha) 
         = \frac{1}{4} \cdot \left[ \frac{1}{2} \cdot (2+8)+ \frac{1}{2} \cdot (1+3)+\frac{1}{2} \cdot (5+4) + \frac{1}{2} \cdot (6+7)   \right]   = \frac{9}{2}
    \end{split}
\end{equation*}
where the same weights $\frac{1}{8}$, despite the heterogeneity in pairwise spillover effects, lead to the equivalence of $\tau^{out}(\alpha)$ and $\tau^{in}(\alpha)$. 
\end{example}

\begin{example}[Example for Condition \ref{ave_equivalence_cond_3}]
\label{examp2_ave_equivalence_cond_3}
    Consider  $K=1$ cluster with a graph structure represented in Figure \ref{fig:exA4}. 
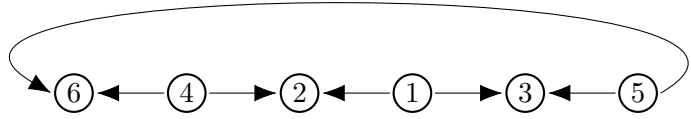
\begin{figure}[H]
\centering
        \begin{tikzpicture}
  \draw[thick] (-2.4,0) circle (0.25cm); 
   \node at (-2.4,0) {$6$};
    \draw[{Latex[length=3mm]}-] (-2.1,0) -- (-1.2,0);

  \draw[thick] (-0.9,0) circle (0.25cm);   
  \node at (-0.9,0) {$4$};
  \draw[-{Latex[length=3mm]}] (-0.6,0) -- (0.3,0);
  
  \draw[thick] (0.6,0) circle (0.25cm); 
  \node at (0.6,0) {$2$};
  \draw[{Latex[length=3mm]}-] (0.9,0) -- (1.8,0);
  
  \draw[thick] (2.1,0) circle (0.25cm); 
  \node at (2.1,0) {$1$};
   \draw[-{Latex[length=3mm]}] (2.4,0) -- (3.3,0);
  
  \draw[thick] (3.6,0) circle (0.25cm); 
  \node at (3.6,0) {$3$};
  \draw[{Latex[length=3mm]}-] (3.9,0) -- (4.8,0);
  
  \draw[thick] (5.1,0) circle (0.25cm); 
   \node at (5.1,0) {$5$};
  \draw [{Latex[length=3mm]}-] (-2.7,0) to [out=150,in=30] (5.4,0);

    
\end{tikzpicture}
\caption{Directed graph, with directed edges representing the presence and direction of links between nodes.}
\label{fig:exA4}
\end{figure}
\noindent Based on the graph, for each $jk$ such that  $|\mathcal{N}^{out}_{jk}|>0$,  we have
$$ \frac{|\mathcal{N}^{in}_{21}|}{|\mathcal{N}^{out}_{11}|}=\frac{|\mathcal{N}^{in}_{31}|}{|\mathcal{N}^{out}_{11}|}=\frac{2}{2}=1, \ \ \frac{|\mathcal{N}^{in}_{21}|}{|\mathcal{N}^{out}_{41}|}=\frac{|\mathcal{N}^{in}_{61}|}{|\mathcal{N}^{out}_{41}|}=\frac{2}{2}=1, \ \ \frac{|\mathcal{N}^{in}_{31}|}{|\mathcal{N}^{out}_{51}|}=\frac{|\mathcal{N}^{in}_{71}|}{|\mathcal{N}^{out}_{51}|}=\frac{2}{2}=1 \ \ , \frac{ {N}^{out} }{ {N}^{in} } = \frac{3}{3}=1.$$
Therefore, the graph structure in this example satisfies Condition \ref{ave_equivalence_cond_3}, which leads to $\tau^{out}(\alpha)=\tau^{in}(\alpha)$, regardless of the value of the pairwise spillover effects. 
\end{example}

\begin{example}[Homogeneous star graphs across clusters]
\label{homo_star_graphs}
Consider  \( K = 2000 \) clusters, each comprising \( n_k = 10 \) units. In each cluster, we assume a directed star graph where the central node $1$ has only outward edges (Figure \ref{fig:exA5.7}). Under this setup, $V^c(\hat{\tau}^{out}(\alpha))=V^c(\hat{\tau}^{in}(\alpha))$. This result also holds if the directed star graph is modified such that the central node has only inward edges (Figure \ref{fig:exA5.8}).\\  \\
\begin{minipage}{0.48\textwidth}
\centering
    \begin{tikzpicture}
        \node[circle, draw, fill=white, minimum size=0.5cm, inner sep=0, text=black] (center1) at (0,0) {1};

        \foreach \i in {1,...,9} {
            \pgfmathtruncatemacro{\label}{\i+1}
            \node[circle, draw, fill=white, minimum size=0.5cm, inner sep=0, text=black] (outer\i) at ({360/9 * (\i-1)}:2) {\label};
            \draw[-{Latex[length=2mm]}] (center1) -- (outer\i);
        }
    \end{tikzpicture}
    
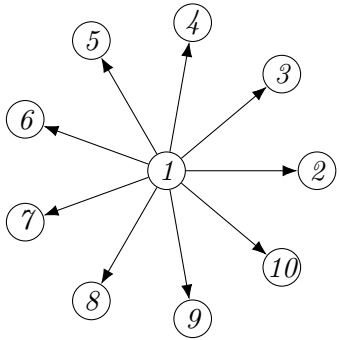
\captionof{figure}{This figure depicts a realization of an outward star graph where unit $1$ has $9$ outward edges.}
    \label{fig:exA5.7}
\end{minipage}%
\hfill 
\begin{minipage}{0.48\textwidth}
\centering
    \begin{tikzpicture}
        \node[circle, draw, fill=white, minimum size=0.5cm, inner sep=0, text=black] (center2) at (0,0) {1};

        \foreach \i in {1,...,9} {
            \pgfmathtruncatemacro{\label}{\i+1}
            \node[circle, draw, fill=white, minimum size=0.5cm, inner sep=0, text=black] (outer\i) at ({360/9 * (\i-1)}:2) {\label};
            \draw[{Latex[length=2mm]}-] (center2) -- (outer\i);
        }
    \end{tikzpicture}
    
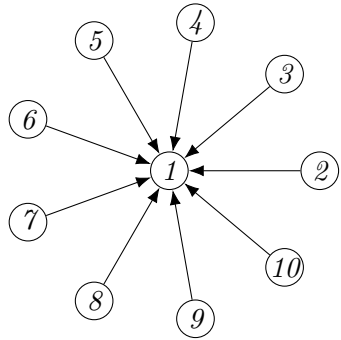
\captionof{figure}{This figure depicts a realization of an inward star graph where unit $1$ has $9$ inward edges.}
        \label{fig:exA5.8}

\end{minipage}
\end{example}
\begin{proof}[Proof for Example \ref{homo_star_graphs}]
    we first consider the setting with outward star graphs and the notations utilized in Proposition \ref{V_out_in_compare}. { Let $S^{out}_{ik,jk}$ and $S^{in}_{ik,jk}$ be as in Theorem \ref{difference_out_in}.} Since for each $k \in [K]$, 
    \begin{equation*}
        \begin{split}
       & { \sum_{ik\in\mathcal{N}^{out}_{1k}}
(S^{out}_{ik,1k}-S^{in}_{ik,1k})Y_{ik}}= \sum_{ik \in \mathcal{N}^{out}_{1k}} \left( \frac{1}{2000 \cdot 9}- \frac{1}{9\cdot 2000 \cdot 1}  \right) Y_{ik}  =0 
        \end{split}
    \end{equation*}
where only unit $1k$ has out-neighbors, then $V^c(\hat{\tau}^{out}(\alpha))=V^c(\hat{\tau}^{in}(\alpha))$. Similarly, for the setting with inward star graphs, we have for $jk \in \{ 2, \cdots ,10 \}$ and $k \in [K]$, 
 \begin{equation*}
        \begin{split}
{ \sum_{ik\in\mathcal{N}^{out}_{jk}}
(S^{out}_{ik,jk}-S^{in}_{ik,jk})Y_{ik}}= \sum_{ik \in \mathcal{N}^{out}_{jk}} \left( \frac{1}{ 9\cdot 2000 \cdot 1 }- \frac{1}{2000 \cdot 9}  \right) Y_{ik}  =0,
        \end{split}
    \end{equation*}
then it results to $V^c(\hat{\tau}^{out}(\alpha))=V^c(\hat{\tau}^{in}(\alpha))$. 
\end{proof}

{ 
\section{A Sufficient Condition for the Non-equivalence of \(\tau^{out}(\alpha)\) and \(\tau^{in}(\alpha)\)}
\label{append:tau_out_tau_in_unequal}

\begin{corollary}
\label{corol:suff_cond3_not_equal_relaxed}
Under the notation of Theorem~\ref{difference_out_in}, suppose
that $L_2=\emptyset$ and that there exists at least one pair
$(ik,jk)\in L_1$ such that $\Delta_{ik,jk}\tau_{ik,jk}(\alpha)>0$. Then $\tau^{out}(\alpha)>\tau^{in}(\alpha)$. Similarly, if $L_1=\emptyset$ and there exists at least one pair
$(ik,jk)\in L_2$, then $\tau^{out}(\alpha)<\tau^{in}(\alpha)$.
\end{corollary}

The proof of Corollary \ref{corol:suff_cond3_not_equal_relaxed} appears in Section \ref{append_comp_out_in_spillover}. Corollary~\ref{corol:suff_cond3_not_equal_relaxed} gives an interpretable
sufficient condition for non-equivalence. The condition requires sign alignment
between the graph-induced weight difference,
$\Delta_{ik,jk}=S^{out}_{ik,jk}-S^{in}_{ik,jk}$, and the pairwise spillover
effect $\tau_{ik,jk}(\alpha)$. If all nonzero products
$\Delta_{ik,jk}\tau_{ik,jk}(\alpha)>0$, then receiver--sender pairs
with positive pairwise spillover effects receive relatively larger outward
than inward weights, while pairs with negative pairwise spillover effects
receive relatively larger inward than outward weights. Consequently, $\tau^{out}(\alpha)>\tau^{in}(\alpha)$. Conversely, if all nonzero products
$\Delta_{ik,jk}\tau_{ik,jk}(\alpha)$ are negative, then pairs with positive
pairwise spillover effects receive relatively smaller outward than inward
weights, while pairs with negative pairwise spillover effects receive
relatively larger outward than inward weights. Consequently, $\tau^{out}(\alpha)<\tau^{in}(\alpha)$. Thus, the relative magnitude of the two estimands is governed by how the
weight difference $S^{out}_{ik,jk}-S^{in}_{ik,jk}$ co-varies with the pairwise spillover effect $\tau_{ik,jk}(\alpha)$ across receiver-sender pairs.
}

{ 
\section{Linear combinations of the estimators}
\label{append:comb_estimators}
In this section, we consider linear combinations of $\hat{\tau}^{out}(\alpha)$ and $\hat{\tau}^{in}(\alpha)$. Such combinations may improve efficiency relative to using either estimator alone. Specifically, for $\lambda\in[0,1]$, define
\[
\hat{\tau}_{\lambda}(\alpha)
=
\lambda \hat{\tau}^{out}(\alpha)
+
(1-\lambda)\hat{\tau}^{in}(\alpha).
\]
When $\tau^{out}(\alpha)=\tau^{in}(\alpha)$, $\hat{\tau}_{\lambda}(\alpha)$ is unbiased for the common estimand. We define the optimal value of $\lambda$ as the minimizer of the conservative variance,
\[
\lambda^*
=
\arg\min_{\lambda\in[0,1]}
V^c\left(
\lambda \hat{\tau}^{out}(\alpha)
+
(1-\lambda)\hat{\tau}^{in}(\alpha)
\right).
\]
Proposition~\ref{prop:opt_lambda} gives the analytical form of $\lambda^*$. Lemma~\ref{suf_cond_lambda_non_trivial} then provides a sufficient condition, involving both the graph structure and the outcomes, under which the optimal weight is strictly interior.

\begin{proposition}
\label{prop:opt_lambda}
Consider the class of estimators
$\hat{\tau}_{\lambda}(\alpha)
=
\lambda \hat{\tau}^{out}(\alpha)
+
(1-\lambda)\hat{\tau}^{in}(\alpha)$ where $\lambda\in[0,1]$.
The estimator in this class with minimum conservative variance is $\hat{\tau}_{\lambda^*}(\alpha)$ where $\lambda^*
=
\arg\min_{\lambda\in[0,1]}
V^c\left[
\lambda \hat{\tau}^{out}(\alpha)
+
(1-\lambda)\hat{\tau}^{in}(\alpha)
\right]$. Let $$\lambda_0
=
\frac{
V^c(\hat{\tau}^{in}(\alpha))
-
C(\hat{\tau}^{out}(\alpha),\hat{\tau}^{in}(\alpha))
}{
V^c(\hat{\tau}^{out}(\alpha))
+
V^c(\hat{\tau}^{in}(\alpha))
-
2C(\hat{\tau}^{out}(\alpha),\hat{\tau}^{in}(\alpha))
}.$$ where $
        C(\hat{\tau}^{out}(\alpha), \hat{\tau}^{in}(\alpha)):=\sum_{k=1}^K \sum_{jk \in \mathcal{N}^{out}_k} |\mathcal{N}^{out}_{k}| \   \mathbb{E} \left[ W^2_{jk}(\mathbf{Z}_k) (Z_{jk}+(1-Z_{jk}))(N^{out}N^{in})^{-1} \bar{Y}^{out}_{jk} \widetilde{Y}^{out}_{jk} \right]$, $\bar{Y}^{out}_{jk}$ and $\widetilde{Y}^{out}_{jk}$ are defined in Definition \ref{est_tau_out} and Proposition \ref{V_out_V_in}, respectively. Then the optimal weight is 
\[
\lambda^*
=
\begin{cases}
0, & \lambda_0<0,\\
\lambda_0, & 0\leq \lambda_0\leq 1,\\
1, & \lambda_0>1.
\end{cases}
\]
\end{proposition}
The proof is in Section \ref{Supp_sec_Comp_var} of the Supplementary Material. The expression for $\lambda_0$ in Proposition~\ref{prop:opt_lambda} shows how the optimal rule balances the two estimators. The estimator with the larger conservative variance is down-weighted. In particular, if $V^c(\hat{\tau}^{in}(\alpha))$ is large relative to $C(\hat{\tau}^{out}(\alpha),\hat{\tau}^{in}(\alpha))$, then $\lambda_0$ is closer to $1$ and more weight is placed on $\hat{\tau}^{out}(\alpha)$. Conversely, if $V^c(\hat{\tau}^{out}(\alpha))$ is large relative to $C(\hat{\tau}^{out}(\alpha),\hat{\tau}^{in}(\alpha))$, then $\lambda_0$ is closer to $0$ and more weight is placed on $\hat{\tau}^{in}(\alpha)$. Under Condition~\ref{ave_equivalence_cond_3}, however, all estimators in the class $\{\hat{\tau}_{\lambda}(\alpha):\lambda\in[0,1]\}$ have the same conservative variance; see Corollary~\ref{V_lambda_homo_cond_3} in this section.  

The resulting conservative variance is given in \eqref{linear_comb_V_c}. It comprises
\(V^c(\hat{\tau}^{out}(\alpha))\),
\(V^c(\hat{\tau}^{in}(\alpha))\), and the cross term
\(C(\hat{\tau}^{out}(\alpha),\hat{\tau}^{in}(\alpha))\), defined in
\eqref{covc_in_formula}. Consistent estimators of
\(V^c(\hat{\tau}^{out}(\alpha))\) and
\(V^c(\hat{\tau}^{in}(\alpha))\) are given in
Proposition~\ref{est_V_out_V_in}, while \(C(\hat{\tau}^{out}(\alpha),\hat{\tau}^{in}(\alpha))\) can be consistently estimated by its empirical analogue, obtained by replacing the expectation in \eqref{covc_in_formula} with the corresponding observed sample quantity. We next state the sufficient condition under which the optimal weight is strictly interior.

\begin{lemma}
\label{suf_cond_lambda_non_trivial}
Let $\lambda^*
=
\arg\min_{\lambda\in[0,1]}
V^c\left[
\lambda \hat{\tau}^{out}(\alpha)
+
(1-\lambda)\hat{\tau}^{in}(\alpha)
\right]$. Suppose there are $K$ clusters, each of which is either an in-star graph or an out-star graph. Assume that:
\begin{enumerate}
    \item[(a)] among the $K$ star graphs, $t$ are in-star graphs with degree $m\geq 2$, and $K-t$ are out-star graphs with degree $m\geq 2$;

    \item[(b)] with \begin{small}$R
    = {
    \sum_{k=1}^{K-t}
    \sum_{jk\in\mathcal{N}^{out}_k}
    \mathbb{E}
    \left[
    W^2_{jk}(\mathbf{Z}_k)
    \left(
    \sum_{ik\in\mathcal{N}^{out}_{jk}}Y_{ik}
    \right)^2
    \right]
    } \left\lbrace
    \sum_{k=K-t+1}^{K}
    \sum_{jk\in\mathcal{N}^{out}_k}
    \mathbb{E}
    \left[
    W^2_{jk}(\mathbf{Z}_k)
    \left(
    \sum_{ik\in\mathcal{N}^{out}_{jk}}Y_{ik}
    \right)^2
    \right]\right\rbrace^{-1}$\end{small}, we have
    \[
    \frac{K-t}{t}\frac{1}{m}<R<\frac{(K-t)}{t} m.
    \]
\end{enumerate}
Then $\lambda^*\in(0,1)$.
\end{lemma}
The proof is in Section \ref{Supp_sec_Comp_var} of the Supplementary Material. Lemma~\ref{suf_cond_lambda_non_trivial} shows that, under suitable conditions on the graph structure and the outcomes, a non-trivial linear combination of $\hat{\tau}^{out}(\alpha)$ and $\hat{\tau}^{in}(\alpha)$ achieves the smallest conservative variance within the class. 

\begin{corollary}
\label{V_lambda_homo_cond_3}
Under Condition~\ref{ave_equivalence_cond_3},
\[
V^c(\hat{\tau}_{\lambda}(\alpha))
=
V^c(\hat{\tau}^{out}(\alpha))
=
V^c(\hat{\tau}^{in}(\alpha)),
\qquad \lambda\in[0,1].
\]
\end{corollary}
The proof is in Section \ref{Supp_sec_Comp_var} of the Supplementary Material.
}

\section{Lower bound between $V^c(\hat{\tau}^{\cdot}(\alpha))$ and $V(\hat{\tau}^{\cdot}(\alpha))$}
\label{append:lower_bound_V_c_V}
To clarify the factors driving the difference between
\(V^c(\hat{\tau}^{\cdot}(\alpha))\) and
\(V(\hat{\tau}^{\cdot}(\alpha))\), we derive a lower bound for the bias of
\(V^c(\hat{\tau}^{\cdot}(\alpha))\) relative to
\(V(\hat{\tau}^{\cdot}(\alpha))\). The reason for obtaining the lower bounds is that their expressions are easy to interpret and provide insights into the factors driving the bias. 
\begin{proposition}[Lower bound between $V^c(\hat{\tau}^{\cdot}(\alpha))$ and $V(\hat{\tau}^{\cdot}(\alpha))$] 
\label{dif_V_var}
Based on  Proposition \ref{V_out_V_in}, the bias between $V^c(\hat{\tau}^{out}(\alpha))$ and $V(\hat{\tau}^{out} (\alpha))$ is 
\begin{small}
\begin{equation*}
    \begin{split}
       &    V^c(\hat{\tau}^{out}(\alpha) )- V(\hat{\tau}^{out}(\alpha))   \geq  \frac{1}{(N^{out})^2} \sum_{k=1}^K \sum_{jk \in \mathcal{N}^{out}_{k}}  { |\mathcal{N}^{out}_k| } \left[ { \frac{1}{|\mathcal{N}^{out}_{jk}|} \mbox{$\sum_{ik \in \mathcal{N}^{out}_{jk} }$ } \left( \bar{Y}_{ik}(Z_{jk}=1,\alpha)- \bar{Y}_{ik}(Z_{jk}=0,\alpha)  \right)  } \right]^2   
    \end{split}
\end{equation*}
\end{small}
and the bias between $V^c(\hat{\tau}^{in}(\alpha))$ and $V(\hat{\tau}^{in}(\alpha))$ is 
\begin{small}
\begin{equation*}
    \begin{split}
       &  V^c(\hat{\tau}^{in}(\alpha))- V(\hat{\tau}^{in}(\alpha))  \geq  \frac{1}{(N^{in})^2} \sum_{k=1}^K \sum_{jk \in \mathcal{N}^{out}_{k}}  {  |\mathcal{N}^{out}_k| } \left[ { \mbox{ $\sum_{ik \in \mathcal{N}^{out}_{jk} } $} \frac{1}{|\mathcal{N}^{in}_{ik}|} \left( \bar{Y}_{ik}(Z_{jk}=1,\alpha)- \bar{Y}_{ik}(Z_{jk}=0,\alpha)  \right)  } \right]^2.   
    \end{split}
\end{equation*}
\end{small}
\end{proposition}
The proof is in \ref{append_sec:Inference}. Proposition \ref{dif_V_var} demonstrates that the difference between the conservative variance and the true variance of our estimators, beyond being influenced by \(N^{out}\) and \(N^{in}\), also depends on different weighted average spillover effects on out-neighbors and the sizes of $\mathcal{N}^{out}_k$. As the spillover effects or $|\mathcal{N}^{out}_k|$ increase, the discrepancy between \(V^c(\hat{\tau}^{\cdot}(\alpha))\) and \( V(\hat{\tau}^{\cdot}(\alpha))\) also grows. Conversely, as the number of clusters \( K \) increases, this difference diminishes. Simulation studies that validate this proposition are presented in Supplementary Material \ref{sim_comp_V_c_V}.


\section{Conditional spillover effects}
\label{sec_cond_spill_over}
\subsection{Causal Estimands}
\label{append_sec_estimand}
We first define a covariate \( X_{jk} \), representing the source of heterogeneity for pairwise spillover effects. 
We then formally define the conditional outward and inward spillover effects, conditional on senders with covariate $X_k=x$.



\begin{definition}[Conditional outward spillover effect]
\label{cond_out_spillover}
Consider a unit \( jk \) with covariate value \( X_{jk}=x \) and \( |\mathcal{N}^{out}_{jk}| > 0 \). Let \( \mu^{out}_{jk}(z, \alpha) \) denote the average outcome among the out-neighbors of unit \( jk \) when its treatment is set to \( z \in \{0, 1\} \), while the rest of the cluster takes a random experiment with parameter \( \alpha \), i.e.,
\[
\mu^{out}_{jk}(z, \alpha) := \frac{1}{|\mathcal{N}^{out}_{jk}|} \sum_{ik \in \mathcal{N}^{out}_{jk}} \bar{Y}_{ik}( Z_{jk} = z, \alpha).
\]
The individual outward spillover effect for unit \( jk \), whose covariate value equals \( x \), is defined as:
\begin{equation*}
\tau^{out}_{jk}(\alpha) = \mu^{out}_{jk}( 1, \alpha) - \mu^{out}_{jk}(0, \alpha).
\end{equation*}
Then the conditional outward spillover effect, averaged across all units with covariate value \( x \), is defined as
\[
\tau^{out}(x, \alpha) = \frac{1}{N^{out}(x)} \sum_{k=1}^K \sum_{jk \in \mathcal{N}_k^{out}(x)} \left( \mu^{out}_{jk}( 1, \alpha) - \mu^{out}_{jk}(0, \alpha) \right),
\]
where
$
\mathcal{N}^{out}_k(x) = \left\{ jk \in \mathcal{N}_k : |\mathcal{N}^{out}_{jk}| > 0, X_{jk} = x \right\}$ and $
N^{out}(x) = \sum_{k=1}^K |\mathcal{N}^{out}_k(x)|$. 
\end{definition}


\begin{definition}[Conditional inward spillover effect]
\label{cond_in_spillover}
Consider a unit \( ik \) with \( |\mathcal{N}^{in}_{ik}(x)| > 0 \), where \( \mathcal{N}^{in}_{ik}(x) = \{ jk \in \mathcal{N}^{in}_{ik} : X_{jk} = x \} \) denotes the set of in-neighbors of unit \( ik \) whose covariate value equals \( x \). Let \( \mu^{in}_{ik}(x, z, \alpha) \) represent the average potential outcome of unit \( ik \) when the treatment of one of its in-neighbors with covariate value \( x \) is set to \( z \in \{0,1\} \), while the rest of the cluster takes a random experiment with parameter \( \alpha \), i.e.,
\[
\mu^{in}_{ik}(x, z, \alpha) := \frac{1}{|\mathcal{N}^{in}_{ik}(x)|} \sum_{jk \in \mathcal{N}^{in}_{ik}(x)} \bar{Y}_{ik}( Z_{jk} = z, \alpha).
\]
The individual inward spillover effect, measuring the impact of treatments of \( ik \)'s in-neighbors with covariate value \( x \) on \( ik \)'s outcome, is defined as
\[
\tau^{in}_{ik}(x, \alpha) := \mu^{in}_{ik}(x, 1, \alpha) - \mu^{in}_{ik}(x, 0, \alpha).
\]
Then the conditional inward spillover effect, averaged across all units with in-neighbors having covariate value \( x \), is
\[
\tau^{in}(x, \alpha) = \frac{1}{N^{in}(x)} \sum_{k=1}^K \sum_{ik \in \mathcal{N}^{in}_k(x)} \left[ \mu^{in}_{ik}(x, 1, \alpha) - \mu^{in}_{ik}(x, 0, \alpha) \right],
\]
where $
\mathcal{N}^{in}_k(x) = \{ ik \in \mathcal{N}_k : |\mathcal{N}^{in}_{ik}(x)| > 0 \}$ and 
$N^{in}(x) = \sum_{k=1}^K |\mathcal{N}^{in}_k(x)|$.
\end{definition}
While the conditional outward spillover effect adds a restriction regarding the senders' covariate, the individual outward spillover effect remains consistent with Definition \ref{out_spillover}.
In contrast, the individual inward spillover effect introduces a covariate-specific restriction to the set of in-neighbors, which differs from Definition \ref{in_spillover}.

\subsection{Comparison between conditional outward and inward spillover effects}
\label{comp_cond_out_in_spillover} 
We provide a sufficient and necessary condition for $\tau^{out}(x,\alpha)\neq \tau^{in}(x, \alpha)$ and sufficient conditions for $\tau^{out}(x,\alpha)= \tau^{in}(x, \alpha)$, similar to those for average spillover effects. 
\begin{proposition}
\label{difference_cond_out_in}
$\tau^{out}(x,\alpha) \neq \tau^{in}(x,\alpha)$ if and only if \ $\sum_{k=1}^K \sum_{jk \in \mathcal{N}^{out}_{k}(x) } \sum_{ik \in \mathcal{N}^{out}_{jk} } A_{ik,jk} \cdot F_{ik, jk} \neq 0 $ \ where $$A_{ik,jk}= {(N^{out}(x) \cdot |\mathcal{N}^{out}_{jk}|)^{-1}} - {(N^{in}(x) \cdot |\mathcal{N}^{in}_{ik}(x)|)^{-1} }$$ and $$F_{ik,jk}=   \bar{Y}_{ik}( Z_{jk}=1,\alpha)- \bar{Y}_{ik}( Z_{jk}=0,\alpha).$$
\end{proposition}
Proposition \ref{difference_cond_out_in} is harder to satisfy comparing to Theorem \ref{difference_out_in} mainly because of the difficulty of achieving \( A_{ik,jk} = 0 \) across all \( ik \) in \( \mathcal{N}^{out}_{jk} \) and \( jk \) in \( \mathcal{N}^{out}_k(x) \). Even when the graph is undirected and regular, where $\tau^{out}(\alpha)=\tau^{in}(\alpha)$ based on Condition \ref{ave_equivalence_cond_3} , it remains non-trivial to assign \( x \) such that \( A_{ik,jk} = 0 \). We present Example \ref{examp_cond_sender_dif}, which satisfies the condition for \(\tau^{out}(x, \alpha) \neq \tau^{in}(x, \alpha)\) as outlined in Proposition \ref{difference_cond_out_in}.
Now, we introduce conditions leading to \( \tau^{out}(x,\alpha) = \tau^{in}(x,\alpha) \):
\begin{condition}
\label{cond_equivalence_cond_1}
For $ik \in \mathcal{N}^{out}_{jk}$, $jk \in \mathcal{N}^{out}_{k}(x)$ and $k \in [K]$, the pairwise spillover effect $\bar{Y}_{ik}(Z_{jk}=1,\alpha)-\bar{Y}_{ik}( Z_{jk}=0,\alpha)=c_{k,x}$ where $c_{k,x}$ is a constant indexed by $k$ and $x$. Meanwhile, for each $k \in [K]$, $\frac{N^{in}(x)}{N^{out}(x)}= \frac{ |\mathcal{N}^{out}_{k}(x)| }{|\mathcal{N}^{in}_{k}(x) | }$. 
\end{condition}
\begin{condition}
\label{cond_equivalence_cond_2}
For $ik \in \mathcal{N}^{out}_{jk}$, $jk \in \mathcal{N}^{out}_{k}(x)$ and $k \in [K]$, 
$\bar{Y}_{ik}( Z_{jk}=1,\alpha)-\bar{Y}_{ik}(Z_{jk}=0,\alpha)=c_x$ where $c_x$ is a constant indexed by $x$. 
\end{condition}
\begin{condition}
\label{cond_equivalence_cond_3}
For $ik \in \mathcal{N}^{out}_{jk}$, $jk \in \mathcal{N}^{out}_{k}(x)$ and $k \in [K]$, $\frac{N^{in}(x)}{N^{out}(x) }= \frac{ |\mathcal{N}^{out}_{jk}| }{|\mathcal{N}^{in}_{ik}(x)|}$.  
\end{condition}
These conditions are analogous to those for average spillover effects but are restricted to the units possessing the covariate value \(x\) and the units who have at least one neighbor with covariate value $x$. Condition \ref{cond_equivalence_cond_1} imposes an additional constraint on the ratio between the cardinalities of units with covariate value  \(x\) and those units where at least one of their neighbors has covariate \(x\). Condition \ref{cond_equivalence_cond_2} asserts that once homogeneity of pairwise spillover effect is achieved among neighbors with covariate \(x\), then \(\tau^{out}(x,\alpha) = \tau^{in}(x,\alpha)\) holds irrespective of the graph structure. In Condition \ref{cond_equivalence_cond_3}, since \(jk\) is in \(\mathcal{N}^{in}_{ik}\) and \(X_{jk}=x\), \(|\mathcal{N}^{in}_{ik}(x)|\) is necessarily greater than zero, ensuring that $|\mathcal{N}^{out}_{jk}|$ over $|\mathcal{N}^{in}_{ik}(x)|$ is well-defined. Moreover, while Condition \ref{cond_equivalence_cond_1} introduces a milder ratio condition on ${N^{in}(x)}/{N^{out}(x)}$, it requires homogenous pairwise spillover effects compared to Condition \ref{cond_equivalence_cond_3}. 

\subsection{\texorpdfstring{Examples satisfying the conditions in Section \ref{comp_cond_out_in_spillover}}{} }
In this section, we first provide an example satisfying the condition of Proposition \ref{difference_cond_out_in}. We then provide an example for each condition where one and only one of Conditions \ref{cond_equivalence_cond_1}, \ref{cond_equivalence_cond_2} and \ref{cond_equivalence_cond_3} is satisfied.

\begin{example}[Example for Proposition \ref{difference_cond_out_in}]
\label{examp_cond_sender_dif}
    Let us revisit Example \ref{examp_uncond_dif} with additional conditions: $X_{11}= X_{21}=x, \  X_{31}\neq x$, $X_{41}\neq x$ 
    for any value of $x$. 
    Under these conditions and according to Definition \ref{cond_out_spillover}, the conditional outward spillover effect is
    \begin{equation*}
        \begin{split}
        &\sum_{k=1}^K \sum_{jk \in \mathcal{N}^{out}_k(x) } \sum_{ik \in \mathcal{N}^{out}_{jk}} \frac{ 1   }{N^{out}(x) \cdot  |\mathcal{N}^{out}_{jk}|} \left(  \bar{Y}_{ik}(Z_{jk}=1,\alpha)- \bar{Y}_{ik}( Z_{jk}=0,\alpha)  \right) \\
        & = \frac{1}{2\cdot 1 } \cdot 10 \cdot 1 + \frac{1}{2\cdot 2} (3+3) \cdot 1= \frac{13}{2}. \\        \end{split}
    \end{equation*}
Based on Definition \ref{cond_in_spillover}, the conditional inward spillover effect is 
\begin{equation*}
    \begin{split}
    & \sum_{k=1}^K  \sum_{ik \in \mathcal{N}^{in}_k(x) } \sum_{jk \in \mathcal{N}^{in}_{ik}(x)} \frac{ 1 }{N^{in} (x) \cdot |\mathcal{N}^{in}_{ik}(x)|} \cdot \left( \bar{Y}_{ik}(Z_{jk}=1,\alpha)- \bar{Y}_{ik}( Z_{jk}=0,\alpha)  \right)  \\
    & =  \frac{1}{3\cdot 1} \cdot 3+ \frac{1}{3 \cdot 1} \cdot 10 + \frac{1}{3 \cdot 1} 3  = \frac{16}{3}. \\
    \end{split}
\end{equation*}
Therefore, in this example, the conditional outward spillover effect is different from the conditional inward spillover effect even if the potential outcomes do not depend on covariate $x$. 
\end{example}
We then consider an example which satisfies Condition \ref{cond_equivalence_cond_1} but not \ref{cond_equivalence_cond_2} or \ref{cond_equivalence_cond_3}. 
\begin{example}[Example for Condition \ref{cond_equivalence_cond_1}]
Consider a network comprising two clusters (\( K=2 \)), with gender (F: female, M: male) serving as the covariate of interest, and a graph structure and covariate distribution represented in Figure \ref{fig:exB2}. In cluster 1, we have one central male unit (\( X_{41} = M \)) surrounded by two female units (\( X_{j1} = F \) for \( j \in \{1,2\} \)). Additionally, there is a central female unit (\( X_{31} = F \)) encircled by two male units (\( X_{j1} = M \) for \( j \in \{5,6\} \)). Cluster 2 mirrors this gender distribution, replicating the network structure and demographic setup of cluster $1$. \\ \\ 
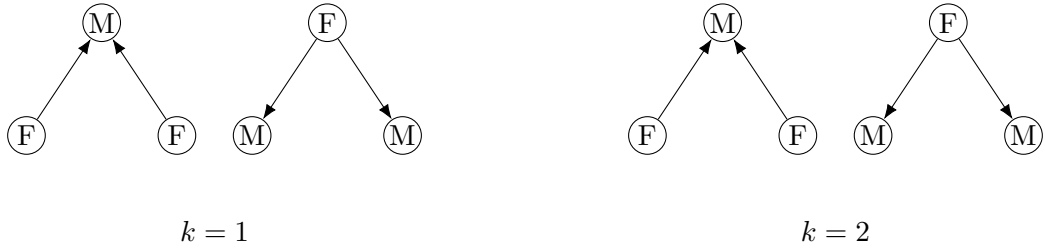
\begin{figure}[H]
\begin{minipage}{0.5\textwidth}
\centering
    \begin{tikzpicture}
        \node[circle, draw, fill=white, minimum size=0.5cm, inner sep=0, text=black] (center1) at (0,0) {M};
        \node[circle, draw, fill=white, minimum size=0.5cm, inner sep=0, text=black] (center2) at (3,0) {F};

        \foreach \i in {0,2} {
            \node[circle, draw, fill=white, minimum size=0.5cm, inner sep=0, text=black] (outer1\i) at (\i*1 - 1, -1.5) {F};
            \draw[{Latex[length=2mm]}-] (center1) -- (outer1\i);
        }

         \foreach \i in {3,5} {
            \node[circle, draw, fill=white, minimum size=0.5cm, inner sep=0, text=black] (outer2\i) at (\i*1 - 1, -1.5) {M};
            \draw[-{Latex[length=2mm]}] (center2) -- (outer2\i);
        }
        \node[below] at (1.5, -2.5) {$k=1$};
    \end{tikzpicture}
\end{minipage}%
\begin{minipage}{0.5\textwidth}
\centering
    \begin{tikzpicture}
        \node[circle, draw, fill=white, minimum size=0.5cm, inner sep=0, text=black] (center1) at (0,0) {M};
        \node[circle, draw, fill=white, minimum size=0.5cm, inner sep=0, text=black] (center2) at (3,0) {F};

        \foreach \i in {0,2} {
            \node[circle, draw, fill=white, minimum size=0.5cm, inner sep=0, text=black] (outer1\i) at (\i*1 - 1, -1.5) {F};
            \draw[{Latex[length=2mm]}-] (center1) -- (outer1\i);
        }

         \foreach \i in {3,5} {
            \node[circle, draw, fill=white, minimum size=0.5cm, inner sep=0, text=black] (outer2\i) at (\i*1 - 1, -1.5) {M};
            \draw[-{Latex[length=2mm]}] (center2) -- (outer2\i);
        }
        \node[below] at (1.5, -2.5) {$k=2$};
    \end{tikzpicture}
\end{minipage}
    \caption{Directed graph for $K=2$ clusters, with directed edges representing the presence and direction of links between nodes.}
    \label{fig:exB2}
\end{figure}
The structural model of potential outcome models is specified as:
\begin{equation*}
    Y_{ik}(\mathbf{z}_{k})= \beta_0 + \beta_1 z_{ik}  + \beta_{2k} \sum_{jk \in \mathcal{N}^{in}_{ik} } g(X_{jk}) \cdot z_{jk} + \epsilon_{ik},
\end{equation*}
where \( \beta_{2k} \) varies by cluster, taking the value of $1$ for \( k=1 \) and $2$ for \( k=2 \). The function \( g(X_{jk}) \) is defined by \( g(X_{jk}) = 1 \) if \( X_{jk} = F \) and \( g(X_{jk}) = -1 \) if \( X_{jk} = M \).
Based on the structural model of potential outcomes, for each $jk\in \mathcal{N}_k(F)$, we have 
\begin{equation*}
    \bar{Y}_{ik}( Z_{jk}=1, \alpha) - \bar{Y}_{ik}( Z_{jk}=0, \alpha) = \left\{
    \begin{array}{ll}
        1, & \text{if } k = 1 \\
        2, & \text{if } k = 2
    \end{array}
    \right.
\end{equation*}
which is homogeneous within each cluster. Meanwhile, 
\begin{equation*}
    \begin{split}
        \frac{N^{in}(F)}{N^{out}(F)}= \frac{6}{6}
    \end{split}
\end{equation*}
which is the same as that in each cluster, i.e.,
\begin{equation*}
    \begin{split}
        \frac{N_1^{in}(F)}{N_1^{out}(F)}= \frac{N_2^{in}(F)}{N_2^{out}(F)}= \frac{3}{3}. 
    \end{split}
\end{equation*}
Therefore, the example satisfies Condition \ref{cond_equivalence_cond_1}, which leads to the conclusion that \(\tau^{out}(F,\alpha) = \tau^{in}(F,\alpha)\). We can verify this equivalence by computing  the conditional spillover effects  based on Definitions \ref{cond_out_spillover} and \ref{cond_in_spillover} as follows. 
\begin{equation*}
    \begin{split}
        \tau^{out}(F,\alpha)= \frac{1}{6} \left[   \left( \frac{1}{1} \cdot 1 + \frac{1}{1} \cdot 1 + \frac{1}{2} \cdot 2 \right)+ \left( \frac{1}{1} \cdot 2 + \frac{1}{1} \cdot 2 + \frac{1}{2} \cdot 2\cdot 2   \right)  \right] = \frac{11}{6}
    \end{split}
\end{equation*}
\begin{equation*}
    \begin{split}
        \tau^{in}(F,\alpha)= \frac{1}{6} \left[   \left(\frac{1}{2} \cdot 2+ \frac{1}{1} \cdot 1 + \frac{1}{1} \cdot 1  \right)+ \left( \frac{1}{2} \cdot 2\cdot 2 +\frac{1}{1} \cdot 2 + \frac{1}{1} \cdot 2    \right)  \right] = \frac{11}{6}
    \end{split}
\end{equation*}
In this example, Condition \ref{cond_equivalence_cond_2} is not satisfied, as the pairwise spillover effects differ between clusters $1$ and $2$. Moreover, Condition \ref{cond_equivalence_cond_3} is not met, as evidenced by the unequal ratios:
\begin{equation*}
    \frac{ |\mathcal{N}^{out}_{11}| }{ |\mathcal{N}^{in}_{41}(F)| } = 1 \mathrm{\ and \ } \frac{ |\mathcal{N}^{out}_{31}| }{ |\mathcal{N}^{in}_{51}(F)| } = 2. 
\end{equation*}
\end{example}
We now present an example that satisfies Condition \ref{cond_equivalence_cond_2} but does not satisfy Conditions \ref{cond_equivalence_cond_1} or \ref{cond_equivalence_cond_3}.
\begin{example}[Example for Condition \ref{cond_equivalence_cond_2}]
Consider an undirected network consisting of two clusters (i.e., \( K=2 \)), with gender (F: female, M: male) as the covariate of interest, and a graph structure and covariate distribution represented in Figure \ref{fig:exB3}. In cluster 1, the central unit is female (\( X_{11} = F \)) surrounded by male units (\( X_{i1} = M \) for \( i \in \{2, \cdots, 10\} \)). Conversely, in cluster 2, the central unit is male (\( X_{12} = M \)), and the surrounding units are female (\( X_{i2} = F \) for \( i \in \{2, \cdots, 10\} \)). \\  \\
\begin{figure}[H]
\begin{minipage}{0.48\textwidth}
\centering
    \begin{tikzpicture}
        \node[circle, draw, fill=white, minimum size=0.5cm, inner sep=0, text=black] (center1) at (0,0) {F};

        \foreach \i in {1,...,9} {
            \node[circle, draw, fill=white, minimum size=0.5cm, inner sep=0, text=black] (outer\i) at ({360/9 * (\i-1)}:2) {M};
            \draw[-] (center1) -- (outer\i);
        }
        \node[align=center, font=\large, yshift=-3cm] at (center1) {k=1};
    \end{tikzpicture}
\end{minipage}%
\hfill 
\begin{minipage}{0.48\textwidth}
\centering
    \begin{tikzpicture}
        \node[circle, draw, fill=white, minimum size=0.5cm, inner sep=0, text=black] (center2) at (0,0) {M};

        \foreach \i in {1,...,9} {
            \node[circle, draw, fill=white, minimum size=0.5cm, inner sep=0, text=black] (outer\i) at ({360/9 * (\i-1)}:2) {F};
            \draw[-] (center2) -- (outer\i);
        }
        \node[align=center, font=\large, yshift=-3cm] at (center1) {k=2}; 
    \end{tikzpicture}
\end{minipage}
\caption{Undirected graph for $K=2$ clusters, with edges representing the presence of links between nodes.}
    \label{fig:exB3}
\end{figure}
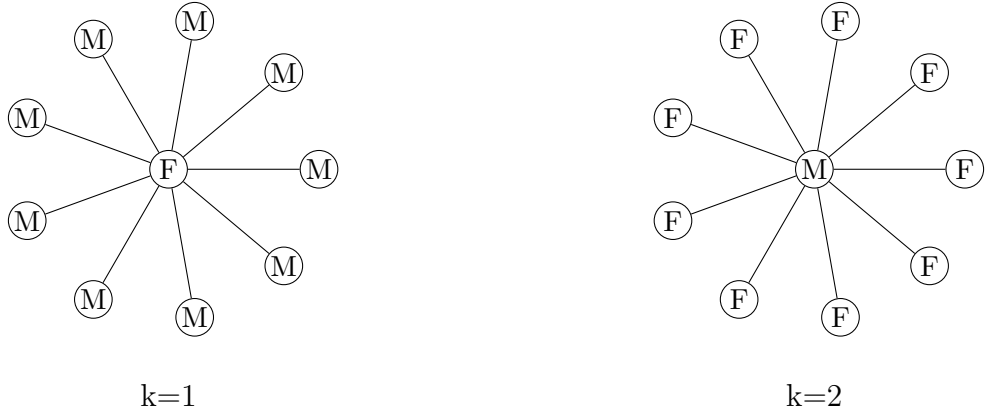
The structural model for potential outcomes for all $ik$, with $k=1,2$ and $i=1, \dots, 10$, is defined as follows
\begin{equation*}
    Y_{ik}(\mathbf{z}_{jk})= \beta_0 + \beta_1 z_{ik}  + \sum_{jk \in \mathcal{N}_{ik} } g(X_{jk}) \cdot z_{jk} + \epsilon_{ik},
\end{equation*}
where \(g(X_{jk}) = 1\) if \(X_{jk} = F\) and \(g(X_{jk}) = -1\) if \(X_{jk} = M\). From this model, the pairwise spillover effects are as follows
\[
\bar{Y}_{ik}(Z_{jk}=1, \alpha) - \bar{Y}_{ik}( Z_{jk}=0, \alpha) = 1,\quad \forall ik\in \mathcal{N}_k^{in}, k=1, 2, \forall jk\in \mathcal{N}^{in}_{ik}(F), 
\]
\[
\bar{Y}_{ik}(Z_{jk}=1, \alpha) - \bar{Y}_{ik}( Z_{jk}=0, \alpha) = -1, \quad \forall ik\in \mathcal{N}_k^{in}, k=1, 2, \forall jk\in \mathcal{N}^{in}_{ik}(M).
\]
Then using Definition \ref{cond_out_spillover}, the conditional outward spillover effects are calculated as:
\begin{equation*}
    \begin{split}
        \tau^{out} (F,\alpha)= \frac{1}{1+(10-1)} \left[ \frac{1}{9}\cdot 9+(10-1)\cdot \frac{1}{1} \cdot 1 \right] =1 
    \end{split}
\end{equation*}
\begin{equation*}
    \begin{split}
        \tau^{out} (M,\alpha)= \frac{1}{(10-1)+1} \left[ (10-1)\cdot \frac{1}{1} \cdot (-1)+\frac{1}{9}\cdot (-9) \right] =-1 
    \end{split}
\end{equation*}
Similarly, from Definition \ref{cond_in_spillover}, the conditional inward spillover effects are:
\begin{equation*}
    \begin{split}
        \tau^{in} (F,\alpha)= \frac{1}{10-1+ 1} \cdot \left[ 9 \cdot \frac{1}{1} \cdot 1 + 1 \cdot \frac{1}{9} \cdot 9  \right]  =1 
    \end{split}
\end{equation*}
\begin{equation*}
    \begin{split}
        \tau^{in} (M,\alpha)= \frac{1}{1+(10-1)} \cdot \left[ \frac{1}{9} \cdot (-9)+ 9 \cdot \frac{1}{1} \cdot (-1) \right] =-1  
    \end{split}
\end{equation*}
These calculations confirm that the outward and inward spillover effects are equal for each gender, which is expected since the setting of the example satisfies Condition \ref{cond_equivalence_cond_2}, i.e., the pairwise outward and inward spillover effects are homogeneous among females and males respectively. 

Additionally, when evaluating the average spillover effects in this finite population, where genders are equally distributed, we have
\begin{equation*}
    \begin{split}
        \tau^{out}(\alpha)= \frac{10}{20} \cdot \tau^{out}(F,\alpha)+ \frac{10}{20}\cdot  \tau^{out}(M,\alpha)= 0
    \end{split}
\end{equation*}
\begin{equation*}
    \tau^{in}(\alpha) = \frac{10}{20} \tau^{in}(M,\alpha) + \frac{10}{20} \tau^{in}(F,\alpha) = 0.
\end{equation*}
These results indicate that the average spillover effects from both genders perfectly offset each other due to both the graph structure and the distribution of gender.

Furthermore, the following ratios are observed for nodes with covariate \( F \):
\[
\frac{N^{in}(F)}{N^{out}(F)} = \frac{9+1}{1+9} = 1, \quad \frac{|\mathcal{N}^{out}_1(F)|}{|\mathcal{N}^{in}_1(F)|} = \frac{1}{9}, \quad \frac{|\mathcal{N}^{out}_2(F)|}{|\mathcal{N}^{in}_2(F)|} = \frac{9}{1}.
\]
The corresponding ratios for nodes with covariate \( M \) are similar. Consequently, Condition \ref{cond_equivalence_cond_1} is not satisfied in this example. Additionally, the following ratios are observed for nodes with covariate \( F \), where $i\in \{2, \cdots, 10\}$ and $j=1$:
\[
\frac{N^{in}(F)}{N^{out}(F)} = \frac{9+1}{1+9} = 1, \quad \frac{|\mathcal{N}^{}_{11}|}{|\mathcal{N}^{}_{i1}(F)|} = \frac{9}{1}, \quad \frac{|\mathcal{N}^{}_{j2}|}{|\mathcal{N}^{}_{12}(F)|} = \frac{1}{9}.
\]
The corresponding ratios for nodes with covariate \( M \) are similar. As a result, Condition \ref{cond_equivalence_cond_3} is also not satisfied.
\end{example}
We conclude with an example that satisfies Condition \ref{cond_equivalence_cond_3} but does not satisfy Conditions \ref{cond_equivalence_cond_1} and \ref{cond_equivalence_cond_2}.
\begin{example}[Example for Condition \ref{cond_equivalence_cond_3}]
\label{example_cond_equivalence_cond_3}
Consider a directed graph comprising a single cluster with  structure represented in Figure \ref{fig:exB4}.  
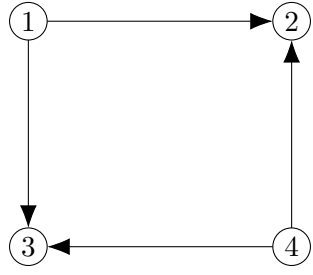
\begin{figure}[H]
\centering
\begin{tikzpicture}
\draw[-{Latex[length=3mm]}] (0,4) -- (3,4);
\draw (-0.25,4) circle (0.25 cm);
\node at (-0.25,4) {$1$};
\draw (3.25,4) circle (0.25 cm);
\node at (3.25,4) {$2$};
\draw[-{Latex[length=3mm]}] (-0.25,3.75) -- (-0.25,1.25);
\draw (-0.25,1) circle (0.25 cm);
\node at (-0.25,1) {$3$};
\draw[-{Latex[length=3mm]}] (3.25,1.25) -- (3.25,3.75);
\draw (3.25,1) circle (0.25 cm);
\node at (3.25,1) {$4$};
\draw[{Latex[length=3mm]}-] (0,1) -- (3,1);
\end{tikzpicture}
\caption{Directed graph for $K=1$ cluster, with directed edges representing the presence and direction of links between nodes.}
\label{fig:exB4}
\end{figure}
\noindent where $X_{11}=X_{41}=F$, $X_{21}=X_{31}=M$. The structural model for potential outcomes for all $ik$, with $k=1$ and $i=1, \dots, 4$, is defined as follows:
\begin{equation*}
    Y_{ik}(\mathbf{z}_{k})= \beta_0 + \beta_1 z_{ik} + \sum_{jk \in \mathcal{N}^{in}_{ik}} \beta_{ik,jk} \cdot g(X_{jk}) \cdot z_{jk} + \epsilon_{ik},
\end{equation*}
where \(g(X_{jk}) = 1\) if \(X_{jk} = F\) and \(g(X_{jk}) = -1\) if \(X_{jk} = M\). Coefficients are set to \(\beta_{21,11}=3\), \(\beta_{21,41}=4\), \(\beta_{31,11}=10\), and \(\beta_{31,41}=8\). 

Based on the graph structure and the distribution of the gender, we have 
$$ \frac{N^{in}(F)}{ N^{out}(F)}= \frac{|\mathcal{N}^{out}_{11}|}{|\mathcal{N}^{in}_{21}(F)|}= \frac{|\mathcal{N}^{out}_{11}|}{|\mathcal{N}^{in}_{31}(F)|} = \frac{|\mathcal{N}^{out}_{41}|}{|\mathcal{N}^{in}_{21}(F)|}= \frac{|\mathcal{N}^{out}_{41}|}{|\mathcal{N}^{in}_{31}(F)|} = \frac{2}{2}=1.$$
Thus, Condition \ref{cond_equivalence_cond_3} is met, suggesting \(\tau^{out}(F,\alpha) = \tau^{in}(F,\alpha)\). We also verify the equivalence by computing these two spillover effects based on Definitions \ref{cond_out_spillover} and \ref{cond_in_spillover}. Based on the structural model of potential outcomes, the pairwise spillover effects are as follows: \begin{small}
$$\bar{Y}_{21}( z_{11}=1,\alpha)-\bar{Y}_{21}(z_{11}=0,\alpha)=3, \quad \bar{Y}_{21}(z_{41}=1,\alpha)-\bar{Y}_{21}( z_{41}=0,\alpha)=4,$$
$$\bar{Y}_{31}(z_{11}=1,\alpha)-\bar{Y}_{31}(z_{11}=0,\alpha)=10, \quad \bar{Y}_{31}( z_{41}=1,\alpha)-\bar{Y}_{31}(z_{41}=0,\alpha)=8.$$
\end{small}
Then the conditional outward spillover effect is \begin{equation*}
    \begin{split}
        \tau^{out}(F, \alpha)& =\frac{1}{2} \sum_{j1 \in \{ 11,41 \}}  \frac{1}{  |\mathcal{N}^{out}_{j1}|} \sum_{i1 \in \mathcal{N}^{out}_{j1}} \left[  \bar{Y}_{i1}( z_{j1}=1,\alpha)- \bar{Y}_{i1}( z_{j1}=0,\alpha)  \right]  \\
        & = \frac{1}{2} \cdot \left[ \frac{1}{2} \cdot \left( 3+10 \right)+ \frac{1}{2} \cdot \left(  4+8  \right)   \right]=\frac{25}{4}. \\
    \end{split}
\end{equation*}
For the conditional inward spillover effect, we have 
\begin{equation*}
    \begin{split}
       \tau^{in}(F,\alpha) &= \frac{1}{2} \sum_{i1 \in \{21,31\}} \frac{1}{|\mathcal{N}^{in}_{i1}(F)|}  \sum_{j1 \in \mathcal{N}^{in}_{i1}(F) }  \left[  \bar{Y}_{i1}( z_{j1}=1,\alpha)- \bar{Y}_{i1}(z_{j1}=0,\alpha)  \right] \\
      & = \frac{1}{2} \cdot \left[ \frac{1}{2} \cdot \left( 3+4 \right)+ \frac{1}{2} \cdot \left(  10+8  \right)   \right]=\frac{25}{4} \\
    \end{split}
\end{equation*}
which verifies $\tau^{out}(F,\alpha)= \tau^{in}(F,\alpha)$ under Condition \ref{cond_equivalence_cond_3}, while Conditions \ref{cond_equivalence_cond_1} and \ref{cond_equivalence_cond_2} do not apply in this example due to heterogeneous pairwise spillover effects.
\end{example}

\section{Simulation results}
\label{Simulations}
{  
\subsection{\texorpdfstring{Difference between $\tau^{out}(\alpha)$ and $\tau^{in}(\alpha)$}{Difference between tau-out and tau-in}}
\label{Append:sim_dif_tau_out_tau_in}

This section provides simulation evidence on when $\tau^{out}(\alpha)$ differs from $\tau^{in}(\alpha)$. We first examine three representative settings with commonly used network-generation mechanisms and heterogeneous spillover effects, showing that $\tau^{out}(\alpha)-\tau^{in}(\alpha)$ can be large under realistic graph structures and spillover-effect heterogeneity. Across all three settings, potential outcomes are generated according to
\[
Y_{ik}(\mathbf z_k)
=
\beta_0+\beta_1 z_{ik}
+
\beta_{2,ik}
\sum_{jk\in\mathcal N^{in}_{ik}}
\beta_{2,jk}z_{jk}
+
\epsilon_{ik},
\]
where $\epsilon_{ik}\sim N(0,\sigma^2), \sigma=0.05, \beta_0=0.1, \beta_1=0.2$. Under this specification, the pairwise spillover effect from unit $jk$ to unit $ik$ is
\[
\tau_{ik,jk}(\alpha)
=
\beta_{2,ik}\beta_{2,jk}.
\]
Thus the pairwise spillover effects are symmetric in the ordered pair $(ik,jk)$ and heterogeneous across pairs through the unit-level coefficients $\beta_{2,ik}$ and $\beta_{2,jk}$. For each simulation design, we vary the number of clusters over $K\in\{20,40,80,100,150,200,250\}$, with fixed cluster size $n_k=20$. 
We next describe the within-cluster graph-generating mechanisms used in the three settings.

\begin{enumerate}
    \item \textit{Setting 1.} Within each cluster, we generate a scale-free network from the Barab\'asi--Albert preferential attachment model with attachment power equal to one. This model yields a degree distribution with power-law decay proportional to $d^{-3}$, where $d$ denotes the degree of an existing node. In this setting, we take the spillover coefficients to depend on both in- and out-degrees. Specifically, we set $$ \beta_{2,ik}
=
\frac{|\mathcal N^{in}_{ik}|}{1.5|\bar{\mathcal N}^{in}|}, \quad \beta_{2,jk}
=
\frac{|\mathcal N^{out}_{jk}|}{1.5|\bar{\mathcal N}^{out}|} $$ where $|\bar{\mathcal N}^{in}|
=
{(N^{in})}^{-1}
\sum_{k=1}^K
\sum_{ik\in\mathcal N^{in}_k}
|\mathcal N^{in}_{ik}|$ and $|\bar{\mathcal N}^{out}|
=
{(N^{out})^{-1}}
\sum_{k=1}^K
\sum_{jk\in\mathcal N^{out}_k}
|\mathcal N^{out}_{jk}|$. Thus units with larger in-degree have larger receiver-side spillover coefficients, while units with larger out-degree have larger sender-side spillover coefficients.
\item \textit{Setting 2.}
Within each cluster, we generate a directed stochastic block model with two blocks and edge-probability matrix $$\begin{pmatrix}
0.05 & 0.30 \\
0.02 & 0.05
\end{pmatrix}.$$
The asymmetric off-diagonal probabilities induce substantially more edges from block 1 to block 2 than from block 2 to block 1. Consequently, units in block 1 tend to have larger out-degrees, whereas units in block 2 tend to have larger in-degrees. In this setting, we isolate sender-side heterogeneity by setting
\[
\beta_{2,ik}=1,
\qquad
\beta_{2,jk}
=
\frac{|\mathcal N^{out}_{jk}|}{1.5|\bar{\mathcal N}^{out}|}.
\]
where $|\bar{\mathcal{N}}^{in}|$ is as defined in Setting $1$. 
\item \textit{Setting 3.} 
The graph-generating mechanism is the same directed stochastic block model as in Setting~2. In contrast to Setting~2, we isolate receiver-side heterogeneity by setting
\[
\beta_{2,ik}
=
\frac{|\mathcal N^{in}_{ik}|}{1.5|\bar{\mathcal N}^{in}|},
\qquad
\beta_{2,jk}=1.
\]
where $|\bar{\mathcal{N}}^{out}|$ is as defined in Setting $1$.
\end{enumerate}
The simulation results for these three settings are reported in Tables~\ref{setting_1:sim_tau_out_dif_tau_in}-\ref{setting_3:sim_tau_out_dif_tau_in}. 

\begin{table}[H]
\centering
\caption{Simulation results for setting $1$ across $K$ \protect\footnotemark}
\label{setting_1:sim_tau_out_dif_tau_in}
\small
\begin{tabular}{cccccc}
\hline
$K$ & positive part & negative part & difference
& $\tau^{out}(\alpha)$
& $\tau^{in}(\alpha)$ \\
\hline
20  & 0.139 & -1.031 & -0.891 & 1.071 & 1.962 \\
40  & 0.145 & -1.070 & -0.925 & 1.056 & 1.981 \\
80  & 0.140 & -1.095 & -0.954 & 1.033 & 1.988 \\
100 & 0.140 & -1.096 & -0.955 & 1.038 & 1.994 \\
150 & 0.140 & -1.086 & -0.947 & 1.025 & 1.972 \\
200 & 0.139 & -1.061 & -0.921 & 1.027 & 1.948 \\
250 & 0.140 & -1.068 & -0.928 & 1.030 & 1.958 \\
\hline
\end{tabular}
\end{table}
\footnotetext{The second and third columns report the positive and negative components, respectively, of the decomposition in \eqref{theo_difference_out_in_int_1} of Theorem~\ref{difference_out_in}; these are given by $\sum_{(ik,jk)\in L_h}\Delta_{ik,jk}\tau_{ik,jk}(\alpha)$ for $h=1$ and $h=2$, respectively.}

\begin{table}[H]
\centering
\caption{Simulation results for setting $2$ across $K$}
\label{setting_2:sim_tau_out_dif_tau_in}
\small
\begin{tabular}{cccccc}
\hline
$K$ & positive part & negative part & difference
& $\tau^{out}(\alpha)$
& $\tau^{in}(\alpha)$ \\
\hline
20  & 0.170 & -0.480 & -0.310 & 0.932 & 1.242 \\
40  & 0.168 & -0.472 & -0.303 & 0.920 & 1.223 \\
80  & 0.169 & -0.483 & -0.313 & 0.905 & 1.219 \\
100 & 0.172 & -0.493 & -0.321 & 0.908 & 1.229 \\
150 & 0.172 & -0.487 & -0.315 & 0.904 & 1.219 \\
200 & 0.173 & -0.490 & -0.317 & 0.911 & 1.229 \\
250 & 0.173 & -0.487 & -0.314 & 0.914 & 1.228 \\
\hline
\end{tabular}
\end{table}

\begin{table}[H]
\centering
\caption{Simulation results for setting $3$ across $K$}
\label{setting_3:sim_tau_out_dif_tau_in}
\small
\begin{tabular}{cccccc}
\hline
$K$ & positive part & negative part & difference
& $\tau^{out}(\alpha)$
& $\tau^{in}(\alpha)$ \\
\hline
20  & 0.457 & -0.176 & 0.280 & 1.196 & 0.916 \\
40  & 0.474 & -0.175 & 0.300 & 1.221 & 0.921 \\
80  & 0.480 & -0.177 & 0.303 & 1.218 & 0.916 \\
100 & 0.483 & -0.175 & 0.307 & 1.219 & 0.911 \\
150 & 0.478 & -0.173 & 0.305 & 1.212 & 0.907 \\
200 & 0.481 & -0.174 & 0.307 & 1.217 & 0.910 \\
250 & 0.481 & -0.174 & 0.307 & 1.217 & 0.910 \\
\hline
\end{tabular}
\end{table}Tables~\ref{setting_1:sim_tau_out_dif_tau_in}--\ref{setting_3:sim_tau_out_dif_tau_in} show that the net difference induced by these two components is substantial relative to the magnitudes of \(\tau^{out}(\alpha)\) and \(\tau^{in}(\alpha)\), and does not diminish as the number of clusters increases.

We now explain why Settings~2 and~3 produce positive and negative components of similar magnitude but opposite sign, as shown in the second and third columns of Tables \ref{setting_2:sim_tau_out_dif_tau_in} and \ref{setting_3:sim_tau_out_dif_tau_in}. 
In Setting~2, the pairwise contribution satisfies
\[
\begin{split}
\Delta_{ik,jk}\tau_{ik,jk}(\alpha)
&=
\left(
\frac{1}{N^{out}|\mathcal{N}^{out}_{jk}|}
-
\frac{1}{N^{in}|\mathcal{N}^{in}_{ik}|}
\right)
\frac{|\mathcal{N}^{out}_{jk}|}{1.5|\bar{N}^{out}|}  \\
&=
\frac{1}{
1.5\sum_{k=1}^K
\sum_{jk\in\mathcal{N}^{out}_k}
|\mathcal{N}^{out}_{jk}|
}
-
\frac{
N^{out}|\mathcal{N}^{out}_{jk}|
}{
1.5N^{in}|\mathcal{N}^{in}_{ik}|
\sum_{k=1}^K
\sum_{jk\in\mathcal{N}^{out}_k}
|\mathcal{N}^{out}_{jk}|
}.
\end{split}
\]
In Setting~3, the corresponding contribution is
\[
\begin{split}
\Delta_{ik,jk}\tau_{ik,jk}(\alpha)
&=
\left(
\frac{1}{N^{out}|\mathcal{N}^{out}_{jk}|}
-
\frac{1}{N^{in}|\mathcal{N}^{in}_{ik}|}
\right)
\frac{|\mathcal{N}^{in}_{ik}|}{1.5|\bar{N}^{in}|} \\
&=
\frac{
N^{in}|\mathcal{N}^{in}_{ik}|
}{
1.5N^{out}|\mathcal{N}^{out}_{jk}|
\sum_{k=1}^K
\sum_{ik\in\mathcal{N}^{in}_k}
|\mathcal{N}^{in}_{ik}|
}
-
\frac{1}{
1.5\sum_{k=1}^K
\sum_{ik\in\mathcal{N}^{in}_k}
|\mathcal{N}^{in}_{ik}|
}.
\end{split}
\]
In this stochastic block model setting, $N^{in}\approx N^{out}$. Moreover, within each cluster $k$, the
out-degree of a unit in block~1 is matched by the in-degree of a corresponding
unit in block~2; that is, there are pairs for which
$|\mathcal{N}^{out}_{jk}|$ in block~1 is equal to
$|\mathcal{N}^{in}_{ik}|$ in block~2. Since within-block connections are rare,
we also have $\sum_{k=1}^K
\sum_{ik\in\mathcal{N}^{in}_k}
|\mathcal{N}^{in}_{ik}|
\approx
\sum_{k=1}^K
\sum_{jk\in\mathcal{N}^{out}_k}
|\mathcal{N}^{out}_{jk}|$. It follows that, for each contribution
$\Delta_{ik,jk}\tau_{ik,jk}(\alpha)$ arising in Setting~2, there is a
corresponding contribution
$\Delta_{i'k,j'k}\tau_{i'k,j'k}(\alpha)$ in Setting~3 such that
\[
\Delta_{i'k,j'k}\tau_{i'k,j'k}(\alpha)
\approx
-\Delta_{ik,jk}\tau_{ik,jk}(\alpha).
\]
Thus, the positive and negative components in the decomposition are
approximately interchanged between Settings~2 and~3. This explains why \(\tau^{out}(\alpha)-\tau^{in}(\alpha)\) has opposite signs but similar magnitudes in the two settings.

We next consider settings in which the pairwise spillover effects are independent of the graph-induced weight differences \(\Delta_{ik,jk}\). We retain the graph-generating mechanisms used in Settings~1 and~2, but independently draw $\beta_{2,ik}\sim \Gamma(1.5,\frac{1.5}{1.2})$ and $
\beta_{2,hk}\sim \Gamma(1.5,\frac{1.5}{0.5})$, where the Gamma distribution is parametrized by its shape and rate parameters. Under this construction, the data-generating mechanism for
$\tau_{ik,jk}(\alpha)$ is independent of that for $\Delta_{ik,jk}$.

The results are reported in Tables~\ref{tab:powerlaw_ind_coef_K_setting1} and
\ref{tab:powerlaw_ind_coef_K_setting2}. They show that when $\tau_{ik,jk}(\alpha)$ and $\Delta_{ik,jk}$ are generated independently, The absolute difference between $\tau^{out}(\alpha)$ and $\tau^{in}(\alpha)$ is still nonzero but shrinks as $K$ increases. This follows from the independence of the data-generating processes for $\tau_{ik,jk}(\alpha)$ and $\Delta_{ik,jk}$ and is consistent with the lack of systematic alignment between pairwise spillover effects and graph-induced weight differences.

\begin{table}[H]
\centering
\caption{Simulation results for estimands under independent $\Delta_{ik,jk}$ and $\tau_{ik,jk}(\alpha)$ with graph structures in setting 1}
\label{tab:powerlaw_ind_coef_K_setting1}
\small
\begin{tabular}{cccccc}
\hline
$K$ & positive part & negative part & difference
& $\tau^{out}(\alpha)$
& $\tau^{in}(\alpha)$ \\
\hline
20  & 0.429 & -0.368 &  0.061 & 1.020 & 0.959 \\
40  & 0.203 & -0.258 & -0.055 & 0.529 & 0.585 \\
80  & 0.263 & -0.288 & -0.025 & 0.711 & 0.736 \\
100 & 0.285 & -0.255 &  0.029 & 0.706 & 0.677 \\
150 & 0.184 & -0.201 & -0.018 & 0.524 & 0.541 \\
200 & 0.198 & -0.183 &  0.015 & 0.519 & 0.504 \\
250 & 0.254 & -0.243 &  0.011 & 0.645 & 0.634 \\
\hline
\end{tabular}
\end{table}

\begin{table}[H]
\centering
\caption{Simulation results for estimands under independent $\Delta_{ik,jk}$ and $\tau_{ik,jk}(\alpha)$ with graph structures in setting 2}
\label{tab:powerlaw_ind_coef_K_setting2}
\small
\begin{tabular}{cccccc}
\hline
$K$ & positive part & negative part & difference
& $\tau^{out}(\alpha)$
& $\tau^{in}(\alpha)$ \\
\hline
20  & 0.120 & -0.287 & -0.167 & 0.701 & 0.868 \\
40  & 0.215 & -0.293 & -0.078 & 0.754 & 0.832 \\
80  & 0.254 & -0.210 &  0.044 & 0.753 & 0.709 \\
100 & 0.157 & -0.248 & -0.091 & 0.563 & 0.654 \\
150 & 0.180 & -0.193 & -0.013 & 0.591 & 0.604 \\
200 & 0.180 & -0.187 & -0.007 & 0.584 & 0.591 \\
250 & 0.229 & -0.191 &  0.038 & 0.661 & 0.623 \\
\hline
\end{tabular}
\end{table}
}
\subsection{\texorpdfstring{Comparison between $V^c(\hat{\tau}^{\cdot})$ and $V(\hat{\tau}^{\cdot})$}{} }
\label{sim_comp_V_c_V}
In this section, we consider a directed Erd\H{o}s R\'enyi graph comprising $2000$ clusters, each containing $10$ units, with an edge formation probability of $0.4$. The outcome model is specified in \eqref{spec_outcome_model}, with \( \beta_0 = 0.8 \) and \( \beta_1 = 2 \), and includes noise terms drawn from a normal distribution with mean $0$ and standard deviation $0.2$. The simulation is based on $500$ Monte Carlo repetitions, with fixed noise terms across repetitions to align with design-based uncertainty. Both the hypothetical and realized treatment assignments follow independent and identical Bernoulli distributions, with probabilities \( \alpha = 0.6 \) and \( \beta = 0.4 \), respectively.

Given the homogeneity of these parameters, we can simply vary the scale of $\beta_2$ to control the intensity of the pairwise spillover effects. The choice of such directed Erd\H{o}s R\'enyi graph allows for the approximation of $N^{out}$ and $N^{in}$ being roughly equal. In this way, the influence to $V^c(\hat{\tau}^{\cdot})-V(\hat{\tau}^{\cdot})$ mainly comes from pairwise spillover effects and cluster sizes. 
\begin{table}[H]
    \centering
\begin{center}
\caption{Simulation results for $V^c(\hat{\tau}^{\cdot}(\alpha))-V(\hat{\tau}^{\cdot}(\alpha) )$ by varying scales of spillover effects}
  \label{tab_vary_scale_spillover}
\begin{tabular}{ c|c|c } 
 \hline
  $\beta_2$ &  $V^c(\hat{\tau}^{out}(\alpha))-V(\hat{\tau}^{out}(\alpha) )$  & $V^c(\hat{\tau}^{in}(\alpha))-V(\hat{\tau}^{in}(\alpha) )$ \\ 
 \hline
 0  & $0.0369$ & $0.0466$ \\ 
 1 & $0.1839$  & $0.1943$ \\
  4 &$1.3329$  & $1.2068$   \\ 
  20 & $25.4030$  & $21.8151$ \\
 \hline
\end{tabular}
\end{center}
\end{table}
Based on Table \ref{tab_vary_scale_spillover}, we observe a monotonic increase in the difference $V^c(\hat{\tau}^{\cdot}(\alpha))-V(\hat{\tau}^{\cdot}(\alpha) )$ as $\beta_2$ increases. This validate a part of Proposition \ref{dif_V_var}.

We then conduct a simulation study for the relationship between cluster sizes and the difference \( V^c(\hat{\tau}^{\cdot}(\alpha)) - V(\hat{\tau}^{\cdot}(\alpha)) \). Using the same settings as for Table \ref{tab_vary_scale_spillover}, we change the number of clusters to $4000$ and maintaining \( \beta_2 = 1 \). We then vary the cluster sizes from $10$ to $50$ and the simulation results are as follows.
\begin{table}[H]
    \centering
\begin{center}
\caption{Simulation results for $V^c(\hat{\tau}^{\cdot}(\alpha))-V(\hat{\tau}^{\cdot}(\alpha) )$ by varying cluster sizes}
  \label{tab_vary_cluster_size}
\begin{tabular}{ c|c|c } 
 \hline
  $|\mathcal{N}^{out}_k|$ &  $V^c(\hat{\tau}^{out}(\alpha))-V(\hat{\tau}^{out}(\alpha) )$  & $V^c(\hat{\tau}^{in}(\alpha))-V(\hat{\tau}^{in}(\alpha) )$ \\ 
 \hline
 $10$  & $0.0918$ & $0.0924$ \\ 
 $30$ & $3.6012$  & $3.0675$ \\ 
  $50$ &$13.4354$  & $12.8587$   \\ 
  \hline
\end{tabular}
\end{center}
\end{table}
Based on Table \ref{tab_vary_cluster_size}, we note the difference between $V^c(\hat{\tau}^{\cdot}(\alpha))$ and $V(\hat{\tau}^{\cdot}(\alpha) )$ increases when $|\mathcal{N}^{out}_k|$ increases. This observation supports the remaining aspects of Proposition \ref{dif_V_var} in addition to the impact from pairwise spillover effects.

{  
\subsection{\texorpdfstring{Relation between estimand-weight distributions and estimator efficiency}{Relation between estimand-weight distributions and estimator efficiency}}
\label{append:estimand_weight_efficiency}

In this section, we supplement Lemma \ref{suffi_cond3_example} to show how the quantities in Proposition \ref{general_suff_cond_sender_type_3} vary with the proportion of in-star graphs $t$, and use Figure \ref{fig:estimand-weight-density-all} to illustrate how the distributions of $S^{out}_{ik,jk}$ and $S^{in}_{ik,jk}$ change as $t$ increases under the setting in Section \ref{estimand_weight_efficiency}.

\begin{lemma}
\label{suffi_cond3_example}
Suppose there are $K$ clusters, a proportion $t$ of which are in-star graphs
with degree $m_{in}\geq 2$, and a proportion $1-t$ of which are out-star graphs
with degree $m_{out}\geq 2$. Thus, there are $tK$ in-star graphs and
$(1-t)K$ out-star graphs. Let $R_s(t)$,
$|H_1(t)|/|H_2(t)|$ and $\Delta_{H_1}(t)/\Delta_{H_2}(t)$ be defined as in
Proposition~\ref{general_suff_cond_sender_type_3}, now parametrized by
$t\in(0,1)$. Then $R_s(t)$ decreases in $t$,
$|H_1(t)|/|H_2(t)|$ increases in $t$, and
$\Delta_{H_1}(t)/\Delta_{H_2}(t)$ decreases in $t$.
\end{lemma}
The proof of Lemma~\ref{suffi_cond3_example} is provided in Section~\ref{append_proof_sec_estimand_weight_efficiency} of the Supplementary Material.

\begin{figure}[htbp]
    \centering

    \begin{subfigure}[t]{0.325\textwidth}
        \centering
        \includegraphics[width=\textwidth]{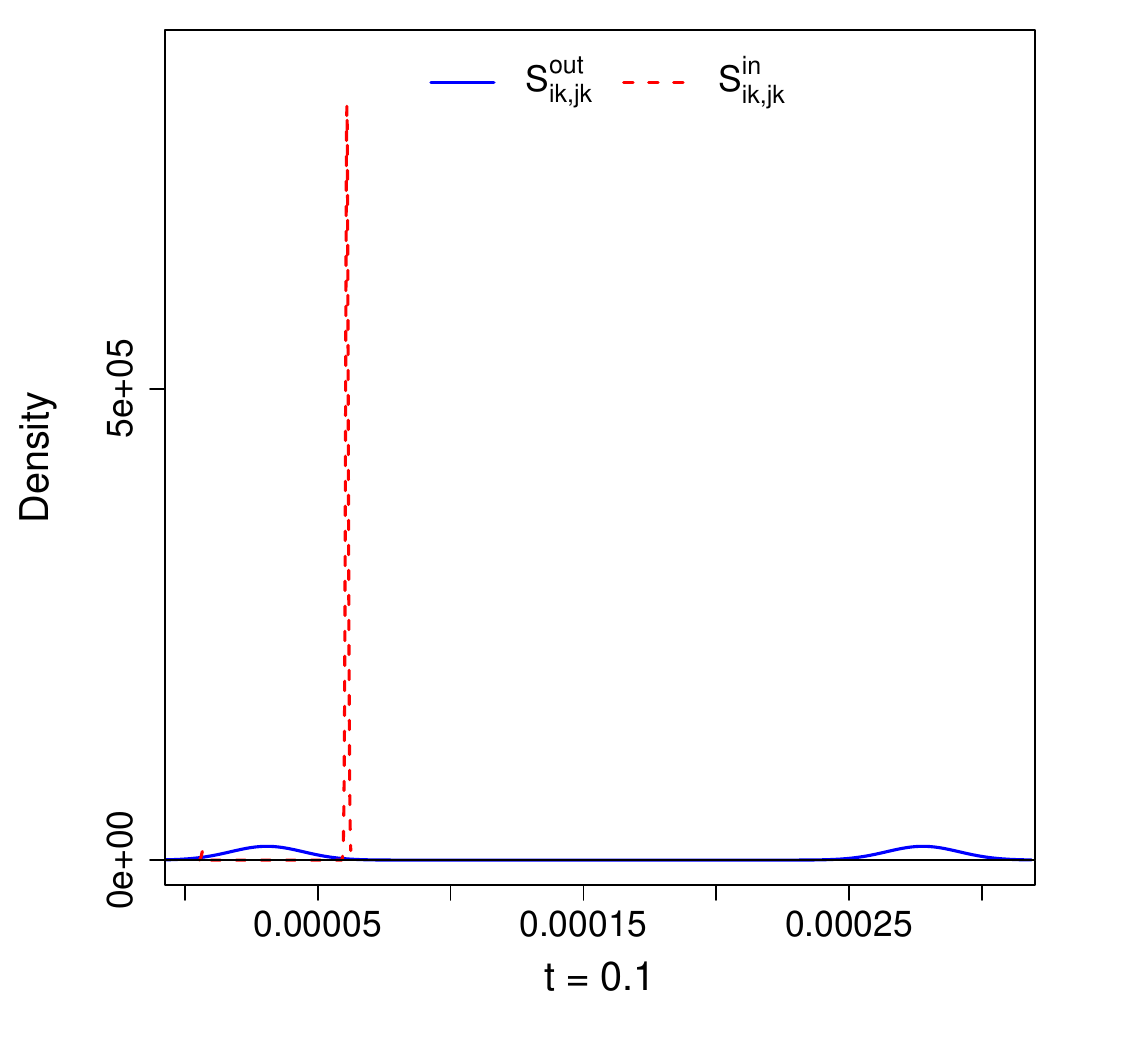}
        \label{fig:estimand-weight-t-01}
    \end{subfigure}
    \hfill
    \begin{subfigure}[t]{0.325\textwidth}
        \centering
        \includegraphics[width=\textwidth]{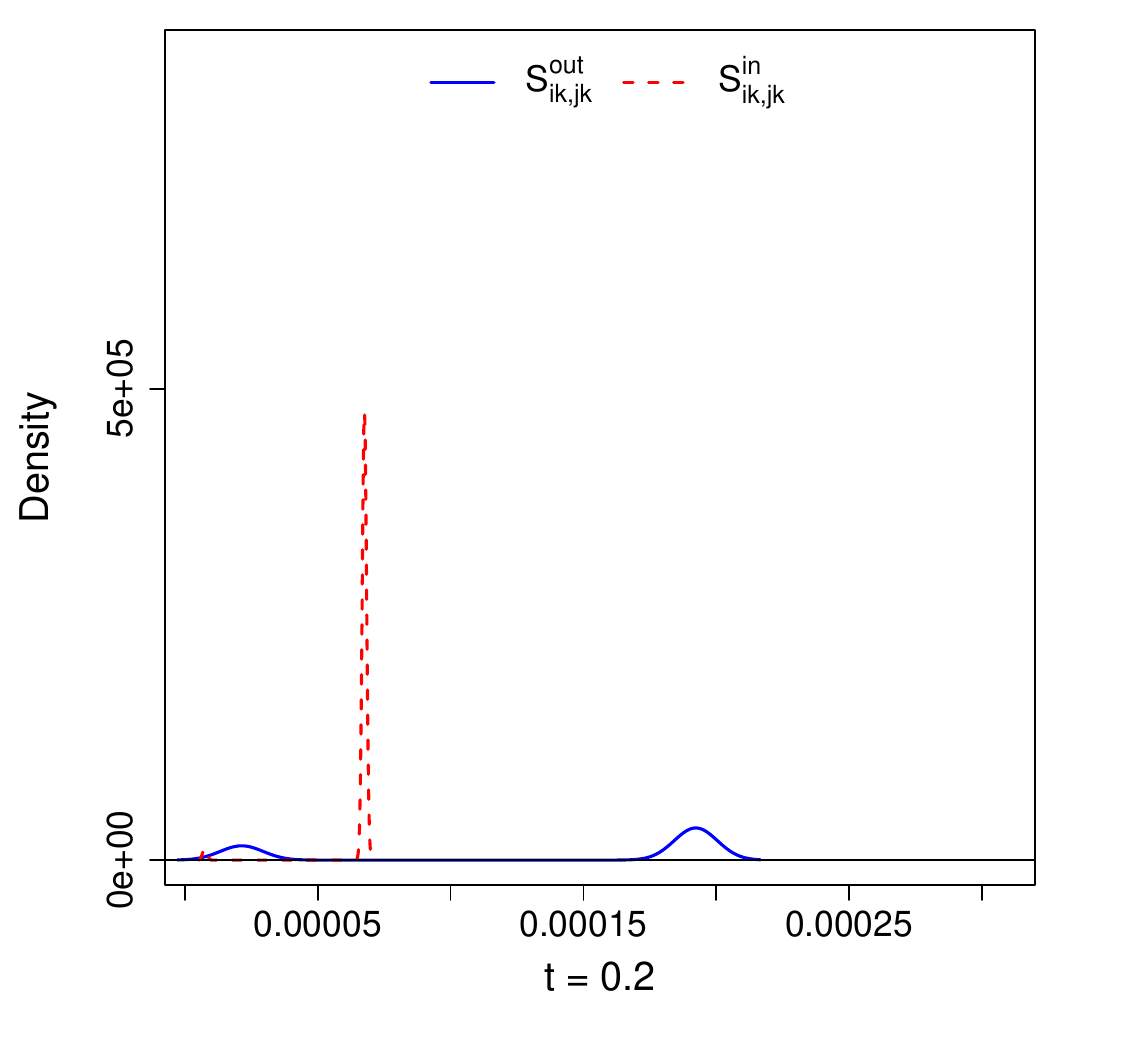}
        \label{fig:estimand-weight-t-02}
    \end{subfigure}
    \hfill
    \begin{subfigure}[t]{0.325\textwidth}
        \centering
        \includegraphics[width=\textwidth]{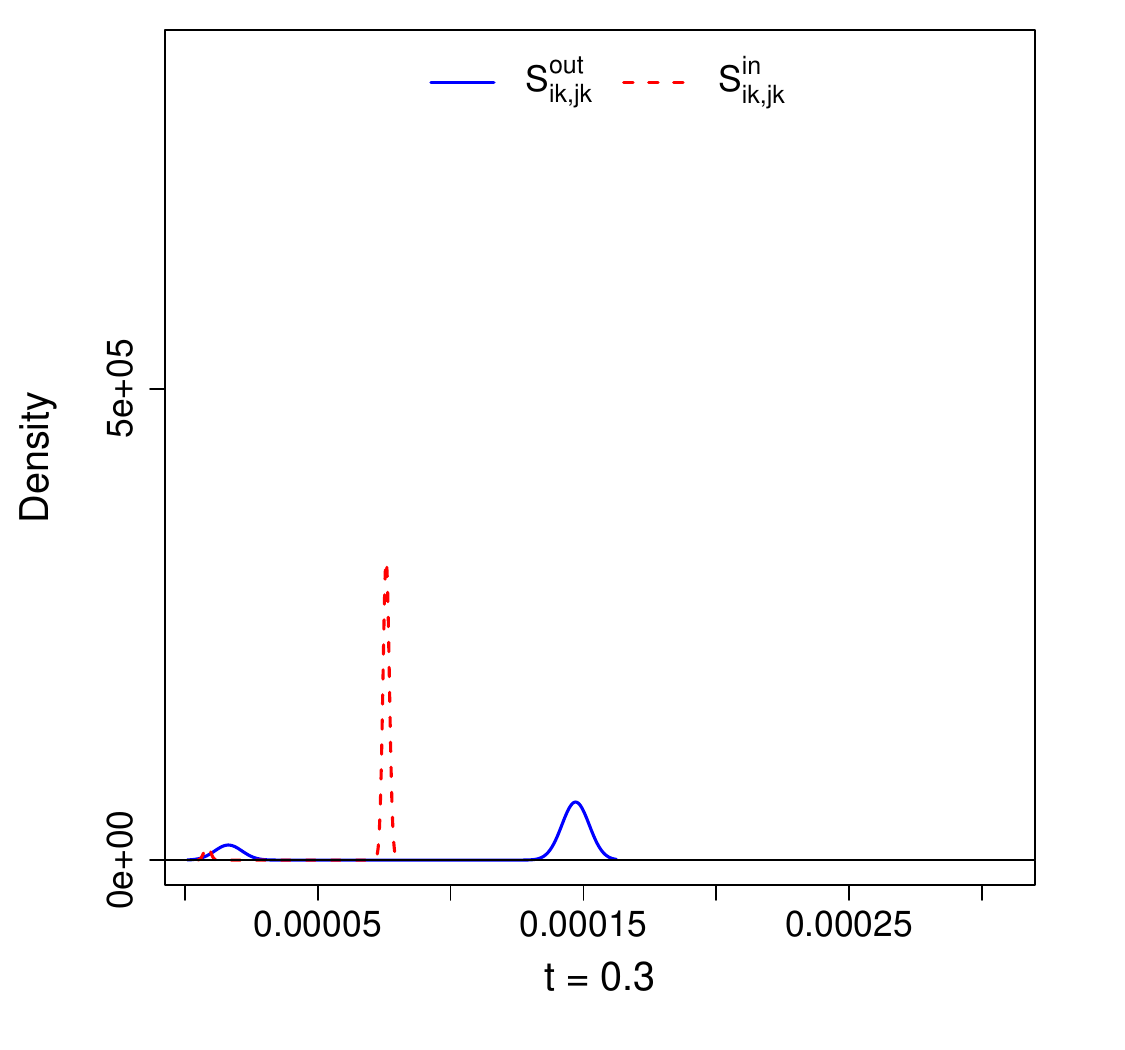}
        \label{fig:estimand-weight-t-03}
    \end{subfigure}

    \vspace{0.5em}

    \begin{subfigure}[t]{0.325\textwidth}
        \centering
        \includegraphics[width=\textwidth]{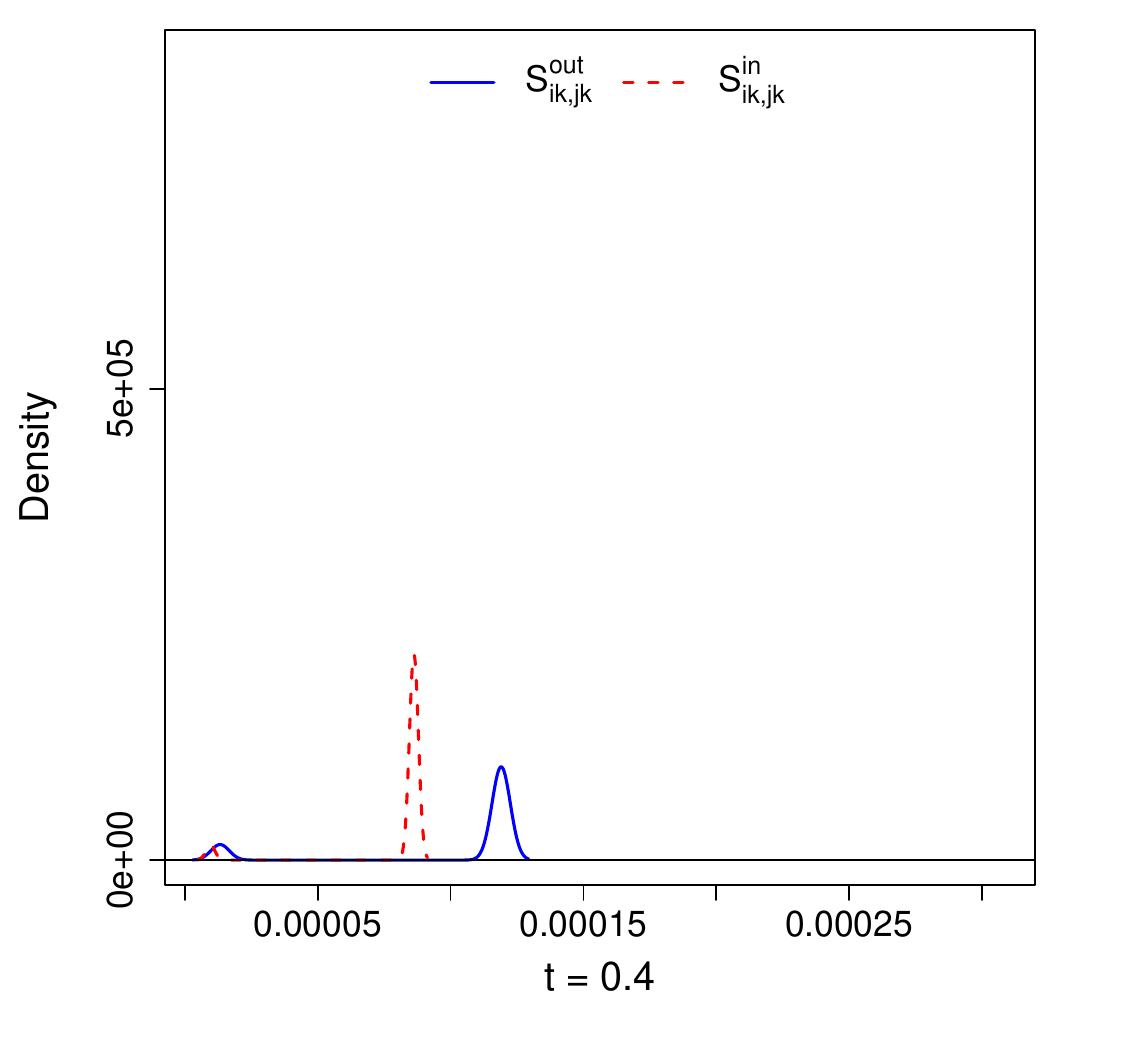}
        \label{fig:estimand-weight-t-04}
    \end{subfigure}
    \hfill
    \begin{subfigure}[t]{0.325\textwidth}
        \centering
        \includegraphics[width=\textwidth]{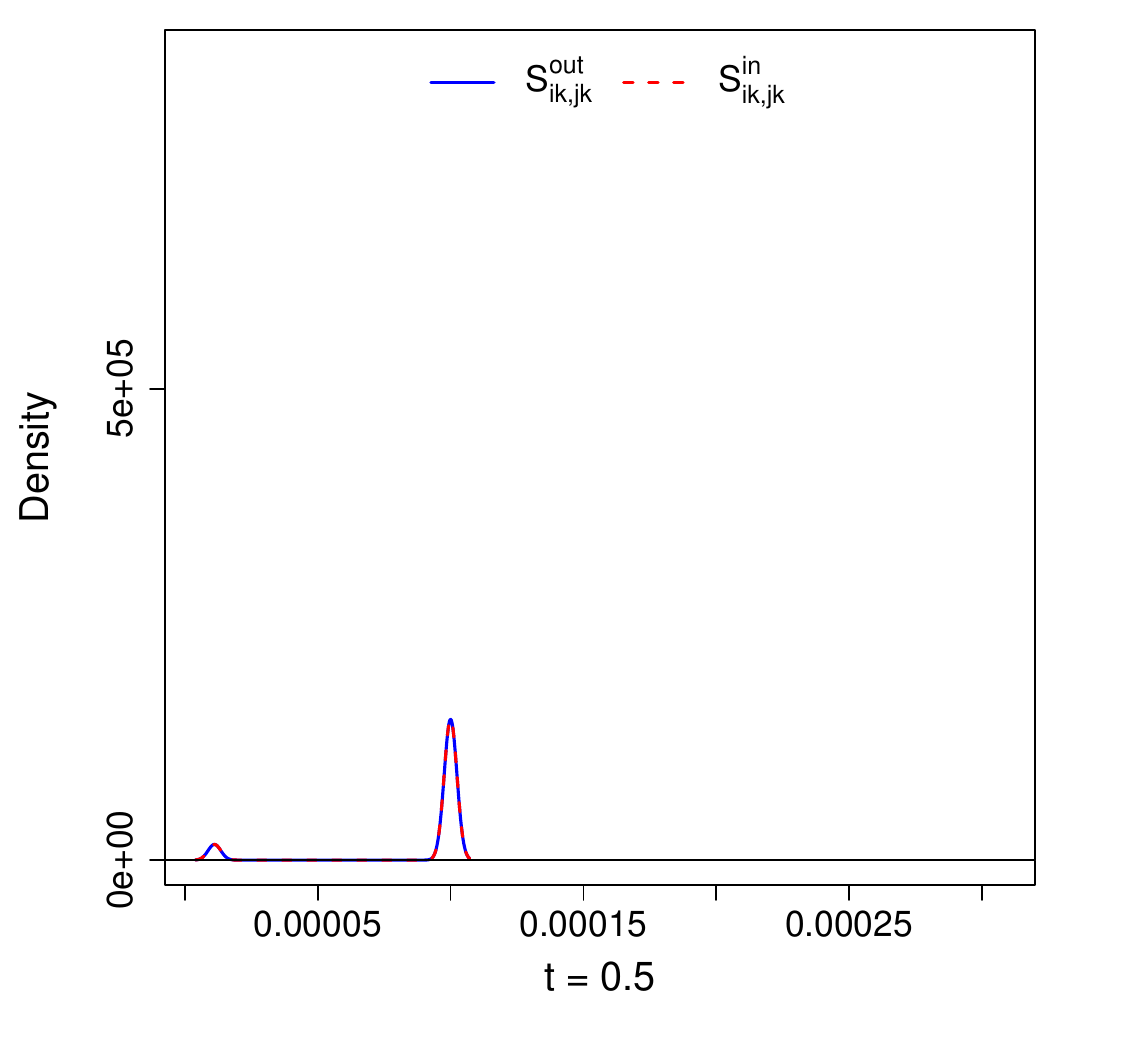}
        \label{fig:estimand-weight-t-05}
    \end{subfigure}
    \hfill
    \begin{subfigure}[t]{0.325\textwidth}
        \centering
        \includegraphics[width=\textwidth]{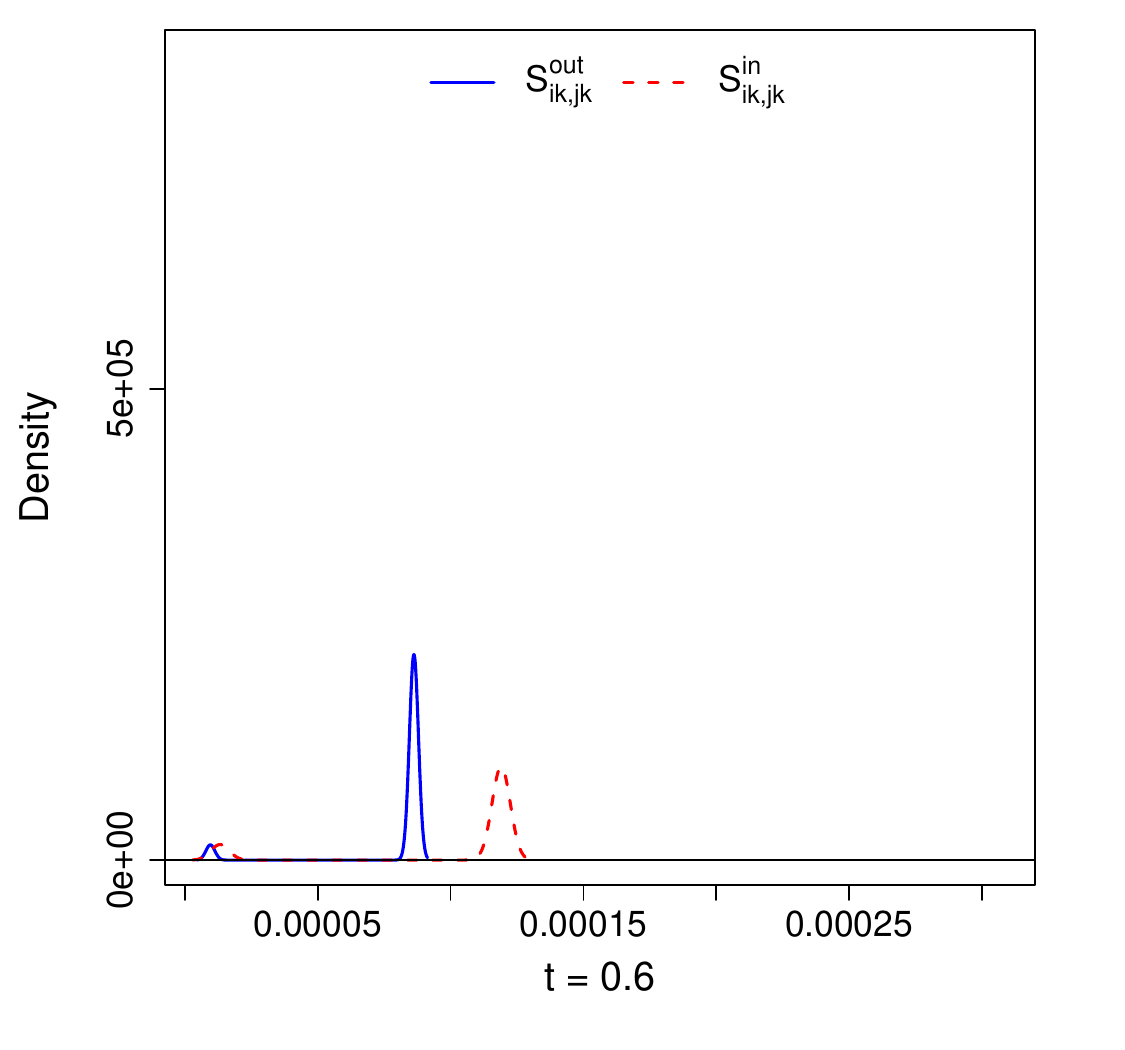}
        \label{fig:estimand-weight-t-06}
    \end{subfigure}

    \vspace{0.5em}

    \begin{subfigure}[t]{0.325\textwidth}
        \centering
        \includegraphics[width=\textwidth]{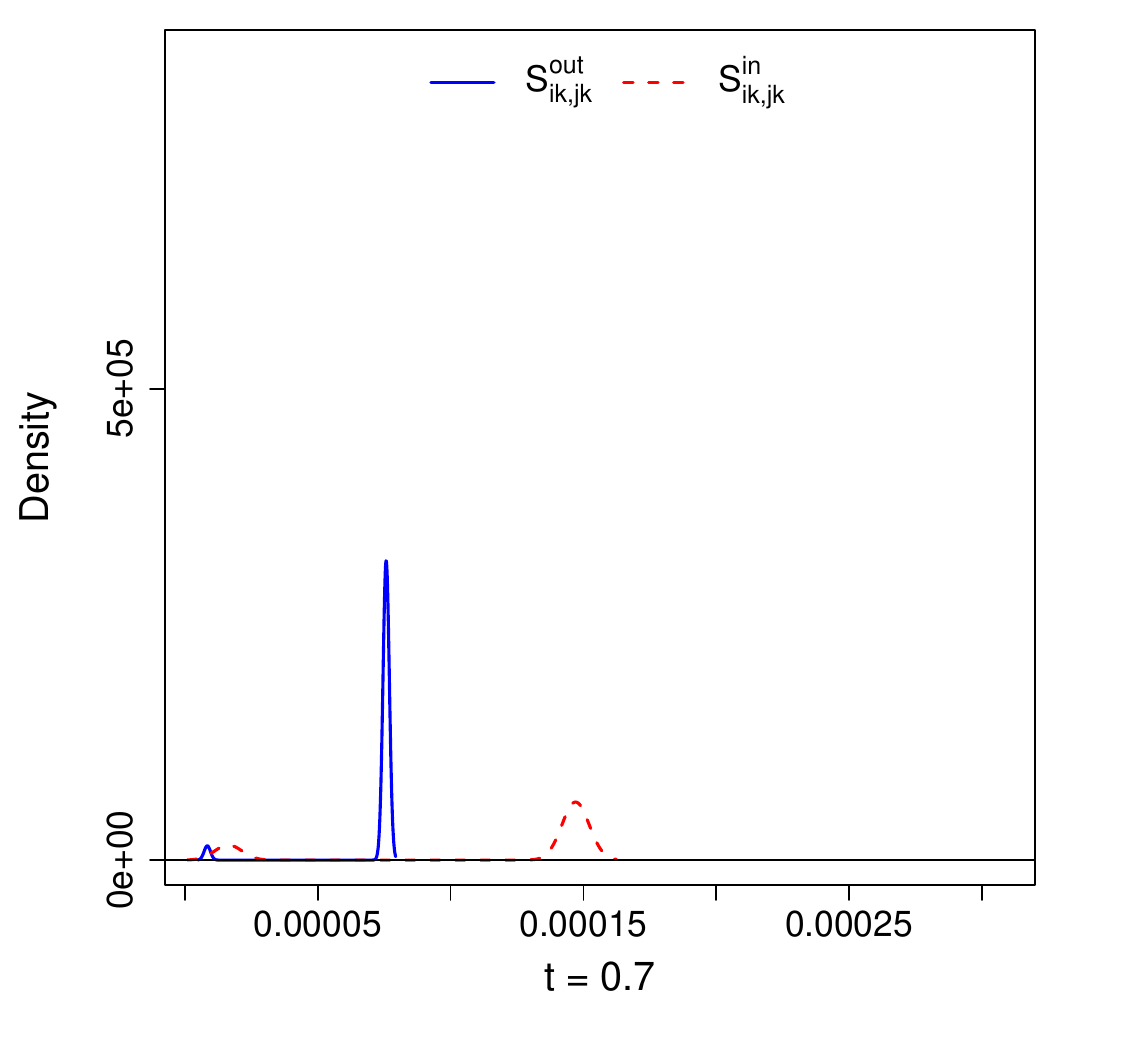}
        \label{fig:estimand-weight-t-07}
    \end{subfigure}
    \hfill
    \begin{subfigure}[t]{0.325\textwidth}
        \centering
        \includegraphics[width=\textwidth]{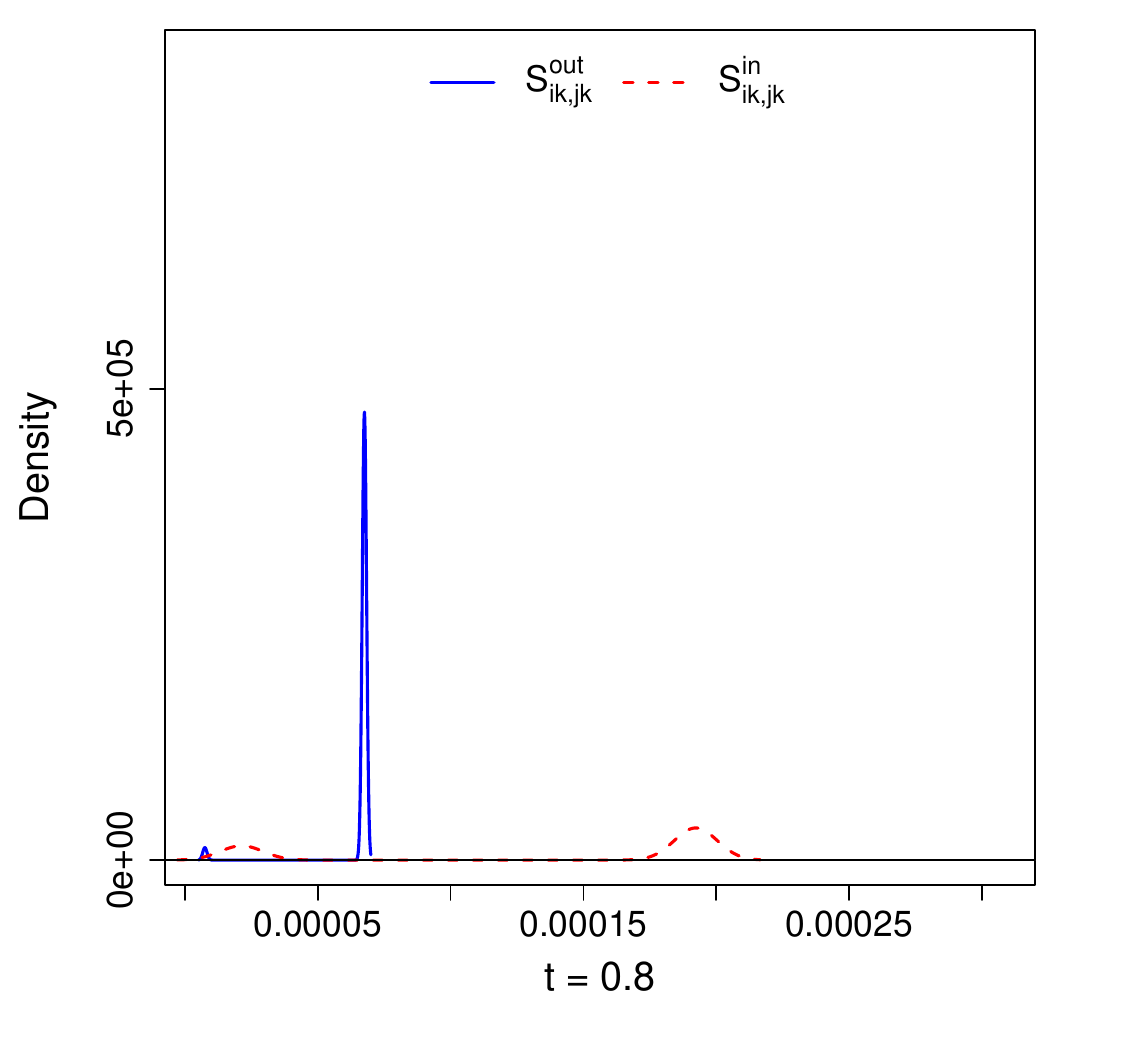}
        \label{fig:estimand-weight-t-08}
    \end{subfigure}
    \hfill
    \begin{subfigure}[t]{0.325\textwidth}
        \centering
        \includegraphics[width=\textwidth]{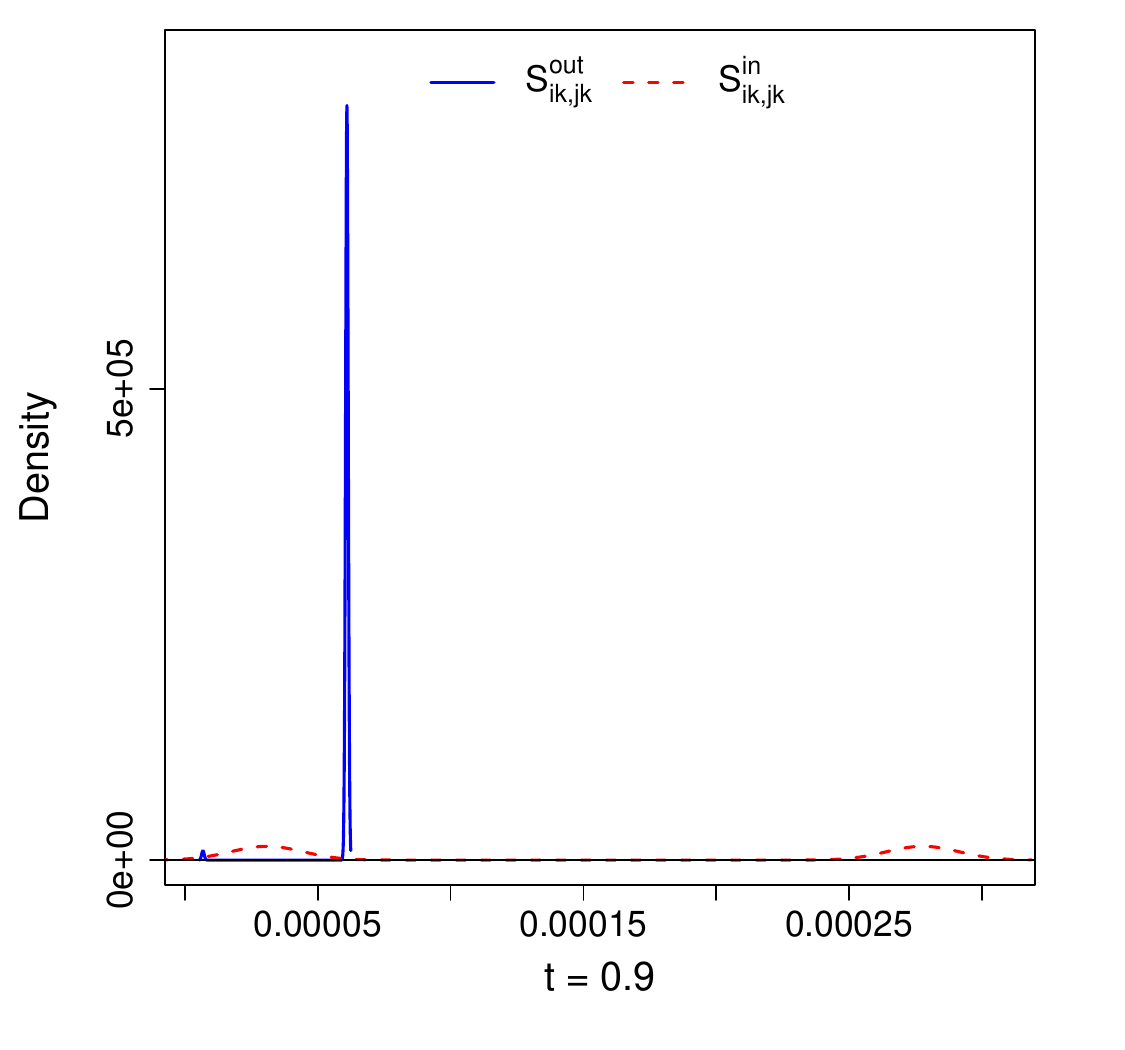}
        \label{fig:estimand-weight-t-09}
    \end{subfigure}

    \caption{Distributions of estimand weights \(S_{ik,jk}^{out}\) and \(S_{ik,jk}^{in}\) for different values of \(t\).}
    \label{fig:estimand-weight-density-all}
\end{figure}
}

{  
\subsection{Comparisons of Exact and Conservative Variances}
\label{Append:simulation_var_compare}
In this section, we consider three simulation settings to compare the conservative and exact variance differences. The first follows the setting in Section~\ref{estimand_weight_efficiency}, except that the hypothetical and realized treatment assignments are generated independently from Bernoulli distributions with probabilities \(\alpha=0.5\) and \(\beta=0.4\), respectively. The results are reported in Table~\ref{tab:prop_in_star_alpha_not_equal_beta}. 

\begin{table}[H]
\centering
\small
\setlength{\tabcolsep}{2.5pt}
\renewcommand{\arraystretch}{1.15}
\caption{Comparison of the conservative and exact variances, together with a decomposition of the exact variance difference, across different proportions of in-star graphs.\protect\footnotemark}
\label{tab:prop_in_star_alpha_not_equal_beta}
\resizebox{\textwidth}{!}{
\begin{tabular}{c C{1.4cm} c c c C{1.9cm} c c C{1.7cm} c c c c}
\hline
$t$
& ${[\mathbb{E}(S^{out}_{ik,jk})}-{\mathbb{E}(S^{in}_{ik,jk})}]\times 10^{5}$ & $R_s$
& $V^c(\hat{\tau}^{out}(\alpha))$
& $V^c(\hat{\tau}^{in}(\alpha))$
& $V^c(\hat{\tau}^{out}(\alpha))
   -V^c(\hat{\tau}^{in}(\alpha))$
& $V(\hat{\tau}^{out}(\alpha))$
& $V(\hat{\tau}^{in}(\alpha))$
& $V(\hat{\tau}^{out}(\alpha))
   -V(\hat{\tau}^{in}(\alpha))$
& $(a)\times 10^{3}$
& $(b)\times 10^{3}$
& $(c)\times 10^{6}$
\\
\hline
0.1 & 9.401 & 3.098 & 0.067 & 0.018 &  0.048 & 0.014 & 0.018 & -0.003 &  -6.649 & 3.055 &  -2.644 \\
0.2 & 7.377 &2.247 & 0.061 & 0.020 &  0.042 & 0.011 & 0.019 & -0.008 & -11.259 & 3.011 &  -3.996 \\
0.3 & 4.745 & 1.688 & 0.053 & 0.022 &  0.031 & 0.009 & 0.021 & -0.012 & -15.022 & 2.568 &  -5.042 \\
0.4 &2.301  & 1.294 & 0.046 & 0.025 &  0.021 & 0.008 & 0.025 & -0.018 & -18.641 & 2.149 &  -6.040 \\
0.5 & 0.002 &1.000 & 0.040 & 0.028 &  0.013 & 0.007 & 0.027 & -0.020 & -22.434 & 1.837 &  -7.111 \\
0.6 & -2.301 & 0.773 & 0.036 & 0.032 &  0.004 & 0.006 & 0.030 & -0.025 & -26.732 & 1.630 &  -8.341 \\
0.7 & -4.745 & 0.592 & 0.032 & 0.037 & -0.005 & 0.005 & 0.036 & -0.031 & -31.653 & 1.443 &  -9.786 \\
0.8 & -7.377 & 0.445 & 0.029 & 0.043 & -0.014 & 0.005 & 0.042 & -0.037 & -36.904 & 1.249 & -11.372 \\
0.9 & -9.401 & 0.323 & 0.027 & 0.048 & -0.022 & 0.004 & 0.044 & -0.040 & -39.335 & 0.940 & -12.045 \\
\hline
\end{tabular}
}
\end{table}
\footnotetext{The quantity $\mathbb{E}(S^{out}_{ik,jk})-\mathbb{E}(S^{in}_{ik,jk})$ is averaged over pairs $(ik,jk)$ satisfying $N^{out}|\mathcal{N}^{out}_{jk}|>0$ and $N^{in}|\mathcal{N}^{in}_{ik}|>0$. The quantity $R_s$ is defined in Proposition~\ref{general_suff_cond_sender_type_3}. The conservative variance difference, $V^c(\hat{\tau}^{out}(\alpha))-V^c(\hat{\tau}^{in}(\alpha))$, is defined in Proposition~\ref{V_out_in_compare}; the exact variance difference, $V(\hat{\tau}^{out}(\alpha))-V(\hat{\tau}^{in}(\alpha))$, and its components (a), (b) and (c) are defined in Proposition~\ref{dif_var_out_var_in}.}

Second, we vary the number of clusters $K$, while keeping all other simulation settings fixed as in Section \ref{estimand_weight_efficiency}; the results are reported in Table~\ref{tab:increase_num_clusters}.
\begin{table}[H]
\centering
\small
\setlength{\tabcolsep}{3pt}
\renewcommand{\arraystretch}{1.15}
\caption{Comparison of the conservative and exact variances, together with a decomposition of the exact variance difference, across different numbers of clusters.}
\label{tab:increase_num_clusters}
\resizebox{\textwidth}{!}{
\begin{tabular}{c C{1.4cm} c c c C{1.9cm} c c C{1.7cm} c c c}
\hline
$K$
& ${[\mathbb{E}(S^{out}_{ik,jk})}-{\mathbb{E}(S^{in}_{ik,jk})}]\times 10^{5}$
& $R_s$
& $V^c(\hat{\tau}^{out}(\alpha))$
& $V^c(\hat{\tau}^{in}(\alpha))$
& $V^c(\hat{\tau}^{out}(\alpha))
   -V^c(\hat{\tau}^{in}(\alpha))$
& $V(\hat{\tau}^{out}(\alpha))$
& $V(\hat{\tau}^{in}(\alpha))$
& $V(\hat{\tau}^{out}(\alpha))
   -V(\hat{\tau}^{in}(\alpha))$
& $(a)\times 10^{3}$
& $(b)\times 10^{3}$
& $(c)\times 10^{6}$
\\
\hline
100  & -147.533 & 0.445 & 0.397 & 0.583 & -0.187 & 0.044 & 0.613 & -0.569 & -503.296 & 0.453 & -227.446 \\
300  & -49.178 & 0.445 & 0.132 & 0.196 & -0.064 & 0.015 & 0.187 & -0.172 & -169.450 & 0.113 &  -75.815 \\
500  & -29.507 & 0.445 & 0.079 & 0.117 & -0.038 & 0.009 & 0.116 & -0.108 & -101.418 & 0.042 &  -45.489 \\
700  & -21.076 & 0.445 & 0.057 & 0.084 & -0.027 & 0.007 & 0.077 & -0.070 &  -72.351 & 0.028 &  -32.492 \\
1000 & -14.753  & 0.445 & 0.040 & 0.059 & -0.019 & 0.005 & 0.054 & -0.050 &  -50.619 & 0.012 &  -22.745 \\
2000 & -7.377 & 0.445 & 0.020 & 0.029 & -0.010 & 0.002 & 0.027 & -0.025 &  -25.365 & 0.004 &  -11.372 \\
3000 & -4.918 & 0.445 & 0.013 & 0.020 & -0.006 & 0.001 & 0.019 & -0.017 &  -16.896 & 0.003 &   -7.582 \\
\hline
\end{tabular}
}
\end{table}
As $K$ increases, both $N^{out}$ and $N^{in}$ increase, whereas $|\mathcal{N}^{out}_{jk}|$ and $|\mathcal{N}^{in}_{ik}|$ remain fixed. Consequently, the magnitudes of $S^{out}_{ik,jk}$ and $S^{in}_{ik,jk}$ decrease with $K$, leading to smaller absolute differences between both the conservative variances and the exact variances; see the sixth and ninth columns of Table~\ref{tab:increase_num_clusters}.

In the third simulation, we consider a more realistic, although less analytically
tractable, network model based on directed preferential attachment. As in
Section~\ref{estimand_weight_efficiency}, we set \(K=2000\) and \(n_k=10\) for
\(k=1,\dots,K\). Within each cluster, we simulate a preferential attachment graph that favors either in-degree or out-degree attachment. Specifically, we use the \texttt{sample\_pa} function in the \texttt{igraph}
package in \texttt{R}, setting the preferential attachment power to \(1\),
the number of edges added at each step to \(1\), and the attractiveness of
vertices with no adjacent edges to \(1\). This procedure generates a preferential attachment graph favoring in-degree attachment; reversing all edge directions produces the corresponding graph favoring out-degree attachment. 

Across clusters, we independently generate preferential attachment graphs favoring in-degree attachment. We retain this orientation for a proportion \(t\in\{0.1,\dots,0.9\}\) of the clusters and reverse all edge directions in the remaining proportion \(1-t\), thereby obtaining graphs favoring out-degree attachment. The hypothetical and realized treatment
probabilities are \(\alpha=0.5\) and \(\beta=0.4\), respectively. The
structural potential-outcome model is the same as in
\eqref{spec_outcome_model}. The results are reported in
Table~\ref{tab:prop_pref_attachment}.

\begin{table}[H]
\centering
\small
\setlength{\tabcolsep}{3pt}
\renewcommand{\arraystretch}{1.15}
\caption{Comparison of the conservative and exact variances, together with a decomposition of the exact variance difference, across different proportions of in-degree preferential attachment graphs.}
\label{tab:prop_pref_attachment}
\resizebox{\textwidth}{!}{
\begin{tabular}{c C{1.4cm} c c c C{1.9cm} c c C{1.7cm} c c c}
\hline
$t$ & ${[\mathbb{E}(S^{out}_{ik,jk})}-{\mathbb{E}(S^{in}_{ik,jk})}]\times 10^{5}$
& $R_s$
& $V^c(\hat{\tau}^{out}(\alpha))$
& $V^c(\hat{\tau}^{in}(\alpha))$
& $V^c(\hat{\tau}^{out}(\alpha))
   -V^c(\hat{\tau}^{in}(\alpha))$
& $V(\hat{\tau}^{out}(\alpha))$
& $V(\hat{\tau}^{in}(\alpha))$
& $V(\hat{\tau}^{out}(\alpha))
   -V(\hat{\tau}^{in}(\alpha))$
& $(a)\times 10^{3}$
& $(b)\times 10^{3}$
& $(c)\times 10^{6}$
\\
\hline
0.1 & 2.611  &1.900 & 0.021 & 0.025 & -0.003 & 0.005 & 0.008 & -0.003 & -3.063 & 0.375 & -0.959 \\
0.2 & 2.080 & 1.587 & 0.023 & 0.026 & -0.003 & 0.005 & 0.008 & -0.003 & -3.582 & 0.459 & -1.138 \\
0.3 & 1.393 &1.362 & 0.023 & 0.027 & -0.003 & 0.005 & 0.008 & -0.004 & -4.017 & 0.454 & -1.278 \\
0.4 & 0.664 & 1.180 & 0.023 & 0.028 & -0.004 & 0.004 & 0.009 & -0.004 & -4.394 & 0.400 & -1.394 \\
0.5 &-0.002  &1.002 & 0.023 & 0.029 & -0.006 & 0.004 & 0.008 & -0.004 & -4.641 & 0.340 & -1.468 \\
0.6 & -0.718 &0.855 & 0.022 & 0.029 & -0.007 & 0.004 & 0.008 & -0.004 & -4.734 & 0.293 & -1.492 \\
0.7 &  -1.425 & 0.734 & 0.022 & 0.030 & -0.008 & 0.004 & 0.008 & -0.004 & -4.544 & 0.231 & -1.437 \\
0.8 &-1.995 &0.628 & 0.021 & 0.030 & -0.009 & 0.004 & 0.008 & -0.004 & -4.004 & 0.166 & -1.274 \\
0.9 &-2.617 &0.523 & 0.020 & 0.029 & -0.009 & 0.003 & 0.006 & -0.003 & -2.885 & 0.062 & -0.929 \\
\hline
\end{tabular}
}
\end{table}

Here, the distribution of the estimand weights is similar to that in Figure~\ref{fig:estimand-weight-density-all}. Accordingly, the mean values of $S^{out}_{ik,jk}$ and $S^{in}_{ik,jk}$, together with $R_s$, exhibit a pattern similar to that in Table~\ref{tab:prop_in_star}. As $t$ increases, the relative magnitude shifts from larger $S^{out}_{ik,jk}$ to larger $S^{in}_{ik,jk}$, as shown in columns~2 and~3 of Table~\ref{tab:prop_pref_attachment}, while $R_s$ decreases. 

This pattern, however, does not translate into a monotone increase in $V^c(\hat{\tau}^{out}(\alpha))$ and $V(\hat{\tau}^{out}(\alpha))$, or a monotone decrease in $V^c(\hat{\tau}^{in}(\alpha))$ and $V(\hat{\tau}^{in}(\alpha))$, in contrast to those in Table~\ref{tab:prop_in_star} of the manuscript. One possible explanation is that the differences between the outward and inward estimand weights are smaller than those in Table~\ref{tab:prop_in_star}. The resulting weight imbalance is therefore less pronounced, so the relative magnitudes of $V^c(\hat{\tau}^{out}(\alpha))$ and $V^c(\hat{\tau}^{in}(\alpha))$ depend more strongly on the interaction among the outcomes of out-neighbors. Consequently, both the conservative variances and the exact variances remain relatively stable as $t$ varies, and their outward-minus-inward differences remain negative throughout the range of $t$ considered. We do not connect this example directly to Proposition~\ref{general_suff_cond_sender_type_3}, because Assumption~(ii) of that proposition is not satisfied in this simulation.
}

\section{ H\'{a}jek estimators }
\label{hajek}
Based on the Horvitz-Thompson estimator $\hat{\tau}^{out}(\alpha)$ in Definitions \ref{est_tau_out}, the corresponding H\'{a}jek estimator for outward spillover effects is 
\begin{equation*}
    \begin{split}
       \hat{\tau}^{out}_{hj}(\alpha)= \frac{\sum_{k=1}^K \sum_{jk \in \mathcal{N}^{out}_{k}} \ W_{jk}(\mathbf{Z}_k) \cdot Z_{jk} \cdot \bar{Y}^{out}_{jk} }{ \sum_{k=1}^K \sum_{jk \in \mathcal{N}^{out}_k} \ W_{jk}(\mathbf{Z}_k) \cdot Z_{jk} }-  \frac{\sum_{k=1}^K \sum_{jk \in \mathcal{N}^{out}_{k}} \ W_{jk}(\mathbf{Z}_k) \cdot (1-Z_{jk}) \cdot \bar{Y}^{out}_{jk} }{ \sum_{k=1}^K \sum_{jk \in \mathcal{N}^{out}_k} \ W_{jk}(\mathbf{Z}_k) \cdot (1-Z_{jk})  }. 
    \end{split}
\end{equation*}
Based on the Horvitz-Thompson estimator $\tilde{W}^{1}_{ik}(\mathbf{Z}_k)$ and $\tilde{W}^{0}_{ik}(\mathbf{Z}_k)$ for $ik \in \mathcal{N}^{in}_{k}$ and $k = 1\cdots,K$ and that of $\hat{\tau}^{in}(\alpha)$ in Definition \ref{est_tau_in}, the corresponding H\'{a}jek estimator for inward spillover effect is 
\begin{equation}
\label{est_in_spillover_natural_estimand}
    \begin{split}
        \hat{\tau}^{in}_{hj} (\alpha) & = \frac{\sum_{k=1}^K \sum_{ik \in  \mathcal{N}^{in}_k}  \tilde{W}^1_{ik}(\mathbf{Z}_k) Y_{ik} }{ \sum_{k=1}^K \sum_{ik \in  \mathcal{N}^{in}_k}  \tilde{W}^1_{ik}(\mathbf{Z}_k) } - \frac{ \sum_{k=1}^K \sum_{ik \in  \mathcal{N}^{in}_{k}} \tilde{W}^0_{jk}(\mathbf{Z}_k) {Y}_{ik}}{\sum_{k=1}^K \sum_{ik \in  \mathcal{N}^{in}_{k}} \tilde{W}^0_{jk}(\mathbf{Z}_k)}. \\
    \end{split}
\end{equation}
Then by applying the proofs in $A.2.1$ and $A.4.2$ in \cite{wang2024designbasedinferencespatialexperiments} to $\hat{\tau}^{out}_{hj}(\alpha)$, we have the conservative variances for $\hat{\tau}^{out}_{hj}(\alpha)$ as
\begin{equation}
\label{V_out_hj}
    \begin{split}
        V^c \left( \hat{\tau}^{out}_{hj} (\alpha) \right) & = \frac{1}{N^{out \ 2 }} \sum_{k=1}^K \sum_{jk \in \mathcal{N}^{out}_{k}} |\mathcal{N}^{out}_{k}| \cdot \mathbb{E} \left\lbrace Z_{jk} W_{jk} (\mathbf{Z}_k)  \left( \bar{Y}^{out}_{jk}- \mu^{out}(1, \alpha)   \right)    \right\rbrace^2  \\
       & + \frac{1}{N^{out \ 2 }} \sum_{k=1}^K \sum_{jk \in \mathcal{N}^{out}_{k}} |\mathcal{N}^{out}_{k}| \cdot \mathbb{E} \left\lbrace (1-Z_{jk}) W_{jk}(\mathbf{Z}_{k}) \left( \bar{Y}^{out}_{jk}- \mu^{out}(0, \alpha)   \right)    \right\rbrace^2
    \end{split}
\end{equation}
where $\mu^{out}(z,\alpha)=\frac{1}{N^{out}} \sum_{k=1}^K  \sum_{jk \in \mathcal{N}_k^{out} } \mu^{out}_{jk}(Z_{jk}=z,\alpha) $ and $z$ takes values in $\{0,1\}$. \(\mu^{out}_{jk}(z_{jk} = z, \alpha)\) is specified in Definition \ref{out_spillover}. 

To derive $V^c\left(\hat{\tau}^{in}_{hj}(\alpha)\right)$, we first write $\hat{\tau}^{in}_{hj}(\alpha)$ in \eqref{est_in_spillover_natural_estimand} as 
\begin{small}
\begin{equation}
\label{est_in_spillover_change_sum_estimand}
\hat{\tau}^{in}_{hj}(\alpha)= \frac{\sum_{k=1}^K \sum_{jk \in \mathcal{N}^{out}_{k}} W_{jk}(\mathbf{Z}_k) \cdot Z_{jk} \cdot \widetilde{Y}^{out}_{jk}   }{ \sum_{k=1}^K \sum_{jk  \in \mathcal{N}^{out}_k} W_{jk}(\mathbf{Z}_k) Z_{jk} \left( \sum_{ik \in \mathcal{N}^{out}_{jk}} \frac{1}{|\mathcal{N}^{in }_{ik}|}   \right)   }- \frac{\sum_{k=1}^K \sum_{jk \in \mathcal{N}^{out}_{k}} \ W_{jk}(\mathbf{Z}_k) \cdot (1-Z_{jk}) \cdot \widetilde{Y}^{out}_{jk} }{ \sum_{k=1}^K \sum_{jk \in \mathcal{N}^{out}_k} \ W_{jk}(\mathbf{Z}_k) \cdot (1-Z_{jk}) \left( \sum_{ik \in \mathcal{N}^{out}_{jk}} \frac{1}{|\mathcal{N}^{in}_{ik}|}   \right) }.
\end{equation}
\end{small}
\noindent Then by applying the proofs detailed in Sections $A.2.1$ and $A.4.2$ of \cite{wang2024designbasedinferencespatialexperiments} to \eqref{est_in_spillover_change_sum_estimand}, we can derive \(V^c\left(\hat{\tau}^{in}_{hj}(\alpha)\right)\) as shown below: 
\begin{equation}
\label{V_in_hj}
    \begin{split}
        V^c\left(\hat{\tau}^{in}_{hj}(\alpha)\right) & = \frac{1}{(N^{in})^2} \sum_{k=1}^K \sum_{jk \in \mathcal{N}^{out}_k} |\mathcal{N}^{out}_k | \cdot \left[ \mathbb{E} \left\lbrace W_{jk}(\mathbf{Z}_k) Z_{jk} \left( \widetilde{Y}^{out}_{jk}-\sum_{ik \in \mathcal{N}^{out}_{jk}} {|\mathcal{N}^{in}_{ik}|^{-1}} \mu^{in}(1,\alpha)  \right) \right\rbrace^2    \right. \\
        & + \left.  \mathbb{E} \left\lbrace W_{jk}(\mathbf{Z}_k) (1-Z_{jk}) \left( \widetilde{Y}^{out}_{jk}-\sum_{ik \in \mathcal{N}^{out}_{jk}} {|\mathcal{N}^{in}_{ik}|^{-1}} \mu^{in}(0,\alpha)  \right)            \right\rbrace^2            \right] \\
    \end{split}
\end{equation}
where $\mu^{in}(z,\alpha)= \frac{1}{N^{in}} \sum_{k=1}^K \sum_{ik \in \mathcal{N}^{in}_k  }  \mu^{in}_{ik}\left(z ,\alpha \right)$ and $z$ takes values in $\{0, 1 \}$. $\mu^{in}_{ik}\left(z ,\alpha \right)$ is defined in Definition \ref{in_spillover}. 

{  For the difference $V^c(\hat{\tau}^{out}_{hj}(\alpha))-V^c(\hat{\tau}^{in}_{hj}(\alpha))$, it can be expressed using the formula \eqref{V_out_in_compare_eq} of Proposition \ref{V_out_in_compare}, with appropriate modifications to terms $\sum_{ik \in N^{out}_{jk}} (S^{out}_{ik,jk}- S^{in}_{ik,jk}) Y_{ik}$ and $\sum_{ik \in N^{out}_{jk}} (S^{out}_{ik,jk}+ S^{in}_{ik,jk}) Y_{ik}$
, by utilizing the variance formulas \eqref{V_out_hj} and \eqref{V_in_hj}. Specifically, if \( \mu^{out}(z,\alpha) = \mu^{in}(z,\alpha) := \mu(z,\alpha) \) for \( z \in \{0,1\} \), where \( \mu^{out}(z,\alpha) \) and \( \mu^{in}(z,\alpha) \) are as defined in \eqref{V_out_hj} and \eqref{V_in_hj}, these two terms are replaced by $\sum_{ik \in N^{out}_{jk}} (S^{out}_{ik,jk}- S^{in}_{ik,jk})  \left( Y_{ik}-\mu(z,\alpha) \right)$ and $\sum_{ik \in N^{out}_{jk}} (S^{out}_{ik,jk}+ S^{in}_{ik,jk}) \left( Y_{ik}-\mu(z,\alpha) \right)$ where \( z = 1 \) when \( Z_{jk} = 1 \) and \( z = 0 \) when \( Z_{jk} = 0 \). The sign of the difference between these variances is difficult to determine due to the dependence on the subtraction of \( \mu(z,\alpha) \) for each replaced term, even under the same setting as in Section \ref{estimand_weight_efficiency}, where \( Y_{ik} > 0 \) for all \( ik = 1, \dots, n_k \) and \( k =1, \dots, K \).}

\section{Technical details}
\label{sec_tech_details}
\subsection{Technical Details for Section \ref{comp_out_in_spillover}}
\label{append_comp_out_in_spillover}
Throughout this section, let $[K]:= \{1,\cdots, K\}$, $[n]:= \{1,\cdots, n\}$ and $[n_k]:=\{1,\cdots,n_k\}$ for $k=1,\cdots, K$.
\begin{proof}[Proof of Theorem \ref{difference_out_in}]
Let $A_{ij}$ denote the indicator for whether unit $ik$ points to unit $jk$, i.e., $A_{ij}=1$ if $e_{ik, jk}\in E_k$. The inward spillover effect defined in Definition \ref{in_spillover} can be written as
\begin{equation}
\label{prof_difference_out_in_int_1}
    \begin{split}
       & \frac{1}{N^{in }} \sum_{k=1}^K \sum_{ik \in \mathcal{N}^{in}_k } \frac{ 1  }{|\mathcal{N}^{in}_{ik}|} \sum_{jk \in \mathcal{N}^{in}_{ik}} \left( \bar{Y}_{ik} (Z_{jk}=1, \alpha)-  \bar{Y}_{ik} (Z_{jk}=0, \alpha)  \right) \\
       & = \frac{1}{N^{in}} \sum_{k=1}^K \sum_{ik \in \mathcal{N}^{in}_k} \frac{ 1  }{|\mathcal{N}^{in}_{ik}|} \sum_{jk=1}^{n_k} A_{ji}   \left( \bar{Y}_{ik} (Z_{jk}=1, \alpha)-  \bar{Y}_{ik} (Z_{jk}=0, \alpha)  \right) \\
       & = \frac{1}{N^{in}} \sum_{k=1}^K \sum_{jk=1}^{n_k} \sum_{ik \in \mathcal{N}^{in}_k}  A_{ji} \frac{1}{|\mathcal{N}^{in}_{ik}|} \left( \bar{Y}_{ik} (Z_{jk}=1, \alpha)-  \bar{Y}_{ik} (Z_{jk}=0, \alpha)  \right) \\
       & =_{(1)}  \sum_{k=1}^K \sum_{jk=1}^{n_k} \sum_{ik \in \mathcal{N}^{out}_{jk}} \frac{1}{ N^{in} |\mathcal{N}^{in}_{ik}|} \left( \bar{Y}_{ik} (Z_{jk}=1, \alpha)-  \bar{Y}_{ik} (Z_{jk}=0, \alpha)  \right) \\
       & =_{(2)} \sum_{k=1}^K \sum_{jk \in \mathcal{N}^{out}_k} \sum_{ik \in \mathcal{N}^{out}_{jk}} \frac{1}{ N^{in} |\mathcal{N}^{in}_{ik}|} \left( \bar{Y}_{ik} (Z_{jk}=1, \alpha)-  \bar{Y}_{ik} (Z_{jk}=0, \alpha)  \right) \\
    \end{split}
\end{equation}
$(1)$ is by the fact that $A_{ji}=1$ implies that (a) $ik \in \mathcal{N}^{in}_k$, (b) $ik \in \mathcal{N}^{out}_{jk}$. Furthermore, for any $ik \in \mathcal{N}^{out}_{jk}$, $ik \in \mathcal{N}^{in}_k$. Therefore, $\mathcal{N}^{out}_{jk} \subset \mathcal{N}^{in}_k$. Then $\mathcal{N}^{in}_k \cap \mathcal{N}^{out}_{jk}=\mathcal{N}^{out}_{jk}$. $(2)$ is by the fact that for $jk \not\in \mathcal{N}^{out}_k$, $\mathcal{N}^{out}_{jk}=\emptyset$. Then 
{ 
\begin{equation}
\label{prof_difference_out_in_int_2}
    \begin{split}
   & \tau^{out}(\alpha)-\tau^{in}(\alpha) \\
   &=_{(1)} \sum_{k=1}^K \sum_{jk \in \mathcal{N}^{out}_{k} } \sum_{ik \in \mathcal{N}^{out}_{jk} }\left( \frac{1}{N^{out} \ |\mathcal{N}^{out}_{jk}|} - \frac{1}{N^{in} \ |\mathcal{N}^{in}_{ik}| }  \right) \left( \bar{Y}_{ik}(Z_{jk}=1,\alpha)- \bar{Y}_{ik}(Z_{jk}=0,\alpha) \right)    \\ 
   & =_{(2)} \sum_{(ik,jk)\in L_1}
\Delta_{ik,jk}\tau_{ik,jk}(\alpha)
+
\sum_{(ik,jk)\in L_2}
\Delta_{ik,jk}\tau_{ik,jk}(\alpha).
    \end{split}
\end{equation}
$(1)$ is by equation \eqref{prof_difference_out_in_int_1} and Definition \ref{out_spillover}. (2) follows directly from the definitions of \(L_1\) and \(L_2\). Therefore, the sufficient and necessary condition of $\tau^{out}(\alpha) \neq \tau^{in}(\alpha)$ is that equation \eqref{prof_difference_out_in_int_2} is not equal to $0$. }
\end{proof}

{ 
\begin{proof}[Proof of Corollary \ref{corol:suff_cond3_not_equal_relaxed}]
By Theorem~\ref{difference_out_in},
\begin{equation*}
\begin{split}
  & \tau^{out}(\alpha)-\tau^{in}(\alpha)
=
\sum_{(ik,jk)\in L_1}
\Delta_{ik,jk}\tau_{ik,jk}(\alpha)
+
\sum_{(ik,jk)\in L_2}
\Delta_{ik,jk}\tau_{ik,jk}(\alpha). \\
\end{split}
\end{equation*}
If $L_2=\emptyset$ and that there exists at least one pair
$(ik,jk)\in L_1$ such that $\Delta_{ik,jk}\tau_{ik,jk}(\alpha)>0$, then 
\begin{equation*}
    \begin{split}
        \tau^{out}(\alpha)-\tau^{in}(\alpha)=\sum_{(ik,jk)\in L_1}
\Delta_{ik,jk}\tau_{ik,jk}(\alpha)>0.
    \end{split}
\end{equation*}
If $L_1=\emptyset$ and there exists at least one pair
$(ik,jk)\in L_2$, then
\begin{equation*}
    \begin{split}
        \tau^{out}(\alpha)-\tau^{in}(\alpha)=\sum_{(ik,jk)\in L_2}
\Delta_{ik,jk}\tau_{ik,jk}(\alpha)<0.
    \end{split}
\end{equation*}
This completes the proof.
\end{proof}
}

\begin{proof}[Proof of Condition \ref{ave_equivalence_cond_1}]
\label{proof_ave_equivalence_cond_1}
Given Condition \ref{ave_equivalence_cond_1} is satisfied, for the outward spillover effect,  we have
\begin{equation}
\label{proof_uncond_sit1_equal_int1}
    \begin{split}
        \tau^{out}(\alpha)=& \frac{1}{N^{out}} \sum_{k=1}^K \sum_{jk \in \mathcal{N}^{out}_k} \sum_{ik \in \mathcal{N}^{out}_{jk}} \frac{ 1 }{|\mathcal{N}^{out}_{jk}|} \left( \bar{Y}_{ik}(Z_{jk}=1, \alpha)- \bar{Y}_{ik}(Z_{jk}=0, \alpha) \right) \\
        &  = \frac{1}{N^{out}} \sum_{k=1}^K c_k \sum_{jk \in \mathcal{N}^{out}_k} \sum_{ik \in \mathcal{N}^{out}_{jk}} \frac{1 }{|\mathcal{N}^{out}_{jk}|}= \frac{1}{N^{out}} \sum_{k=1}^K c_k \cdot |\mathcal{N}^{out}_{k}|. \\
    \end{split}
\end{equation}
Similarly, for the inward spillover effect, 
\begin{equation}
\label{proof_uncond_sit1_equal_int2}
    \begin{split}
        \tau^{in}(\alpha)=& \frac{1}{N^{in}} \sum_{k=1}^K \sum_{ik \in \mathcal{N}^{in}_k} \frac{1}{|\mathcal{N}^{in}_{ik}|} \sum_{jk \in \mathcal{N}^{in}_{ik}} \left( \bar{Y}_{ik} (Z_{jk}=1, \alpha)-  \bar{Y}_{ik} (Z_{jk}=0, \alpha)  \right) = \frac{1}{N^{in}} \sum_{k=1}^K c_k \sum_{ik \in \mathcal{N}^{in}_{k}} \frac{1}{|\mathcal{N}^{in}_{ik}|} \cdot |\mathcal{N}^{in}_{ik}|\\
        &= \frac{1}{N^{in}} \sum_{k=1}^K c_k |\mathcal{N}^{in}_k|=_{(1)} \frac{1}{N^{in }} \sum_{k=1}^K c_k \cdot |\mathcal{N}^{out}_k|=_{(2)} \frac{1}{N^{out }} \sum_{k=1}^K c_k \cdot |\mathcal{N}^{out}_k|.\\ 
    \end{split}
\end{equation}
$(1)$ is based on the fact that the graph is undirected. Then for $ik \in [n_k]$, $|\mathcal{N}^{in}_{ik}|=|\mathcal{N}^{out}_{ik}|$. Therefore, $|\mathcal{N}^{out}_k|=|\mathcal{N}^{in}_k|$ for each $k \in [K]$. $(2)$ is by $N^{out}=\sum_{k=1}^K | \mathcal{N}^{out}_k |=\sum_{k=1}^K | \mathcal{N}^{in}_k |= N^{in}$. Based on \eqref{proof_uncond_sit1_equal_int1} and \eqref{proof_uncond_sit1_equal_int2}, $\tau^{out}(\alpha)=\tau^{in}(\alpha)$. 
\end{proof}

\begin{proof}[Proof of Condition \ref{ave_equivalence_cond_2}]
\label{proof_ave_equivalence_cond_2}
Given that Condition \ref{ave_equivalence_cond_2} holds, we have, for the outward spillover effect, 
\begin{small}
\begin{equation}
\label{proof_uncond_sit2_equal_int1}
    \begin{split}
        & \frac{1}{N^{out}} \sum_{k=1}^K \sum_{jk \in \mathcal{N}^{out}_k} \sum_{ik \in \mathcal{N}^{out}_{jk}} \frac{ 1 }{|\mathcal{N}^{out}_{jk}|} \left( \bar{Y}_{ik}(Z_{jk}=1, \alpha)- \bar{Y}_{ik}(Z_{jk}=0, \alpha) \right) = \frac{c}{N^{out}} \sum_{k=1}^K  \sum_{jk \in \mathcal{N}^{out}_{k}} \sum_{ik \in \mathcal{N}^{out}_{jk}} \frac{1 }{|\mathcal{N}^{out}_{jk}|}=c. \\
    \end{split}
\end{equation}
\end{small}
Similarly, for the inward spillover effect, 
\begin{equation}
\label{proof_uncond_sit2_equal_int2}
    \begin{split}
        & \frac{1}{N^{in}} \sum_{k=1}^K \sum_{i \in \mathcal{N}^{in}_{k} } \frac{1}{|\mathcal{N}^{in}_{ik}|} \sum_{j \in \mathcal{N}^{in}_{ik}} \left( \bar{Y}_{ik} (Z_{jk}=1, \alpha)-  \bar{Y}_{ik} (Z_{jk}=0, \alpha)  \right) = \frac{c}{N^{in}} \sum_{k=1}^K \sum_{i \in \mathcal{N}^{in}_{k}} \frac{1}{|\mathcal{N}^{in}_{ik}|} \cdot |\mathcal{N}^{in}_{ik}|= c \\
    \end{split}
\end{equation}
Based on \eqref{proof_uncond_sit2_equal_int1} and \eqref{proof_uncond_sit2_equal_int2}, $\tau^{out}(\alpha)$ is equivalent to $\tau^{in}(\alpha)$ under Condition \ref{ave_equivalence_cond_2}.    
\end{proof}

\begin{proof}[Proof of Condition \ref{ave_equivalence_cond_3}]
\label{proof_ave_equivalence_cond_3}
Assuming that Condition \ref{ave_equivalence_cond_3} holds, we show that  the outward spillover effect is equal to the inward spillover effect: 
\begin{equation*}
    \begin{split}
      \tau^{out}(\alpha)=& \sum_{k=1}^K \sum_{j \in \mathcal{N}^{out}_k } \sum_{i \in \mathcal{N}^{out}_{jk}}  \frac{1}{ N^{out} \cdot |\mathcal{N}^{out}_{jk}|} \left( \bar{Y}_{ik}(Z_{jk}=1, \alpha)- \bar{Y}_{ik}(Z_{jk}=0, \alpha) \right) \\  
      & =_{(1)}  \sum_{k=1}^K \sum_{j \in \mathcal{N}^{out}_k } \sum_{i \in \mathcal{N}^{out}_{jk}}  \frac{1}{ N^{in} \cdot |\mathcal{N}^{in}_{ik}|} \left( \bar{Y}_{ik}(Z_{jk}=1, \alpha)- \bar{Y}_{ik}(Z_{jk}=0, \alpha) \right)=_{(2)} \tau^{in}(\alpha) \\
    \end{split}
\end{equation*}
$(1)$ is by Condition \ref{ave_equivalence_cond_3}. $(2)$ is by equation \eqref{prof_difference_out_in_int_1}. 
\end{proof}

\subsection{Technical details of Section \ref{sec:Inference}}
\label{append_sec:Inference}
Throughout this section, let $[K]:= \{1,\cdots, K\}$, $[n]:= \{1,\cdots, n\}$ and $[n_k]:=\{1,\cdots,n_k\}$ for $k=1,\cdots, K$.
{ 
\begin{lemma}
\label{est_out_in_equivalence}
   Under Condition \ref{ave_equivalence_cond_3}, $\hat{\tau}^{out}(\alpha)=\hat{\tau}^{in}(\alpha)$.  
\end{lemma}
\begin{proof}[Proof of Lemma \ref{est_out_in_equivalence}]
   The estimators $\hat{\tau}^{out}(\alpha)$ and $\hat{\tau}^{in}(\alpha)$ can be written as 
$$ \hat{\tau}^{out}(\alpha)= \sum_{k=1}^K \sum_{jk \in \mathcal{N}^{out}_{k}} \sum_{ik \in \mathcal{N}^{out}_{jk}} \frac{1}{N^{out} |\mathcal{N}^{out}_{jk}| } W_{jk}(\mathbf{Z}_k) Y_{ik} \left[ Z_{jk}-(1-Z_{jk})  \right] $$ and 
$$ \hat{\tau}^{in}(\alpha)=_{(1)} \sum_{k=1}^K \sum_{jk \in \mathcal{N}^{out}_{k}} \sum_{ik \in \mathcal{N}^{out}_{jk}} \frac{1}{N^{in} |\mathcal{N}^{in}_{ik}| } W_{jk}(\mathbf{Z}_k) Y_{ik} \left[ Z_{jk}-(1-Z_{jk})  \right]. $$
(1) follows from \eqref{prof_difference_out_in_int_1}. Therefore,
\begin{equation*}
    \begin{split}
        \hat{\tau}^{out}(\alpha)-\hat{\tau}^{in}(\alpha)=\sum_{k=1}^K \sum_{jk \in \mathcal{N}^{out}_{k}} \sum_{ik \in \mathcal{N}^{out}_{jk}} \left( \frac{1}{N^{out} |\mathcal{N}^{out}_{jk}| }-\frac{1}{N^{in} |\mathcal{N}^{in}_{ik}|}\right) W_{jk}(\mathbf{Z}_k)Y_{ik}[Z_{jk}-(1-Z_{jk})].
    \end{split}
\end{equation*}
Thus, Condition \ref{ave_equivalence_cond_3} suffices to guarantee the equivalence of $\hat{\tau}^{out}(\alpha)$ and $\hat{\tau}^{in}(\alpha)$.
\end{proof}}

\begin{proof}[Proof of Proposition \ref{unbiasedness}]
\label{proof_unbiasedness}
We begin by establishing the unbiasedness of $\hat{\tau}^{out}(\alpha)$. For each $jk \in \mathcal{N}^{out}_{jk}$ and $k \in [K]$, we have:
\begin{small}
\begin{equation}
\label{proof_unbiasedness_int1}
   \begin{split}
     & \mathbb{E}_{\mathbf{Z}|\beta} \left(W_{jk}(\mathbf{Z}_k)Z_{jk}\bar{Y}^{out}_{jk} \right) \\
     & =_{(1)} \sum_{z_{-jk} \in \{0,1\}^{n_{k}-1} } \frac{\mathbb{P}_{\alpha}(\mathbf{Z}_{-jk}=z_{-jk})\cdot \mathbb{P}_{\beta}(Z_{jk}=1, \mathbf{Z}_{-jk}=z_{-jk}) }{\mathbb{P}_{\beta}(Z_{jk}=1, \mathbf{Z}_{-jk}=z_{-jk})} \cdot 1 \cdot \left( \frac{1}{|\mathcal{N}^{out}_{jk}|} \sum_{ik \in \mathcal{N}^{out}_{jk}} Y_{ik}(Z_{jk}=1,\mathbf{Z}_{-jk}=z_{-jk})  \right)  \\
     & =  \frac{1}{|\mathcal{N}^{out}_{jk}|} \sum_{ik \in \mathcal{N}^{out}_{jk}}         \sum_{z_{-jk} \in \{0,1\}^{n_{k}-1} }  \mathbb{P}_{\alpha}(\mathbf{Z}_{-jk}=z_{-jk}) \cdot Y_{ik}(Z_{jk}=1, Z_{-jk}=z_{-jk}) \\
     &= \frac{1}{|\mathcal{N}^{out}_{jk}|} \sum_{ik \in \mathcal{N}^{out}_{jk}} \bar{Y}_{ik}(1,\alpha) = \mu^{out}_{jk}(1,\alpha). 
   \end{split} 
\end{equation}
\end{small}
$(1)$ is by definition of $W_{jk}(\mathbf{Z}_k)$ and Assumption \ref{unif_bound_weight}. Similarly, 
\begin{equation}
\label{proof_unbiasedness_int2}
    \begin{split}
       \mathbb{E}_{\mathbf{Z}|\beta} \left(W_{jk}(Z_k)(1-Z_{jk})\bar{Y}^{out}_{jk} \right)=\mu^{out}_{jk}(0,\alpha). 
    \end{split}
\end{equation}
Therefore, 
\begin{equation*}
    \begin{split}
    \mathbb{E}(\hat{\tau}^{out}(\alpha)) & = \frac{1}{N^{out}} \left\lbrace \sum_{k=1}^K \sum_{jk \in  \mathcal{N}^{out}_k} \mathbb{E} \left( W_{jk}(\mathbf{Z}_k) Z_{jk}  \bar{Y}^{out}_{jk} \right) -  \sum_{k=1}^K \sum_{jk \in   \mathcal{N}^{out}_k}\mathbb{E}\left(  W_{jk}(\mathbf{Z}_k) (1-Z_{jk} )  \bar{Y}^{out}_{jk}  \right)  \right\rbrace \\
    &=_{(1)} \frac{1}{N^{out}} \sum_{k=1}^K \sum_{jk \in \mathcal{N}^{out}_{k}} \left( \mu^{out}_{jk} (1, \alpha)- \mu^{out}_{jk}(0, \alpha) \right) = \tau^{out}(\alpha)
    \end{split} 
\end{equation*}
$(1)$ is by equations \eqref{proof_unbiasedness_int1} and \eqref{proof_unbiasedness_int2}. We then consider the estimator $\hat{\tau}^{in}(\alpha)$ in Definition \ref{est_tau_in}. For each $ik \in \mathcal{N}^{in}_k$ and $k \in [K]$, 
\begin{small}
\begin{equation}
\label{proof_unbiasedness_int3}
    \begin{split}
      & \mathbb{E}_{\mathbf{Z}|\beta} \left(    \tilde{W}^1_{ik}(\mathbf{Z}_{k}) Y_{ik} \right)  \\
      & =_{(1)}  \frac{1}{|\mathcal{N}^{in}_{ik}|} \sum_{jk \in \mathcal{N}^{in}_{ik}} \sum_{ z_{-jk} \in \{0,1\}^{n_k-1} } \frac{\mathbb{P}_{\alpha}(\mathbf{Z}_{-jk}=z_{-jk}) \cdot \mathbb{P}_{\beta}(Z_{jk}=1, Z_{-jk}=z_{-jk}) }{ \mathbb{P}_{\beta}(Z_{jk}=1, Z_{-jk}=z_{-jk})  } \cdot 1 \cdot Y_{ik}(Z_{jk}=1, \mathbf{Z}_{-jk}=z_{-jk} ) \\
      & = \frac{1}{ |\mathcal{N}^{in}_{ik}| } \sum_{jk \in \mathcal{N}^{in}_{ik}} \bar{Y}_{ik}(Z_{jk}=1,\alpha)= \mu^{in}_{ik}(1, \alpha). \\
    \end{split}
\end{equation}
\end{small}
$(1)$ is by definition of $W_{jk}(\mathbf{Z}_k)$ and Assumption \ref{unif_bound_weight}. Similarly, 
\begin{equation}
\label{proof_unbiasedness_int4}
    \begin{split}
       &  \mathbb{E}_{\mathbf{Z}|\beta} \left( \tilde{W}^0_{ik} (\mathbf{Z}_{k}) Y_{ik} \right)= \mu^{in}_{ik}(0, \alpha). \\
    \end{split}
\end{equation}
Then based on equations \eqref{proof_unbiasedness_int3} and \eqref{proof_unbiasedness_int4}, 
\begin{equation*}
    \begin{split}
    \mathbb{E} (\hat{\tau}^{in}(\alpha)) & = \frac{1}{N^{in}} \sum_{k=1}^K \sum_{ik \in \mathcal{N}^{in}_{k}} \left[ \mathbb{E} \left( \tilde{W}^1_{ik}(\mathbf{Z}_k) Y_{ik}   \right) - \mathbb{E} \left( \tilde{W}^0_{ik}(\mathbf{Z}_k) Y_{ik}   \right)  \right] \\
    & = \frac{1}{N^{in}} \sum_{k=1}^K \sum_{ik \in \mathcal{N}^{in}_k} \left(  \mu^{in}_{ik} (1, \alpha) -\mu^{in}_{ik} (0, \alpha) \right) = \tau^{in}(\alpha).
    \end{split}
\end{equation*}
\end{proof}

\begin{lemma}[Lemma $1$ in \citet{ogburn2022causal} and Lemma A.5 in \citet{wang2024designbasedinferencespatialexperiments}] 
\label{Theorem_CLT_dependence}
Let $U_1, \cdots , U_n $ be bounded, mean-zero random variables with finite fourth moment, and let $D_i$ be the set of units that unit $i$ is correlated to. If $|D_i|\leq c$ for all $i\in[n]$ and $\frac{c^2}{n} \rightarrow 0$, then 
\begin{equation*}
    \begin{split}
        \frac{\sum_{i=1}^n U_i}{\sqrt{var(\sum_{i=1}^n U_i)}} \overset{d}{\rightarrow} N(0,1)
    \end{split}
\end{equation*}
\end{lemma}

\begin{proof}[Proof of Proposition \ref{clt_out_in_spillover}]
\label{proof_clt_out_in_spillover}
For each $jk \in \mathcal{N}^{out}_k$ and $k\in [K]$, let $\hat{\mu}^{out}_{jk}(1,\alpha)= W_{jk}(\mathbf{Z}_k)Z_{jk}\bar{Y}^{out}_{jk}$ and $\hat{\mu}^{out}_{jk}(0,\alpha)= W_{jk}(\mathbf{Z}_k) (1-Z_{jk}) \bar{Y}^{out}_{jk}$ as in Definition \ref{est_tau_out}. Then we set 
$$U_{jk}= \frac{1}{N^{out}} \left[\hat{\mu}^{out}_{jk} (1, \alpha) - \hat{\mu}^{out}_{jk} (0, \alpha)-  \left(   {\mu}^{out}_{jk} (1, \alpha) - {\mu}^{out}_{jk} (0, \alpha)  \right)\right].$$ 
We verify whether $U_{1k},\cdots,U_{n_K K}$ satisfy the conditions in Lemma \ref{Theorem_CLT_dependence}. We have that the variables  $U_{jk}$ for $jk \in \mathcal{N}^{out}_k$ and $k \in [K]$ are bounded and their moments are finite based on definition of the weights $W_{jk}(\mathbf{Z}_k)$ and Assumptions \ref{pos_realized_treatment} and  \ref{unif_bound_pot_out}. $\mathbb{E}(U_{jk})=0$ for $jk \in \mathcal{N}^{out}_k$ and $k \in [K]$ based on equations \eqref{proof_unbiasedness_int1} and \eqref{proof_unbiasedness_int2} in the proof of Proposition \ref{unbiasedness}. In addition, for each unit $jk$, with $jk \in \mathcal{N}^{out}_k$ 
and $k \in [K]$, $U_{jk}$ is correlated to at most $n_k$ units with constant order based on Assumptions \ref{part_intf} and \ref{order1_cluster}. Consequently, there exists a constant $c>0$ such that $n_k \leq c$ for $k \in [K]$ and $\frac{c^2}{n} \rightarrow 0$ holds. Then, based on Lemma \ref{Theorem_CLT_dependence}, we have
\begin{equation*}
    \begin{split}
    \frac{  \sum_{k \in [K]} \sum_{jk \in \mathcal{N}^{out}_{k} } U_{jk}   }{  \sqrt{var\left( \sum_{k \in [K]} \sum_{jk  \in \mathcal{N}^{out}_{k} } U_{jk} \right)  } } \overset{d}{\rightarrow}  N (0, 1) 
    \end{split}
\end{equation*}
when $K \rightarrow \infty$ based on Assumption \ref{one_in_and_out_neigh}. This completes the proof of the Central Limit Theorem (CLT) for $\hat{\tau}^{out}(\alpha)$ since $\hat{\tau}^{out}(\alpha)-\tau^{out}(\alpha)= \sum_{k=1}^K \sum_{jk \in \mathcal{N}^{out}_{k}} U_{jk}$. Similarly, for $\hat{\tau}^{in}(\alpha)$, let $\hat{\mu}^{in}_{ik}(1,\alpha)=\tilde{W}^1_{ik}(\mathbf{Z}_k) Y_{ik}$ and $\hat{\mu}^{in}_{ik}(0,\alpha)=\tilde{W}^0_{ik}(\mathbf{Z}_k) Y_{ik}$ for $ik \in \mathcal{N}^{in}_k$ and $k \in [K]$ as in Definition \ref{est_tau_in}. Then we set  
$$R_{ik}=\frac{1}{N^{in}} \left[\hat{\mu}^{in}_{ik} (1, \alpha)-\hat{\mu}^{in}_{ik} (0, \alpha)- \left( 
{\mu}^{in}_{ik} (1, \alpha) -   {\mu}^{in}_{ik} (0, \alpha) \right)\right].$$ 
Similarly, the variables  \( R_{ik} \) for \( ik \in \mathcal{N}^{in}_k \) and \( k \in [K] \) are bounded, and their moments are finite, based on the definition of \( W_{jk}(\mathbf{Z}_k) \) and Assumptions \ref{pos_realized_treatment} and  \ref{unif_bound_pot_out}. From equations \eqref{proof_unbiasedness_int3} and \eqref{proof_unbiasedness_int4} in the proof of Proposition \ref{unbiasedness}, we have \( \mathbb{E}(R_{ik}) = 0 \) for all \( ik \in \mathcal{N}^{in}_k \) and \( k \in [K] \). Furthermore, each \( R_{ik} \) is correlated with at most \( n_k \) units, where \( n_k \) has constant order, as ensured by Assumptions \ref{part_intf} and \ref{order1_cluster}. Consequently, there exists a constant \( c > 0 \) such that \( n_k \leq c \) for all \( k \in [K] \), and \( \frac{c^2}{n} \to 0 \) as \( K \to \infty \). Then based on Lemma \ref{Theorem_CLT_dependence}, we obtain
\[
\frac{\sum_{k \in [K]} \sum_{ik \in \mathcal{N}^{in}_k} R_{ik}}{\sqrt{\mathrm{var}\left( \sum_{k \in [K]} \sum_{ik \in \mathcal{N}^{in}_k} R_{ik} \right)}} \overset{d}{\rightarrow} N(0, 1),
\]
as \( K \to \infty \), under Assumption \ref{one_in_and_out_neigh}. Thus, the proof of the CLT for \( \hat{\tau}^{in}(\alpha) \) is completed, where \( \hat{\tau}^{in}(\alpha) - \tau^{in}(\alpha) = \sum_{k=1}^K \sum_{ik \in \mathcal{N}^{in}_k} R_{ik} \).
\end{proof}

\begin{lemma}
\label{formula_V_out_V_in}
Let \( V_{jk} = V_{1jk} - V_{0jk} \) and \( S_{jk} = S_{1jk} - S_{0jk} \), where \( V_{zjk} \) and \( S_{zjk} \) for \( z \in \{0,1\} \), \( jk \in \mathcal{N}^{out}_{k} \), and \( k \in \{1, \dots, K\} \) are defined in Proposition \ref{V_out_V_in}. Then, the variances of the estimators for the outward and inward spillover effects are given by $V(\hat{\tau}^{out}(\alpha))=\frac{1}{N^{out \ 2}} \sum_{k=1}^K \sum_{jk \in \mathcal{N}^{out}_{k}} [ var(V_{jk})  + \sum_{ik \in \mathcal{N}^{out}_{k}: i \not{=} j }  cov (V_{jk}, V_{ik}) ]$ and $ V(\hat{\tau}^{in}(\alpha))=\frac{1}{N^{in \ 2}} \sum_{k=1}^K \sum_{jk \in \mathcal{N}^{out}_{k}} [ var(S_{jk})  + \sum_{ik \in \mathcal{N}^{out}_{k}: i \not{=} j }  cov (S_{jk}, S_{ik}) ].$
\end{lemma}

\begin{proof}[Proof of Proposition \ref{V_out_V_in}]
\label{proof_V_out_V_in}
We first consider $V(\hat{\tau}^{out}(\alpha) )$. For $jk \in \mathcal{N}^{out}_k$ and $k \in [K]$, let $$V_{jk}= V_{1jk}-V_{0jk}= W_{jk}(\mathbf{Z}_k)Z_{jk} \bar{Y}^{out}_{jk}- W_{jk}(\mathbf{Z}_k)(1-Z_{jk}) \bar{Y}^{out}_{jk}.$$ 
   Then $V(\hat{\tau}^{out}(\alpha) )$ can be expressed as 
    \begin{equation}
     \label{proof_V_out_V_in_int_1}
        \begin{split}
         &  V \left( \hat{\tau}^{out}(\alpha) \right)=  var \left( \frac{1}{N^{out}}  \sum_{k=1}^K \sum_{jk \in \mathcal{N}^{out}_k} V_{jk}     \right) \\
        &=_{(1)} \frac{1}{N^{out \ 2}} \sum_{k=1}^K \sum_{jk \in \mathcal{N}^{out}_{k}} var(V_{jk})+ \frac{1}{N^{out \ 2}} \sum_{k=1}^K \sum_{jk \in \mathcal{N}^{out}_{k}} \sum_{ik \in \mathcal{N}^{out}_{k}: ik \neq jk} cov(V_{jk},V_{ik})\\
        & \leq_{(2)} \frac{1}{N^{out \ 2}} \sum_{k=1}^K \sum_{jk \in \mathcal{N}^{out}_{k}} var(V_{jk})+ \frac{1}{N^{out \ 2}} \sum_{k=1}^K \sum_{jk \in \mathcal{N}^{out}_{k}} \sum_{ik \in \mathcal{N}^{out}_{k}: ik \neq jk} \frac{1}{2} \left( var(V_{jk})+ var(V_{ik})   \right) \\
        & = \frac{1}{N^{out \ 2}} \sum_{k=1}^K \sum_{jk \in \mathcal{N}^{out}_{k}} var(V_{jk})+ \frac{1}{2N^{out \ 2}} \sum_{k=1}^K \sum_{jk \in \mathcal{N}^{out}_{k}} var(V_{jk}) \left( \sum_{ik \in \mathcal{N}^{out}_{k}: i \neq j} 1 \right) \\
        & + \frac{1}{2 N^{out \ 2}} \sum_{k=1}^K \sum_{jk \in \mathcal{N}^{out}_{k}} \sum_{ik \in \mathcal{N}^{out}_{k}: i \neq j}   var(V_{ik}) \\
        & =_{(3)} \frac{1}{N^{out \ 2}} \sum_{k=1}^K \sum_{jk \in \mathcal{N}^{out}_{k}} var(V_{jk})+ \frac{1}{2N^{out \ 2}} \sum_{k=1}^K \sum_{jk \in \mathcal{N}^{out}_k} var(V_{jk}) (|\mathcal{N}^{out}_k|-1) \\
        & + \frac{1}{2 N^{out \ 2}} \sum_{k=1}^K \sum_{ik \in \mathcal{N }^{out}_k} \sum_{jk \in \mathcal{N }^{out}_k: j \neq i} var(V_{ik}) \\
        & = \frac{1}{N^{out \ 2}} \sum_{k=1}^K \sum_{jk \in \mathcal{N}^{out}_{k}} var(V_{jk})+ \frac{2}{2N^{out \ 2}} \sum_{k=1}^K \sum_{jk \in \mathcal{N}^{out}_k} var(V_{jk}) (|\mathcal{N}^{out}_k|-1) \\
        & = \frac{1}{N^{out \ 2}} \sum_{k=1}^K \sum_{jk \in \mathcal{N}^{out}_{k}} |\mathcal{N}^{out}_{k}|\cdot var(V_{jk})
        \end{split}
    \end{equation}
$(1)$ is by the fact that $cov(V_{jk_1},V_{ik_2})=0$ for $k_1\neq k_2$. $(2)$ is by Cauchy-Schwarz inequality as  
$$ \frac{1}{2} \left( var(V_{jk})+ var(V_{ik})      \right) \geq  var^{\frac{1}{2}}(V_{jk}) \cdot var^{\frac{1}{2}}(V_{ik}) \geq cov(V_{jk},V_{ik})$$
$(3)$ is by considering $n_k$-by-$n_k$ matrices with elements in each row as $\left( var(V_{1k}), \cdots, var(V_{n_kk}) \right)$ for $k \in [K]$. Then for each matrix, the double sum is to firstly sum a row and then across rows. It is also equivalent to firstly sum a column and then across columns. We then consider for each $var(V_{jk})$,  
\begin{equation}
\label{proof_V_out_V_in_int_2}
    \begin{split}
        & var(V_{jk}) = var\left(  V_{1jk}- V_{0jk}       \right) = var(V_{1jk})+ var(V_{0jk}) -2cov(V_{1jk},V_{0jk})\\
        & =_{(1)} \mathbb{E}(V^2_{1jk})- (\mathbb{E}(V_{1jk}))^2+\mathbb{E}(V^2_{0jk})- (\mathbb{E}(V_{0jk}))^2-2 \left( \mathbb{E} (0)- \mathbb{E}(V_{0jk })\cdot \mathbb{E}(V_{1jk }) \right) \\
        & = \mathbb{E}(V^2_{1jk})+ \mathbb{E}(V^2_{0jk})- \left[ (\mathbb{E}(V_{1jk}))^2+  \mathbb{E}(V_{0jk}))^2- 2 \mathbb{E}(V_{0jk}) \cdot  \mathbb{E}(V_{1jk}) \right] \\
        & \leq \mathbb{E}(V^2_{1jk})+ \mathbb{E}(V^2_{0jk}) = \mathbb{E}\left(W_{jk}(\mathbf{Z}_k)Z_{jk} \bar{Y}^{out}_{jk}\right)^2 + \mathbb{E} \left( W_{jk}(\mathbf{Z}_k)(1-Z_{jk}) \bar{Y}^{out}_{jk} \right)^2 
    \end{split}
\end{equation}
$(1)$ is by the fact that $Z_{jk}(1-Z_{jk})=0$ such that $V_{1jk} \cdot V_{0jk}=0$. Combine equations \eqref{proof_V_out_V_in_int_1} and \eqref{proof_V_out_V_in_int_2}, we have 
\begin{small}
\begin{equation*}
    \begin{split}
       V \left(  \hat{\tau}^{out}(\alpha) \right) &  \leq  \frac{1}{N^{out \ 2 }} \sum_{k=1}^K \sum_{jk \in \mathcal{N}^{out}_k} |\mathcal{N}^{out}_k| \cdot var (V_{jk})  \leq \frac{1}{N^{out \ 2} } \sum_{k=1}^K \sum_{jk  \in \mathcal{N}^{out}_k} |\mathcal{N}^{out}_k| \cdot \mathbb{E} \left[ W_{jk}(\mathbf{Z}_k)Z_{jk} \bar{Y}^{out}_{jk}\right]^2\\
      & + \frac{1}{N^{out \ 2} } \sum_{k=1}^K \sum_{jk \in \mathcal{N}^{out}_k} |\mathcal{N}^{out}_{k}| \cdot \mathbb{E} \left[ W_{jk}(\mathbf{Z}_k)(1-Z_{jk}) \bar{Y}^{out}_{jk}\right]^2 = V^c(\hat{\tau}^{out}(\alpha)).
    \end{split}
\end{equation*}
\end{small}
For $V(\hat{\tau}^{in}(\alpha))$, we first consider changing the order of summation for $\hat{\tau}^{in}(\alpha)$. Let $A$ denote the adjacency matrix for the graph \( \mathcal{G} \). 
\begin{small}
\begin{equation}
\label{proof_proof_V_out_V_in_int_5}
    \begin{split}
      & \hat{\tau}^{in}(\alpha) = \frac{1}{N^{in}} \left( \sum_{k=1}^K \sum_{ik \in  \mathcal{N}^{in}_k}  \frac{1}{|\mathcal{N}^{in}_{ik}|} \sum_{jk \in \mathcal{N}^{in}_{ik}} W_{jk}(\mathbf{Z}_k) Z_{jk} Y_{ik}   -  \sum_{k=1}^K \sum_{ik \in  \mathcal{N}^{in}_{k}} \frac{1}{|\mathcal{N}^{in}_{ik}|} \sum_{jk \in \mathcal{N}^{in}_{ik}} W_{jk}(\mathbf{Z}_k) (1-Z_{jk}) {Y}_{ik} \right) \\
      & = \frac{1}{N^{in}} \left( \sum_{k=1}^K \sum_{ik \in  \mathcal{N}^{in}_k}  \frac{1}{|\mathcal{N}^{in}_{ik}|} \sum_{jk =1 }^{n_k} A_{ji} W_{jk}(\mathbf{Z}_k) Z_{jk} Y_{ik}   -  \sum_{k=1}^K \sum_{ik \in  \mathcal{N}^{in}_{k}} \frac{1}{|\mathcal{N}^{in}_{ik}|} \sum_{jk =1}^{n_k} A_{ji} W_{jk}(\mathbf{Z}_k) (1-Z_{jk}) {Y}_{ik} \right) \\
      & = \frac{1}{N^{in}} \left( \sum_{k=1}^K  \sum_{jk =1 }^{n_k} \sum_{ik \in  \mathcal{N}^{in}_k} A_{ji} \frac{1}{|\mathcal{N}^{in}_{ik}|}   W_{jk}(\mathbf{Z}_k) Z_{jk} Y_{ik}   -  \sum_{k=1}^K \sum_{jk =1}^{n_k} \sum_{ik \in  \mathcal{N}^{in}_{k}}  A_{ji}  \frac{1}{|\mathcal{N}^{in}_{ik}|} W_{jk}(\mathbf{Z}_k) (1-Z_{jk}) {Y}_{ik} \right) \\
      & =_{(1)} \frac{1}{N^{in}} \left[ \sum_{k=1}^K \sum_{jk \in  \mathcal{N}^{out}_k} W_{jk}(\mathbf{Z}_k) Z_{jk} \widetilde{Y}^{out}_{jk}   -  \sum_{k=1}^K \sum_{jk \in  \mathcal{N}^{out}_k} W_{jk}(\mathbf{Z}_k) (1-Z_{jk}) \widetilde{Y}^{out}_{jk} \right] \\
    \end{split}
\end{equation}
\end{small}
\noindent where $\widetilde{Y}^{out}_{jk}=\sum_{ik\in \mathcal{N}^{out}_{jk}} {|\mathcal{N}^{in}_{ik}|^{-1}} Y_{ik}$ for $jk \in \mathcal{N}^{out}_k$ and $k \in [K]$. \( (1) \) follows from the following observations: (i) For a unit \( ik \) where $A_{ji}=1$, we have \( ik \in \mathcal{N}^{out}_{jk} \). Therefore, $\sum_{ik \in \mathcal{N}^{in}_{k}} A_{ji} {|\mathcal{N}^{in}_{ik}|^{-1}} Y_{ik}=\sum_{ik \in \mathcal{N}^{out}_{jk}}  {|\mathcal{N}^{in}_{ik}|^{-1}} Y_{ik} $. (ii) when $ jk \not\in \mathcal{N}^{out}_k$, there are no terms to sum. Then by setting 
$$S_{jk}= S_{1jk}-S_{0jk}:= W_{jk}(\mathbf{Z}_k)Z_{jk} \widetilde{Y}^{out}_{jk} - W_{jk}(\mathbf{Z}_k)(1-Z_{jk}) \widetilde{Y}^{out}_{jk}$$
and by following the similar proof in equation \eqref{proof_V_out_V_in_int_1}, we have 
\begin{equation}
\label{proof_V_out_V_in_int_3}
    \begin{split}
    V(\hat{\tau}^{in}(\alpha))\leq \frac{1}{N^{in \ 2 }}  \sum_{k=1}^K \sum_{jk \in \mathcal{N}^{out}_k} |\mathcal{N}^{out}_k| var(S_{jk}).    
    \end{split}
\end{equation}
Meanwhile, by following the similar proof in equation \eqref{proof_V_out_V_in_int_2},  
\begin{equation}
\label{proof_V_out_V_in_int_4}
    \begin{split}
        & var(S_{jk}) \leq \mathbb{E}\left(W_{jk}(\mathbf{Z}_k)Z_{jk} \widetilde{Y}^{out}_{jk} \right)^2 + \mathbb{E} \left( W_{jk}(\mathbf{Z}_k)(1-Z_{jk}) \widetilde{Y}^{out}_{jk} \right)^2.  
    \end{split}
\end{equation}
Combining \eqref{proof_V_out_V_in_int_3} and \eqref{proof_V_out_V_in_int_4}, we have 
\begin{equation*}
    \begin{split}
       V \left(  \hat{\tau}^{in}(\alpha) \right) &  \leq  V^c(\hat{\tau}^{in}(\alpha))
    \end{split}
\end{equation*}    
\end{proof}

\begin{proof}[Proof of Proposition \ref{est_V_out_V_in}]
Let \( V_{jk} \) and \( S_{jk} \) follow the same definitions as in the proof of Proposition \ref{V_out_V_in}. 
The difference between $N^{out} \cdot \hat{V}\left( \hat{\tau}^{out}(\alpha) \right)$ and $N^{out} \cdot V^c(\hat{\tau}^{out}(\alpha))$ is 
 \begin{small}
    \begin{equation*}
        \begin{split}
         &  N^{out} \cdot  \hat{V}\left( \hat{\tau}^{out}(\alpha) \right)- N^{out} \cdot V^c(\hat{\tau}^{out}(\alpha)) \\
         & = \frac{1}{N^{out}} \sum_{k=1}^K \sum_{jk \in \mathcal{N}^{out}_k} |\mathcal{N}^{out}_k| \cdot \left( W_{jk}(\mathbf{Z}_k) Z_{jk} \bar{Y}^{out}_{jk}  \right)^2-\frac{1}{N^{out}} \sum_{k=1}^K \sum_{jk \in \mathcal{N}^{out}_k}  |\mathcal{N}^{out}_k| \cdot \mathbb{E} \left( W_{jk}(\mathbf{Z}_k) Z_{jk} \bar{Y}^{out}_{jk}  \right)^2 \\
         & + \frac{1}{N^{out}} \sum_{k=1}^K \sum_{jk \in \mathcal{N}^{out}_k} |\mathcal{N}^{out}_k| \cdot \left( W_{jk}(\mathbf{Z}_k) (1-Z_{jk}) \bar{Y}^{out}_{jk}  \right)^2-\frac{1}{N^{out}} \sum_{k=1}^K \sum_{jk \in \mathcal{N}^{out}_k}  |\mathcal{N}^{out}_k| \cdot \mathbb{E} \left( W_{jk}(\mathbf{Z}_k) (1-Z_{jk})\bar{Y}^{out}_{jk}  \right)^2. \\
        \end{split}
    \end{equation*}
     \end{small}
 To utilize Chebyshev's inequality to the second and third lines above respectively, we first consider 
\begin{equation}
\label{proof_est_V_out_V_in_int_1}
    \begin{split}
      &   var\left\lbrace \frac{1}{N^{out}} \sum_{k=1}^K \sum_{jk \in \mathcal{N}^{out}_k} |\mathcal{N}^{out}_k| \cdot \left( W_{jk}(\mathbf{Z}_k) Z_{jk} \bar{Y}^{out}_{jk}  \right)^2  \right\rbrace \\
      & =_{(1)} \frac{1}{N^{out \ 2 }}  \left\lbrace \sum_{k=1}^K \sum_{jk \in \mathcal{N}^{out}_k} |\mathcal{N}^{out}_k|^2 var( V^2_{1jk} ) +  \sum_{k=1}^K \sum_{jk \in \mathcal{N}^{out}_{k}} \sum_{hk \in \mathcal{N}^{out}_k \backslash \{jk\} }  |\mathcal{N}^{out}_k|^2 cov(V^2_{1jk}, V^2_{1hk})  \right\rbrace    \\ 
    \end{split}
\end{equation}
$(1)$ is by Assumption \ref{part_intf}. Then for each $V_{1jk}$ for $jk \in \mathcal{N}^{out}_k$ and $k \in [K]$, there exists a constant $C>0$ such that 
\begin{equation}
\label{proof_est_V_out_V_in_int_2}
    \begin{split}
       &  var(V^2_{1jk}) = \mathbb{E}( V^4_{1jk} )-  \left( \mathbb{E}(V^2_{1jk})   \right)^2 \leq \mathbb{E}( V^4_{1jk} ) \leq_{(1)} C \\
    \end{split}
\end{equation}
where \( (1) \) follows from Assumption \ref{unif_bound_pot_out} and the boundedness of the weights \( W_{jk}(\mathbf{Z}_k) \) for \( jk \in \mathcal{N}^{out}_k \) and \( k \in [K] \), which is ensured by the definition of \( W_{jk}(\mathbf{Z}_k) \) and Assumption \ref{pos_realized_treatment}. Moreover, for each term of covariance, there exists a constant $C>0$ such that
\begin{equation}
\label{proof_est_V_out_V_in_int_3}
    \begin{split}
        & cov\left(V^2_{1jk},V^2_{1hk} \right) \leq \mathbb{E}[ W^2_{jk}(\mathbf{Z}_k) W^2_{hk}(\mathbf{Z}_k) Z_{jk} Z_{hk} \bar{Y}^{out \ 2}_{jk} \bar{Y}^{out \ 2}_{hk} ]\leq_{(1)} C \\
    \end{split}
\end{equation}
where \( (1) \) follows by the same reasoning as in \eqref{proof_est_V_out_V_in_int_2}. 
Combine the results from \eqref{proof_est_V_out_V_in_int_1}, \eqref{proof_est_V_out_V_in_int_2} and \eqref{proof_est_V_out_V_in_int_3}, we have 

\begin{equation}
\label{proof_est_V_out_V_in_int_4}
    \begin{split}
        & var\left[  \frac{1}{N^{out}} \sum_{k=1}^K \sum_{jk \in \mathcal{N}^{out}_k} |\mathcal{N}^{out}_k| \cdot \left( W_{jk}(\mathbf{Z}_k) Z_{jk} \bar{Y}^{out}_{jk}  \right)^2  \right] \leq_{(1)} \frac{C}{8} \cdot  \frac{1}{K^2} \cdot K = \frac{C}{8} \cdot \frac{1}{K} 
    \end{split}
\end{equation}
where $(1)$ results from $N^{out}=O(K)$ based on Assumption \ref{one_in_and_out_neigh} and $|\mathcal{N}^{out}_k|=O(1)$ based on Assumption \ref{order1_cluster}. Then by using Chebyshev's inequality, we have
\begin{small}
\begin{equation}
\label{proof_est_V_out_V_in_int_5}
    \begin{split}
    \mathbb{P}\left( \mid \frac{1}{N^{out}} \sum_{k=1}^K \sum_{jk \in \mathcal{N}^{out}_{jk}} |\mathcal{N}^{out}_k| \cdot V^2_{1jk} - \frac{1}{N^{out}} \sum_{k=1}^K \sum_{jk \in \mathcal{N}^{out}_{jk}} |\mathcal{N}^{out}_k| \cdot \mathbb{E} (V^2_{1jk} ) \mid  \geq \frac{1}{2} \sqrt{ \frac{\log K}{K} } \right) \leq_{(1)} \frac{C}{8K}  \cdot \frac{4K}{\log K }= \frac{C}{2\log K}
    \end{split}
\end{equation}
\end{small}
$(1)$ is by \eqref{proof_est_V_out_V_in_int_4}. Similarly, 
\begin{small}
\begin{equation*}
    \begin{split}
    \mathbb{P}\left( \mid  \frac{1}{N^{out}} \sum_{k=1}^K \sum_{jk \in \mathcal{N}^{out}_{jk}} |\mathcal{N}^{out}_k| \cdot V^2_{0jk} - \frac{1}{N^{out}} \sum_{k=1}^K \sum_{jk \in \mathcal{N}^{out}_{jk}} |\mathcal{N}^{out}_k| \cdot \mathbb{E} (V^2_{0jk} ) \mid  \geq \frac{1}{2}\sqrt{ \frac{\log K}{K} } \right) \leq \frac{C}{8K}  \cdot \frac{4K}{\log K }= \frac{C}{2\log K}.
    \end{split}
\end{equation*}
\end{small}
Then with probability at least $1-{C}/{\log K}$, $$ |N^{out} \cdot \hat{V}^c (\hat{\tau}^{out}(\alpha))- N^{out} \cdot V^c(\hat{\tau}^{out}(\alpha)) | \leq \sqrt{\frac{\log K}{K}}.$$
For $\hat{V}^c(\hat{\tau}^{in}(\alpha))$, it follows the same proofs of \eqref{proof_est_V_out_V_in_int_1}, \eqref{proof_est_V_out_V_in_int_2} and \eqref{proof_est_V_out_V_in_int_3} except changing $V_{1jk}$ to $S_{1jk}$ and $N^{out}$ to $N^{in}$. Then 
\begin{equation*}
    \begin{split}
        & var\left[  \frac{1}{N^{in}} \sum_{k=1}^K \sum_{jk \in \mathcal{N}^{out}_k} |\mathcal{N}^{out}_k| \cdot \left( W_{jk}(\mathbf{Z}_k) Z_{jk} \widetilde{Y}^{out}_{jk} \right)^2  \right] \leq_{(1)} \frac{C}{8} \cdot  \frac{1}{K^2} \cdot K = \frac{C}{8} \cdot \frac{1}{K} 
    \end{split}
\end{equation*}
\((1)\) follows from \(N^{in} = O(K)\) based on Assumption \ref{one_in_and_out_neigh} and \(|\mathcal{N}^{out}_k| = O(1)\) based on Assumption \ref{order1_cluster}. Then, applying Chebyshev's inequality,
\begin{small}
\begin{equation*}
    \begin{split}
    \mathbb{P}\left( \mid  \frac{1}{N^{in}} \sum_{k=1}^K \sum_{jk \in \mathcal{N}^{out}_{jk}} |\mathcal{N}^{out}_k| \cdot S^2_{1jk} - \frac{1}{N^{in}} \sum_{k=1}^K \sum_{jk \in \mathcal{N}^{out}_{jk}} |\mathcal{N}^{out}_k| \cdot \mathbb{E} (S^2_{1jk} ) \mid  \geq \frac{1}{2} \sqrt{ \frac{\log K}{K} } \right) \leq_{(1)} \frac{C}{8K}  \cdot \frac{4K}{\log K }= \frac{C}{2\log K}.
    \end{split}
\end{equation*}
\end{small}
Similarly, 
\begin{small}
\begin{equation*}
    \begin{split}
    \mathbb{P}\left( \mid  \frac{1}{N^{in}} \sum_{k=1}^K \sum_{jk \in \mathcal{N}^{out}_{jk}} |\mathcal{N}^{out}_k| \cdot S^2_{0jk} - \frac{1}{N^{in}} \sum_{k=1}^K \sum_{jk \in \mathcal{N}^{out}_{jk}} |\mathcal{N}^{out}_k| \cdot \mathbb{E} (S^2_{0jk} ) \mid  \geq \frac{1}{2} \sqrt{ \frac{\log K}{K} } \right) \leq_{(1)} \frac{C}{8K}  \cdot \frac{4K}{\log K }= \frac{C}{2\log K}.
    \end{split}
\end{equation*}
\end{small}
Then with probability at least $1-{C}/{\log K}$, $$ |N^{in} \cdot \hat{V}^c (\hat{\tau}^{in}(\alpha))- N^{in} \cdot V^c(\hat{\tau}^{in}(\alpha)) | \leq \sqrt{\frac{\log K}{K}}.$$
\end{proof}

\begin{proof}[Proof of Proposition \ref{dif_V_var}]
Let \( V_{jk} \) and \( S_{jk} \) follow the same definitions as in the proof of Proposition \ref{V_out_V_in}. We first examine the difference between \( V^c(\hat{\tau}^{out}(\alpha)) \) and \( V(\hat{\tau}^{out}(\alpha)) \).
\begin{equation}
\label{proof_dif_V_var_int1}
\begin{split}
& N^{out} V^{c}(\hat{\tau}^{out}(\alpha))-N^{out}V(\hat{\tau}^{out}(\alpha) ) =_{(1)} \frac{1}{N^{out}} \sum_{k=1}^K \sum_{jk\in \mathcal{N}^{out}_k}\left[ \mathbb{E}(V^2_{1jk})+ \mathbb{E}(V^2_{0jk}) - var(V_{jk})  \right]\\
& +\frac{1}{N^{out}} \sum_{k=1}^K \sum_{jk\in \mathcal{N}^{out}_k}\left[ \left( \mathbb{E}(V^2_{1jk})+ \mathbb{E}(V^2_{0jk})\right) (|\mathcal{N}^{out}_k|-1) - \sum_{ik \in \mathcal{N}^{out}_k: \ ik \neq jk} cov(V_{jk},V_{ik})  \right] \\
 \end{split}
\end{equation}
where \( (1) \) follows from the formula for \( V^c(\hat{\tau}^{out}(\alpha)) \) in Proposition \ref{V_out_V_in}. Then we simplify the first term in \eqref{proof_dif_V_var_int1}.
\begin{equation}
\label{proof_dif_V_var_int2}
    \begin{split}
      &  \frac{1}{N^{out}} \sum_{k=1}^K \sum_{jk\in \mathcal{N}^{out}_k}\left[ \mathbb{E}(V^2_{1jk})+ \mathbb{E}(V^2_{0jk}) - var(V_{jk})  \right] \\
      &= \frac{1}{N^{out}} \sum_{k=1}^K \sum_{jk\in \mathcal{N}^{out}_k}\left\lbrace \mathbb{E}(V^2_{1jk})+ \mathbb{E}(V^2_{0jk}) - \mathbb{E}[(V_{1jk}-V_{0jk} )^2]+ \left( \mathbb{E}(V_{1jk}-V_{0jk})  \right)^2 \right\rbrace \\
      & =_{(1)} \frac{1}{N^{out}} \sum_{k=1}^K \sum_{jk\in \mathcal{N}^{out}_k}\left\lbrace [\mathbb{E}(V_{1jk})]^2+ [\mathbb{E}(V_{0jk})]^2 - 2 \mathbb{E}(V_{1jk})\cdot \mathbb{E}(V_{0jk})  \right\rbrace \\
      & = \frac{1}{N^{out}} \sum_{k=1}^K \sum_{jk\in \mathcal{N}^{out}_k}\left[ \mu^{out}_{jk}(1,\alpha)-\mu^{out}_{jk}(0,\alpha) \right]^2= \frac{1}{N^{out}} \sum_{k=1}^K \sum_{jk\in \mathcal{N}^{out}_k}\left[ \frac{1}{|\mathcal{N}^{out}_{jk}|} \sum_{ik \in \mathcal{N}^{out}_{jk}}\left( \bar{Y}_{ik}(1,\alpha) -   \bar{Y}_{ik}(0,\alpha) \right)\right]^2
     \end{split}
\end{equation}
$(1)$ is by the fact that $V_{1jk} \cdot V_{0jk}=0$.
We now simplify the second term in \eqref{proof_dif_V_var_int1}, 
\begin{equation}
\label{proof_dif_V_var_int3}
    \begin{split}
     &   \frac{1}{N^{out}} \sum_{k=1}^K \sum_{jk\in \mathcal{N}^{out}_k}\left[ \left( \mathbb{E}(V^2_{1jk})+ \mathbb{E}(V^2_{0jk})\right) (|\mathcal{N}^{out}_k|-1) - \sum_{ik \in \mathcal{N}^{out}_k: \ ik \neq jk} cov(V_{jk},V_{ik})  \right] \\
     & \geq \frac{1}{N^{out}} \sum_{k=1}^K \sum_{jk\in \mathcal{N}^{out}_k}\left\lbrace \left( \mathbb{E}(V^2_{1jk})+ \mathbb{E}(V^2_{0jk})\right) (|\mathcal{N}^{out}_k|-1) - \sum_{ik \in \mathcal{N}^{out}_k: \ ik \neq jk} \frac{1}{2}[var(V_{jk})+var(V_{ik})] \right\rbrace \\
     & =_{(1)}  \frac{1}{N^{out}} \sum_{k=1}^K \sum_{jk\in \mathcal{N}^{out}_k}\left[ \left( \mathbb{E}(V^2_{1jk})+ \mathbb{E}(V^2_{0jk}) -  var(V_{jk})\right)(|\mathcal{N}^{out}_k|-1) \right] \\
     &=  \frac{1}{N^{out}} \sum_{k=1}^K \sum_{jk\in \mathcal{N}^{out}_k} \left[ \mathbb{E}(V^2_{1jk})+ \mathbb{E}(V^2_{0jk}) -  \left( var(V_{1jk})+var(V_{0jk})-2cov(V_{1jk}, V_{0jk}) \right) \right](|\mathcal{N}^{out}_k|-1) \\
     &= \frac{1}{N^{out}} \sum_{k=1}^K \sum_{jk\in \mathcal{N}^{out}_k} \left( \mathbb{E}(V_{1jk})- \mathbb{E}(V_{0jk})  \right)^2(|\mathcal{N}^{out}_k|-1) \\ 
     & = \frac{1}{N^{out}} \sum_{k=1}^K \sum_{jk\in \mathcal{N}^{out}_k}\left[ \frac{1}{|\mathcal{N}^{out}_{jk}|} \sum_{ik \in \mathcal{N}^{out}_{jk}}\left( \bar{Y}_{ik}(1,\alpha) -   \bar{Y}_{ik}(0,\alpha) \right)\right]^2 \cdot (|\mathcal{N}^{out}_k|-1)
    \end{split}
\end{equation}
$(1)$ uses the same steps in \eqref{proof_V_out_V_in_int_1} for dealing $2^{-1}[ var(V_{jk})+ var(V_{ik})]$. Combining \eqref{proof_dif_V_var_int2} and \eqref{proof_dif_V_var_int3}, we obtain
\begin{equation*}
    \begin{split}
        N^{out} V^c(\hat{\tau}^{out}(\alpha))-N^{out} V(\hat{\tau}^{out}(\alpha)) \geq \frac{1}{N^{out}} \sum_{k=1}^K \sum_{jk\in \mathcal{N}^{out}_k} |\mathcal{N}^{out}_k| \left[ \frac{1}{|\mathcal{N}^{out}_{jk}|} \sum_{ik \in \mathcal{N}^{out}_{jk}}\left( \bar{Y}_{ik}(1,\alpha) -   \bar{Y}_{ik}(0,\alpha) \right)\right]^2
    \end{split}
\end{equation*}

For inward spillover effects, following the equivalent form of $\hat{\tau}^{in}(\alpha)$ in equation \eqref{proof_proof_V_out_V_in_int_5}, we can use similar arguments as above:
\begin{equation*}
\begin{split}
& N^{in} V^{c}(\hat{\tau}^{in}(\alpha))-N^{in}V(\hat{\tau}^{in}(\alpha) ) =_{(1)} \frac{1}{N^{in}} \sum_{k=1}^K \sum_{jk\in \mathcal{N}^{out}_k}\left[ \mathbb{E}(S^2_{1jk})+ \mathbb{E}(S^2_{0jk}) - var(S_{jk})  \right]\\
& +\frac{1}{N^{in}} \sum_{k=1}^K \sum_{jk\in \mathcal{N}^{out}_k}\left[ \left( \mathbb{E}(S^2_{1jk})+ \mathbb{E}(S^2_{0jk})\right) (|\mathcal{N}^{out}_k|-1) - \sum_{ik \in \mathcal{N}^{out}_k: \ ik \neq jk} cov(S_{jk},S_{ik})  \right] \\
& \geq_{(1)} \frac{1}{N^{in}} \sum_{k=1}^K \sum_{jk \in \mathcal{N}^{out}_{k}}   \left[ { \sum_{ik \in \mathcal{N}^{out}_{jk} } \frac{1}{|\mathcal{N}^{in}_{ik}|} \left( \bar{Y}_{ik}(Z_{jk}=1,\alpha)- \bar{Y}_{ik}(Z_{jk}=0,\alpha)  \right)  } \right]^2\cdot[ 1+ (|\mathcal{N}^{out}_k|-1) ] 
 \end{split}
\end{equation*}
\( (1) \) follows the same steps as those in equations \eqref{proof_dif_V_var_int2} and \eqref{proof_dif_V_var_int3}, with the substitution of \( V_{0jk} \) and \( V_{1jk} \) by \( S_{0jk} \) and \( S_{1jk} \), respectively, and with the fact that \( \mathbb{E}(S_{zjk}) = \sum_{ik \in \mathcal{N}^{out}_{jk}} |\mathcal{N}^{in}_{ik}|^{-1} \bar{Y}_{ik}(Z_{jk} = z, \alpha) \) for \( z \in \{0, 1\} \).
\end{proof}

\subsection{Technical details of Section \ref{Comp_var}}
\label{Supp_sec_Comp_var}
Throughout this section, let $[K]:= \{1,\cdots, K\}$, $[n]:= \{1,\cdots, n\}$ and $[n_k]:=\{1,\cdots,n_k\}$ for $k=1,\cdots, K$.
\begin{proof}[Proof of Proposition \ref{V_out_in_compare}]
Based on the forms of $V^c(\hat{\tau}^{out}(\alpha))$ and $V^c(\hat{\tau}^{in}(\alpha))$, we have 
\begin{equation*}
	\begin{split}
	&V^c(\hat{\tau}^{out}(\alpha))-V^c(\hat{\tau}^{in}(\alpha)) \\
	&=   \sum_{k=1}^K \sum_{jk \in {N}^{out}_k}  |{N}^{out}_k  |  \left[ \mathbb{E} \left(  W_{jk}(\mathbf{Z}_k) Z_{jk } \frac{1}{ {N}^{out} }  \bar{Y}_{jk}^{out}     \right)^2  - \mathbb{E} \left(  W_{jk}(\mathbf{Z}_k)  Z_{jk } \frac{1}{ {N}^{in} }  \widetilde{Y}_{jk}^{out}    \right)^2 \right] \\
	& + \sum_{k=1}^K \sum_{jk \in \mathcal{N}^{out}_{k}} |{N}^{out}_k| \left[ \mathbb{E} \left(  W_{jk}( \mathbf{Z}_k) (1-Z_{jk}) \frac{1}{ |\mathcal{N}^{out}| }  \bar{Y}_{jk}^{out}    \right)^2  - \mathbb{E} \left(  W_{jk}(\mathbf{Z}_k) (1-Z_{jk }) \frac{1}{ |\mathcal{N}^{out}| }  \widetilde{Y}_{jk}^{out}    \right)^2 \right] \\
    &=_{(1)}  \sum_{k=1}^K  \sum_{jk \in \mathcal{N}^{out}_k}  |\mathcal{N}^{out}_k  | \  \mathbb{E}  \left[  W^2_{jk}(\mathbf{Z}_k) Z_{jk }  B_{jk} D_{jk} \right] +  \sum_{k=1}^K  \sum_{jk \in \mathcal{N}^{out}_k}  |\mathcal{N}^{out}_k  | \  \mathbb{E}  \left[  W^2_{jk}(\mathbf{Z}_k) (1-Z_{jk})  B_{jk} D_{jk} \right] \\
	\end{split}
\end{equation*} 
where $B_{jk}= \sum_{ik \in N^{out}_{jk}} \left( S^{out}_{ik,jk} - S^{in}_{ik,jk}  \right) \cdot Y_{ik}$ and $D_{jk}= \sum_{ik \in N^{out}_{jk}} \left( S^{out}_{ik,jk} + S^{in}_{ik,jk}\right) \cdot  Y_{ik} $ for $jk \in \mathcal{N}^{out}_{k}$ and $k \in [K]$. $(1)$ is by the fact that 
$$ \frac{1}{N^{out \ 2}} \bar{Y}^{out \ 2}_{jk}-\frac{1}{N^{in \ 2}} \widetilde{Y}^{out \ 2}_{jk} = \left( \frac{1}{N^{out}} \bar{Y}^{out}_{jk}-\frac{1}{N^{in}} \widetilde{Y}^{out}_{jk} \right)\cdot  \left( \frac{1}{N^{out}} \bar{Y}^{out}_{jk}+\frac{1}{N^{in}} \widetilde{Y}^{out}_{jk} \right).  $$ 
\end{proof}

{ 
\begin{proof}[Proof of Proposition \ref{general_suff_cond_sender_type_3}]
    For $ik \in H_{1,jk}$, $(S^{in}_{ik,jk})^2 \leq (S^{out}_{ik,jk})^2$, so $\sum_{k=1}^K \sum_{jk \in H_{1k}} \sum_{ik \in \mathcal{N}^{out}_{jk}} \bigl[(S^{out}_{ik,jk})^2 - (S^{in}_{ik,jk})^2\bigr] > 0$. For $ik \in H_{2,jk}$, $(S^{in}_{ik,jk})^2 > (S^{out}_{ik,jk})^2$, so $\sum_{k=1}^K \sum_{jk \in H_{2k}} \sum_{ik \in \mathcal{N}^{out}_{jk}} \bigl[(S^{in}_{ik,jk})^2 - (S^{out}_{ik,jk})^2\bigr] > 0$.
Thus, if $R_s > {b^2}/{a^2}$, we have
  \begin{small}
\begin{equation}
    \label{suff1_Vc_compare_int1}
        \begin{split}
        \sum_{k=1}^K \sum_{jk \in \mathcal{N}^{out}_k} \sum_{ik \in H_{1,jk}} \left[  (N^{out}|\mathcal{N}^{out}_{jk}|)^{-2}- (N^{in}|\mathcal{N}^{in}_{ik}|)^{-2}  \right] a^2 > \sum_{k=1}^K \sum_{jk \in \mathcal{N}^{out}_{k}}\sum_{ik \in H_{2,jk}} \left[ (N^{in}|\mathcal{N}^{in}_{ik}|)^{-2}-(N^{out}|\mathcal{N}^{out}_{jk}|)^{-2} \right] b^2 
        \end{split}
    \end{equation}
      \end{small}  
Furthermore,  
\begin{equation}
 \label{suff1_Vc_compare_int2}
        \begin{split}
        & \sum_{k=1}^K \sum_{jk \in \mathcal{N}^{out}_{k}} \sum_{ik \in H_{1,jk}} \mathbb{E} \left\lbrace \left[  (N^{out}|\mathcal{N}^{out}_{jk}|)^{-2} W^2_{jk}(\mathbf{Z}_k) Y^2_{ik} \right] -  \mathbb{E}\left[(N^{in}|\mathcal{N}^{in}_{ik}|)^{-2} W^2_{jk}(\mathbf{Z}_k) Y^2_{ik} \right]  \right\rbrace \\
        & = \sum_{k=1}^K \sum_{jk \in \mathcal{N}^{out}_{k}} \sum_{ik \in H_{1,jk}}  \left[ (N^{out}|\mathcal{N}^{out}_{jk}|)^{-2}- (N^{in}|\mathcal{N}^{in}_{ik}|)^{-2}  \right] \mathbb{E}(W^2_{jk}(\mathbf{Z}_k) Y^2_{ik})\\
        & \geq_{(1)} \sum_{k=1}^K \sum_{jk \in \mathcal{N}^{out}_{k}} \sum_{ik \in H_{1,jk}}  \left[  (N^{out}|\mathcal{N}^{out}_{jk}|)^{-2}- (N^{in}|\mathcal{N}^{in}_{ik}|)^{-2}  \right] \mathbb{E}[W^2_{jk}(\mathbf{Z}_k)(Z_{jk}+(1-Z_{jk})) ] a^2 \\
        & =_{(2)} \left(\frac{1}{p^2}+\frac{1}{(1-p)^2}   \right) \sum_{k=1}^K \sum_{jk \in \mathcal{N}^{out}_{k}} \sum_{ik \in H_{1,jk}}   \left[  (N^{out}|\mathcal{N}^{out}_{jk}|)^{-2}- (N^{in}|\mathcal{N}^{in}_{ik}|)^{-2}  \right] a^2 
        \end{split}
    \end{equation}
(1) is based on Condition (i). (2) is based on Condition (ii). Similarly, we have 
\begin{equation}
 \label{suff1_Vc_compare_int3}
    \begin{split}
       & \sum_{k=1}^K \sum_{jk \in \mathcal{N}^{out}_{k}} \sum_{ik \in H_{2,jk}} \left[ (N^{in}|\mathcal{N}^{in}_{ik}|)^{-2}-(N^{out}|\mathcal{N}^{out}_{jk}|)^{-2} \right] \mathbb{E}(W^2_{jk}(\mathbf{Z}_k) Y^2_{ik}) \\
       & \leq_{(1)} \left( \frac{1}{p^2}+\frac{1}{(1-p)^2} \right) \sum_{k=1}^K \sum_{jk \in \mathcal{N}^{out}_{k}} \sum_{ik \in H_{2,jk}} \left[ (N^{in}|\mathcal{N}^{in}_{ik}|)^{-2}-(N^{out}|\mathcal{N}^{out}_{jk}|)^{-2} \right]  b^2
    \end{split}
\end{equation}  
(1) is based on Conditions (i) and (ii). Therefore, based on equations \eqref{suff1_Vc_compare_int1}, \eqref{suff1_Vc_compare_int2} and  \eqref{suff1_Vc_compare_int3}, we have 
\begin{equation}
\label{suff1_Vc_compare_int4}
    \begin{split}
        & \sum_{k=1}^K \sum_{jk \in \mathcal{N}^{out}_{k}} \sum_{ik \in H_{1,jk}}  \left\lbrace \mathbb{E}\left[  (N^{out}|\mathcal{N}^{out}_{jk}|)^{-2} W^2_{jk}(\mathbf{Z}_k) Y^2_{ik} \right] -  \mathbb{E}\left[(N^{in}|\mathcal{N}^{in}_{ik}|)^{-2} W^2_{jk}(\mathbf{Z}_k) Y^2_{ik} \right]  \right\rbrace \\
        & +\sum_{k=1}^K \sum_{jk \in \mathcal{N}^{out}_{k}} \sum_{ik \in H_{2,jk}}  \left\lbrace \mathbb{E}  \left[  (N^{out}|\mathcal{N}^{out}_{jk}|)^{-2} W^2_{jk}(\mathbf{Z}_k) Y^2_{ik} \right] -  \mathbb{E}\left[(N^{in}|\mathcal{N}^{in}_{ik}|)^{-2} W^2_{jk}(\mathbf{Z}_k) Y^2_{ik} \right] \right\rbrace \geq 0 \\
    \end{split}
\end{equation}
Therefore, we have 
\begin{equation*}
    \begin{split}
    & V^c(\hat{\tau}^{out}(\alpha))- V^c(\hat{\tau}^{in}(\alpha))=\sum_{k=1}^K \sum_{jk \in \mathcal{N}^{out}_{k}} \sum_{ik \in  \mathcal{N}^{out}_{jk}}  \left\lbrace  \mathbb{E} \left[  (N^{out}|\mathcal{N}^{out}_{jk}|)^{-2} W^2_{jk}(\mathbf{Z}_k) Y^2_{ik} \right] -  \mathbb{E}\left[(N^{in}|\mathcal{N}^{in}_{ik}|)^{-2} W^2_{jk}(\mathbf{Z}_k) Y^2_{ik} \right]  \right\rbrace  \\
        & =_{(1)} \sum_{k=1}^K \sum_{jk \in \mathcal{N}^{out}_k} \sum_{ik \in H_{1,jk}}  \left\lbrace  \mathbb{E} \left[  (N^{out}|\mathcal{N}^{out}_{jk}|)^{-2} W^2_{jk}(\mathbf{Z}_k) Y^2_{ik} \right] -  \mathbb{E}\left[(N^{in}|\mathcal{N}^{in}_{ik}|)^{-2} W^2_{jk}(\mathbf{Z}_k) Y^2_{ik} \right]  \right\rbrace \\
        & + \sum_{k=1}^K \sum_{jk \in \mathcal{N}^{out}_k}\sum_{ik \in H_{2,jk}} \left\lbrace  \mathbb{E} \left[  (N^{out}|\mathcal{N}^{out}_{jk}|)^{-2} W^2_{jk}(\mathbf{Z}_k) Y^2_{ik} \right]- \mathbb{E} \left[(N^{in}|\mathcal{N}^{in}_{ik}|)^{-2} W^2_{jk}(\mathbf{Z}_k) Y^2_{ik} \right]\right\rbrace \geq_{(2)} 0  \\
    \end{split}
\end{equation*}
$(1)$ is by $H_1 \cup H_2= \cup_{k=1}^K \mathcal{N}^{out}_{k}$. $(2)$ is based on \eqref{suff1_Vc_compare_int4}.

Now let us consider $R_s < {a^2}/{b^2}$. Using similar ideas from the proof above, we have
\begin{equation}
\label{suff1_Vc_compare_int5}
        \begin{split}
        \sum_{k=1}^K \sum_{jk \in \mathcal{N}^{out}_{k}} \sum_{ik \in H_{1,jk}} \left[  (N^{out}|\mathcal{N}^{out}_{jk}|)^{-2}- (N^{in}|\mathcal{N}^{in}_{ik}|)^{-2}  \right] b^2 < \sum_{k=1}^K \sum_{jk \in \mathcal{N}^{out}_{k}}\sum_{ik \in H_{2,jk}} \left[ (N^{in}|\mathcal{N}^{in}_{ik}|)^{-2}-(N^{out}|\mathcal{N}^{out}_{jk}|)^{-2} \right] a^2 
        \end{split}
    \end{equation}
Furthermore, we have 
\begin{equation}
 \label{suff1_Vc_compare_int6}
        \begin{split}
        & \sum_{k=1}^K \sum_{jk \in \mathcal{N}^{out}_{k}} \sum_{ik \in H_{1,jk}} \mathbb{E} \left\lbrace \left[  (N^{out}|\mathcal{N}^{out}_{jk}|)^{-2} W^2_{jk}(\mathbf{Z}_k) Y^2_{ik} \right] -  \mathbb{E}\left[(N^{in}|\mathcal{N}^{in}_{ik}|)^{-2} W^2_{jk}(\mathbf{Z}_k) Y^2_{ik} \right]  \right\rbrace \\
        & \leq_{(1)}  \left(\frac{1}{p^2}+\frac{1}{(1-p)^2}   \right) \sum_{k=1}^K \sum_{jk \in \mathcal{N}^{out}_{k}} \sum_{ik \in H_{1,jk}}  \left[  (N^{out}|\mathcal{N}^{out}_{jk}|)^{-2}- (N^{in}|\mathcal{N}^{in}_{ik}|)^{-2}  \right] b^2 
        \end{split}
    \end{equation}
    $(1)$ is based on Conditions (i) and (ii). Furthermore,
\begin{equation}
 \label{suff1_Vc_compare_int7}
    \begin{split}
       & \sum_{k=1}^K \sum_{jk \in \mathcal{N}^{out}_{k}} \sum_{ik \in H_{2,jk}}\left[ (N^{in}|\mathcal{N}^{in}_{ik}|)^{-2}-(N^{out}|\mathcal{N}^{out}_{jk}|)^{-2} \right] \mathbb{E}(W^2_{jk}(\mathbf{Z}_k) Y^2_{ik}) \\
       & \geq_{(1)} \left( \frac{1}{p^2}+\frac{1}{(1-p)^2} \right) \sum_{k=1}^K \sum_{jk \in \mathcal{N}^{out}_{k}} \sum_{ik \in H_{2,jk}} \left[ (N^{in}|\mathcal{N}^{in}_{ik}|)^{-2}-(N^{out}|\mathcal{N}^{out}_{jk}|)^{-2} \right]  a^2.
    \end{split}
\end{equation}
$(1)$ is based on Conditions (i) and (ii). Then based on equations \eqref{suff1_Vc_compare_int5}, \eqref{suff1_Vc_compare_int6} and \eqref{suff1_Vc_compare_int7}, we have 
\begin{equation}
 \label{suff1_Vc_compare_int8}
        \begin{split}
        & \sum_{k=1}^K \sum_{jk \in \mathcal{N}^{out}_{k}} \sum_{ik \in H_{1,jk}} \left\lbrace \mathbb{E} \left[  (N^{out}|\mathcal{N}^{out}_{jk}|)^{-2} W^2_{jk}(\mathbf{Z}_k) Y^2_{ik} \right] -  \mathbb{E}\left[(N^{in}|\mathcal{N}^{in}_{ik}|)^{-2} W^2_{jk}(\mathbf{Z}_k) Y^2_{ik} \right]  \right\rbrace \\
        & \leq  \sum_{k=1}^K \sum_{jk \in \mathcal{N}^{out}_{k}} \sum_{ik \in H_{2,jk}} \left\lbrace \mathbb{E}\left[ (N^{in}|\mathcal{N}^{in}_{ik}|)^{-2} W^2_{jk}(\mathbf{Z}_k) Y^2_{ik}\right]-\mathbb{E}\left[ (N^{out}|\mathcal{N}^{out}_{jk}|)^{-2} W^2_{jk}(\mathbf{Z}_k) Y^2_{ik} \right] \right\rbrace. 
        \end{split}
    \end{equation}
Thus, we obtain
\begin{small}
\begin{equation*}
    \begin{split}
    & V^c(\hat{\tau}^{out}(\alpha))- V^c(\hat{\tau}^{in}(\alpha))=\sum_{k=1}^K \sum_{jk \in \mathcal{N}^{out}_{k}} \sum_{ik \in  \mathcal{N}^{out}_{jk}}  \left\lbrace  \mathbb{E} \left[  (N^{out}|\mathcal{N}^{out}_{jk}|)^{-2} W^2_{jk}(\mathbf{Z}_k) Y^2_{ik} \right] -  \mathbb{E}\left[(N^{in}|\mathcal{N}^{in}_{ik}|)^{-2} W^2_{jk}(\mathbf{Z}_k) Y^2_{ik} \right]  \right\rbrace\\
        & = \sum_{k=1}^K \sum_{jk \in H_{1k}} \sum_{ik \in \mathcal{N}^{out}_{jk}}  \left\lbrace  \mathbb{E} \left[  (N^{out}|\mathcal{N}^{out}_{jk}|)^{-2} W^2_{jk}(\mathbf{Z}_k) Y^2_{ik} \right] -  \mathbb{E}\left[(N^{in}|\mathcal{N}^{in}_{ik}|)^{-2} W^2_{jk}(\mathbf{Z}_k) Y^2_{ik} \right]  \right\rbrace \\
        & - \sum_{k=1}^K \sum_{jk \in H_{2k}}\sum_{ik \in \mathcal{N}^{out}_{jk}} \left\lbrace \mathbb{E} \left[(N^{in}|\mathcal{N}^{in}_{ik}|)^{-2} W^2_{jk}(\mathbf{Z}_k) Y^2_{ik} \right]- \mathbb{E} \left[  (N^{out}|\mathcal{N}^{out}_{jk}|)^{-2} W^2_{jk}(\mathbf{Z}_k) Y^2_{ik} \right]\right\rbrace \leq_{(1)} 0  \\
    \end{split}
\end{equation*}
\end{small}
$(1)$ is by \eqref{suff1_Vc_compare_int8}. 
\end{proof}
}

\begin{proof}[Proof of Proposition \ref{dif_var_out_var_in}]
Based on the same notation in the proof of proposition \ref{V_out_V_in}, we have 
\begin{equation*}
    \begin{split}
       &  V(\hat{\tau}^{out}(\alpha))- V(\hat{\tau}^{in}(\alpha))= \sum_{k=1}^K \sum_{jk \in \mathcal{N}^{out}_{k}} \left( \frac{1}{N^{out \ 2}} var(V_{jk})- \frac{1}{N^{in \ 2 }} var(S_{jk})    \right)   \\
       & + \sum_{k=1}^K \sum_{jk \in \mathcal{N}^{out}_{k}} \sum_{ik \in \mathcal{N}^{out}_{k}: i \not{=} j } \left[  \frac{1}{N^{out \ 2 }} cov (V_{jk}, V_{ik})       - \frac{1}{N^{in \ 2 }} cov (S_{jk}, S_{ik})    \right]. \\
    \end{split}
\end{equation*}
Based on formula \eqref{proof_V_out_V_in_int_2} in the proof of Proposition \ref{V_out_V_in}, 
\begin{equation}
\label{proof_dif_var_out_var_in_int1}
    \begin{split}
        var\left(V_{jk}\right) & = \mathbb{E}(V^2_{1jk})+ \mathbb{E}(V^2_{0jk})- \left[ \mathbb{E}(V_{1jk}))-  \mathbb{E}(V_{0jk})\right]^2\\
        &= \mathbb{E}(V^2_{1jk})+ \mathbb{E}(V^2_{0jk})- \left[ \mu^{out}_{jk}(1,\alpha )- \mu^{out}_{jk}(0,\alpha )  \right]^2=\mathbb{E}(V^2_{1jk})+ \mathbb{E}(V^2_{0jk})- \bar{\tau}_{jk}^2(\alpha). \\
    \end{split}
\end{equation}
Similarly, 
\begin{small}
\begin{equation}
\label{proof_dif_var_out_var_in_int2}
    \begin{split}
     & var\left(S_{jk}\right)= \mathbb{E}(S^2_{1jk})+ \mathbb{E}(S^2_{0jk})- \left[ \mathbb{E}(S_{1jk}))-  \mathbb{E}(S_{0jk})\right]^2 \\
        &= \mathbb{E}(S^2_{1jk})+ \mathbb{E}(S^2_{0jk})- \left[ \sum_{ik \in \mathcal{N}^{out}_{jk}} \frac{1}{ |\mathcal{N}^{in}_{ik}|}  \bar{Y}_{ik}(1,\alpha )- \sum_{ik \in \mathcal{N}^{out}_{jk}} \frac{1}{ |\mathcal{N}^{in}_{ik}|}  \bar{Y}_{ik}(0,\alpha ) \right]^2=\mathbb{E}(S^2_{1jk})+ \mathbb{E}(S^2_{0jk})-\widetilde{\tau}^2_{jk}(\alpha). \\   
    \end{split}
\end{equation}
\end{small}
Therefore, for the following term, we have, based on equations \eqref{proof_dif_var_out_var_in_int1} and  \eqref{proof_dif_var_out_var_in_int2},
\begin{equation*}
    \begin{split}
       & \sum_{k=1}^K \sum_{jk \in \mathcal{N}^{out}_{k}} \left( \frac{1}{N^{out \ 2}} var(V_{jk})- \frac{1}{N^{in \ 2 }} var(S_{jk})    \right) \\
       & =  \sum_{k=1}^K \sum_{jk \in \mathcal{N}^{out}_k } \frac{1}{N^{out \ 2 }} \left[ \mathbb{E}(V^2_{1jk})+ \mathbb{E}(V^2_{0jk}) \right] -  \sum_{k=1}^K \sum_{jk \in \mathcal{N}^{out}_k } \frac{1}{N^{in \ 2 }} \left[ \mathbb{E}(S^2_{1jk})+ \mathbb{E}(S^2_{0jk}) \right] \\
       & -  \sum_{k=1}^K
    \sum_{jk \in \mathcal{N}^{out}_{k}}
    \left[
    \frac{1}{(N^{out})^2}\bar{\tau}^2_{jk}(\alpha)
    -
    \frac{1}{(N^{in})^2}\widetilde{\tau}^2_{jk}(\alpha) \right]=(a)-(c). \\
    \end{split}
\end{equation*}
Then the difference between $V\left( \hat{\tau}^{out}(\alpha) \right)$ and $V(\hat{\tau}^{in}(\alpha) )$ can be written as
\begin{equation*}
    \begin{split}
     & V\left( \hat{\tau}^{out}(\alpha) \right)- V(\hat{\tau}^{in}(\alpha) )= (a) + (b) - (c)\\ 
    \end{split}
\end{equation*}
where the terms (a), (b) and (c) are defined in Proposition \ref{dif_var_out_var_in}. 
\end{proof}

{ 
\begin{proof}[Proof of Proposition \ref{prop:opt_lambda}]
From \eqref{est_tau_out} and \eqref{est_tau_in}, we obtain
\begin{equation*}
    \begin{split}
       & \hat{\tau}_{\lambda}(\alpha)= \sum_{k=1}^K \sum_{jk \in \mathcal{N}^{out}_{k}} \sum_{ik \in \mathcal{N}^{out}_{jk}} \left( \frac{\lambda}{N^{out} |\mathcal{N}^{out}_{jk}| } + \frac{1-\lambda}{N^{in} |\mathcal{N}^{in}_{ik}|} \right) W_{jk}(\mathbf{Z}_{k}) \left[ Z_{jk}-(1-Z_{jk}) \right] Y_{ik} \\
       &: =  \sum_{k=1}^K \sum_{jk \in \mathcal{N}^{out}_{k}} \sum_{ik \in \mathcal{N}^{out}_{jk}} S_{ik,jk}(\lambda) W_{jk}(\mathbf{Z}_k) \left[ Z_{jk}-(1-Z_{jk}) \right] Y_{ik} \\
       & = \sum_{k=1}^K \sum_{jk \in \mathcal{N}^{out}_{k}} \left[ W_{jk}(\mathbf{Z}_k) \left( Z_{jk}-(1-Z_{jk}) \right) \sum_{ik \in \mathcal{N}^{out}_{jk}} S_{ik,jk}(\lambda) Y_{ik} \right]:= \sum_{k=1}^K \sum_{jk \in \mathcal{N}^{out}_{k}} (U_{1jk}-U_{0jk})
    \end{split}
\end{equation*}
where $S_{ik,jk}(\lambda):=  {\lambda}{ (S^{out}_{ik,jk})^{-1} } + {(1-\lambda)}{(S^{in}_{ik,jk})^{-1}}$, $U_{zjk}=W_{jk}(\mathbf{Z}_{k})1 \{Z_{jk}=z\} \sum_{ik \in \mathcal{N}^{out}_{jk}} S_{ik,jk}(\lambda) Y_{ik}$ for $z \in \{0,1\}$. 
Following the proof of Proposition \ref{V_out_V_in} for $V^c(\hat{\tau}^{out}(\alpha))$, the asymptotic conservative variance of $V^c(\hat{\tau}^{out}(\alpha))$ is
\begin{small}
\begin{equation}
\label{V_c_tau_hat_lambda_int1}
    \begin{split}
        & var(\hat{\tau}_{\lambda}(\alpha)) \leq V^c(\hat{\tau}_{\lambda}(\alpha))  = \sum_{k=1}^K \sum_{jk \in \mathcal{N}^{out}_{k}} |\mathcal{N}^{out}_k| \left[ \mathbb{E} \left(U^2_{1jk} \right) +  \mathbb{E} \left(U^2_{0jk} \right) \right] \\
        & = \sum_{k=1}^K \sum_{jk \in \mathcal{N}^{out}_{k}} |\mathcal{N}^{out}_{jk}| \  \mathbb{E} \left[ W^2_{jk}(\mathbf{Z}_k) \left(Z_{jk}+(1-Z_{jk}) \right) \left(\sum_{ik \in \mathcal{N}^{out}_{jk}} S_{ik,jk}(\lambda) Y_{ik} \right)^2    \right]. \\
    \end{split}
\end{equation}
\end{small}
Recall $\bar{Y}^{out}_{jk}=\sum_{ik\in \mathcal{N}^{out}_{jk}} {|\mathcal{N}^{out}_{jk}|^{-1}} Y_{ik}$ and $\widetilde{Y}^{out}_{jk}=\sum_{ik\in \mathcal{N}^{out}_{jk}} {|\mathcal{N}^{in}_{ik}|^{-1}} Y_{ik}$. Then for each $jk \in \mathcal{N}^{out}_{k}$, $k = 1,\dots,K$,
\begin{small}
\begin{equation}
\label{V_c_tau_hat_lambda_int2}
    \begin{split}
    & \left[\sum_{ik \in \mathcal{N}^{out}_{jk}} S_{ik,jk}(\lambda) Y_{ik} \right]^2= \left[\sum_{ik \in \mathcal{N}^{out}_{jk}} \left( \frac{\lambda}{N^{out} |\mathcal{N}^{out}_{jk}| } + \frac{1-\lambda}{N^{in} |\mathcal{N}^{in}_{ik}|} \right)  Y_{ik} \right]^2 \\
    & = \left[ \frac{\lambda}{N^{out}} \bar{Y}^{out}_{jk}+ \frac{1-\lambda}{N^{in}} \widetilde{Y}^{out}_{jk}\right]^2 = \lambda^2 \frac{1}{N^{out \ 2 }} \bar{Y}^{out \ 2}_{jk} + (1-\lambda)^2 \frac{1}{N^{in \ 2 }} \widetilde{Y}^{out \ 2}_{jk}+ 2 \lambda(1-\lambda) \frac{1}{N^{out}N^{in}} \bar{Y}^{out}_{jk}\widetilde{Y}^{out}_{jk}. 
    \end{split}
\end{equation}
\end{small}

Based on Equations \eqref{V_c_tau_hat_lambda_int1} and \eqref{V_c_tau_hat_lambda_int2}, we have 
\begin{small}
\begin{equation}
\label{V_c_hat_tau_lambda_int1}
    \begin{split}
      & V^c(\hat{\tau}_{\lambda}(\alpha))= \sum_{k=1}^K \sum_{jk \in \mathcal{N}^{out}_k} |\mathcal{N}^{out}_{k}| \ \mathbb{E} \left\lbrace W^2_{jk}(\mathbf{Z}_k) Z_{jk} \left[ \lambda^2 \frac{1}{N^{out \ 2 }} \bar{Y}^{out \ 2}_{jk} + (1-\lambda)^2 \frac{1}{N^{in \ 2 }} \widetilde{Y}^{out \ 2}_{jk}+ 2 \lambda(1-\lambda) \frac{1}{N^{out}N^{in}} \bar{Y}^{out}_{jk}\widetilde{Y}^{out}_{jk} \right]  \right\rbrace \\
      & + \sum_{k=1}^K \sum_{jk \in \mathcal{N}^{out}_k} |\mathcal{N}^{out}_{k}| \ \mathbb{E}\left\lbrace W^2_{jk}(\mathbf{Z}_k) (1-Z_{jk}) \left[ \lambda^2 \frac{1}{N^{out \ 2 }} \bar{Y}^{out \ 2}_{jk} + (1-\lambda)^2 \frac{1}{N^{in \ 2 }} \widetilde{Y}^{out \ 2}_{jk}+ 2 \lambda(1-\lambda) \frac{1}{N^{out}N^{in}} \bar{Y}^{out}_{jk}\widetilde{Y}^{out}_{jk} \right]  \right\rbrace \\
      & = \lambda^2 V^c(\hat{\tau}^{out}(\alpha))+(1-\lambda)^2 V^c(\hat{\tau}^{in}(\alpha)) \\
      & + 2 \lambda(1-\lambda) \sum_{k=1}^K \sum_{jk \in \mathcal{N}^{out}_k} |\mathcal{N}^{out}_{k}| \  \left\lbrace \mathbb{E} \left[ W^2_{jk}(\mathbf{Z}_k) Z_{jk}(N^{out}N^{in})^{-1} \bar{Y}^{out}_{jk} \widetilde{Y}^{out}_{jk} \right] + \mathbb{E} \left[ W^2_{jk}(\mathbf{Z}_k) (1-Z_{jk})(N^{out}N^{in})^{-1} \bar{Y}^{out}_{jk} \widetilde{Y}^{out}_{jk} \right]  \right\rbrace \\
      & : = \lambda^2 V^c(\hat{\tau}^{out}(\alpha))+(1-\lambda)^2 V^c(\hat{\tau}^{in}(\alpha)) + 2\lambda (1-\lambda) C(\hat{\tau}^{out}(\alpha) , \hat{\tau}^{in}(\alpha) ) \\
      &= \left[ V^c(\hat{\tau}^{out}(\alpha))+V^c(\hat{\tau}^{in}(\alpha))-2C (\hat{\tau}^{out}(\alpha) , \hat{\tau}^{in}(\alpha))   \right] \lambda^2 + \left[ 2C (\hat{\tau}^{out}(\alpha) , \hat{\tau}^{in}(\alpha))- 2V^c(\hat{\tau}^{in}(\alpha))  \right] \lambda + V^c(\hat{\tau}^{in}(\alpha))
    \end{split}
\end{equation}
\end{small}
where 

\begin{equation}
\label{covc_in_formula}
    \begin{split}
        C(\hat{\tau}^{out}(\alpha), \hat{\tau}^{in}(\alpha)):=\sum_{k=1}^K \sum_{jk \in \mathcal{N}^{out}_k} |\mathcal{N}^{out}_{k}| \   \mathbb{E} \left[ W^2_{jk}(\mathbf{Z}_k) (Z_{jk}+(1-Z_{jk}))(N^{out}N^{in})^{-1} \bar{Y}^{out}_{jk} \widetilde{Y}^{out}_{jk} \right]. 
    \end{split}
\end{equation}

For each $jk \in \mathcal{N}^{out}_{k}$ and $k = 1,\ldots,K$, we have
\begin{equation}
\label{cauch_swartz_int1}
    \begin{split}
       & 2 \mathbb{E} \left[ W^2_{jk}(\mathbf{Z}_k) (Z_{jk}+(1-Z_{jk}))(N^{out}N^{in})^{-1} \bar{Y}^{out}_{jk} \widetilde{Y}^{out}_{jk} \right] \\
& \leq \frac{1}{N^{out \ 2 }}  \mathbb{E}[ W^2_{jk}(\mathbf{Z}_k) (Z_{jk}+(1-Z_{jk}))   \bar{Y}^{out \ 2}_{jk}] + \frac{1}{N^{in \ 2 }}  \mathbb{E}[ W^2_{jk}(\mathbf{Z}_k) (Z_{jk}+(1-Z_{jk}))   \widetilde{Y}^{out \ 2}_{jk}]  
    \end{split}
\end{equation}
since $\mathbb{E}(XY)\leq \frac{1}{2} {[E(X^2)+E(Y^2)]}$. Summing $jk\in\mathcal{N}^{out}_{k}$ over $k = 1,\dots, K$ yields 
$V^c (\hat{\tau}^{out}(\alpha))+V^c(\hat{\tau}^{in}(\alpha))-2 C (\hat{\tau}^{out}(\alpha), \hat{\tau}^{in}(\alpha))\geq 0$. Thus, $V^c(\hat{\tau}_{\lambda}(\alpha))$ is convex in $\lambda$. Taking the derivative with respect to $\lambda$, the unconstrained minimum of \eqref{V_c_hat_tau_lambda_int1} is
\begin{equation}
\label{opt_lambda}
    \begin{split}
        \lambda_0=  \left[ V^c(\hat{\tau}^{out}(\alpha))+V^c(\hat{\tau}^{in}(\alpha))-2C (\hat{\tau}^{out}(\alpha) , \hat{\tau}^{in}(\alpha))   \right]^{-1} \left[ V^c(\hat{\tau}^{in}(\alpha))-C (\hat{\tau}^{out}(\alpha) , \hat{\tau}^{in}(\alpha))  \right].
    \end{split}
\end{equation}
Then the optimal weight under the constraints $0 \leq \lambda \leq 1$ is
\begin{equation}
\label{lambda_star_formular}
    \begin{split}
        \lambda^*
=
\begin{cases}
0, & \lambda_0<0,\\
\lambda_0, & 0\leq \lambda_0\leq 1,\\
1, & \lambda_0>1.
\end{cases}
    \end{split}
\end{equation}

The corresponding conservative variance is
\begin{small}
\begin{equation}
\label{linear_comb_V_c}
    \begin{split}
       & V^c(\hat{\tau}_{\lambda^*}(\alpha))= \min \left( V^c(\hat{\tau}^{out}(\alpha) \ ,  \ V^c(\hat{\tau}^{in}(\alpha)) \ , \  V^c(\hat{\tau}^{in}(\alpha))-\frac{\left[ V^c(\hat{\tau}^{in}(\alpha))-C (\hat{\tau}^{out}(\alpha) \ , \ \hat{\tau}^{in}(\alpha))\right]^2}{\left[ V^c(\hat{\tau}^{out}(\alpha))+V^c(\hat{\tau}^{in}(\alpha))-2C (\hat{\tau}^{out}(\alpha) , \hat{\tau}^{in}(\alpha))   \right]^2}  \right) \\
    \end{split}
\end{equation}
\end{small}
where $V^c(\hat{\tau}^{out}(\alpha))=V^c(\hat{\tau}_{\lambda=1}(\alpha))$ and $V^c(\hat{\tau}^{in}(\alpha)))=V^c(\hat{\tau}_{\lambda=0}(\alpha))$. 


We now show that, in general, $C(\hat{\tau}^{out}(\alpha), \hat{\tau}^{in}(\alpha))$ is not an upper bound for the covariance $cov(\hat{\tau}^{out}(\alpha), \hat{\tau}^{in}(\alpha))$. Since
\begin{small}
\begin{equation}
\label{cov_hat_tau_out_tau_in}
    \begin{split}
       & cov(\hat{\tau}^{out}(\alpha),\hat{\tau}^{in}(\alpha)) = cov \left( \frac{1}{N^{out}} \sum_{k=1}^K \sum_{jk \in \mathcal{N}^{out}_{k}} W_{jk}(\mathbf{Z}_k) (Z_{jk}-(1-Z_{jk}))\bar{Y}^{out}_{jk},\frac{1}{N^{in}} \sum_{k=1}^K \sum_{jk \in \mathcal{N}^{out}_{k}} W_{jk}(\mathbf{Z}_k) (Z_{jk}-(1-Z_{jk}))\widetilde{Y}^{out}_{jk}  \right) \\
       & =_{(1)} \frac{1}{N^{out}N^{in}} \sum_{k=1}^K \sum_{j_1k \in \mathcal{N}^{out}_k} \sum_{j_2 k \in \mathcal{N}^{out}_k }  cov \left( W_{j_1k}(\mathbf{Z}_{k}) (Z_{j_1k}- (1-Z_{j_1 k})) \bar{Y}^{out}_{j_1k}, W_{j_2k}(\mathbf{Z}_{k}) (Z_{j_2k}- (1-Z_{j_2 k})) \widetilde{Y}^{out}_{j_2k}  \right) \\
       & = \frac{1}{N^{out}N^{in}} \sum_{k=1}^K \sum_{jk \in \mathcal{N}^{out}_{k}} cov \left( W_{jk}(\mathbf{Z}_k) (Z_{jk}- (1-Z_{j k})) \bar{Y}^{out}_{jk} , W_{jk}(\mathbf{Z}_k) (Z_{jk}- (1-Z_{j k})) \widetilde{Y}^{out}_{jk}\right)\\
       & + \frac{1}{N^{out}N^{in}} \sum_{k=1}^K \sum_{j_1k \in \mathcal{N}^{out}_{k}} \sum_{j_2k \in \mathcal{N}^{out}_{k}: j_1k \neq j_2k } cov \left( W_{j_1k}(\mathbf{Z}_k) (Z_{j_1k}- (1-Z_{j_1 k})) \bar{Y}^{out}_{j_1k} , W_{j_2k}(\mathbf{Z}_k) (Z_{j_2k}- (1-Z_{j_2 k})) \widetilde{Y}^{out}_{j_2k}\right) \\
       & =_{(2)} \frac{1}{N^{out}N^{in}} \sum_{k=1}^K \sum_{jk \in \mathcal{N}^{out}_{k}} \mathbb{E} \left[W^2_{jk}(\mathbf{Z}_{jk})(Z_{jk} + (1-Z_{jk})) \bar{Y}^{out}_{jk} \widetilde{Y}^{out}_{jk} \right] \\
       & + \frac{1}{N^{out}N^{in}} \sum_{k=1}^K \sum_{j_1k \in \mathcal{N}^{out}_{k}} \sum_{j_2k \in \mathcal{N}^{out}_{k}: j_1k \neq j_2k } \mathbb{E}\left( W_{j_1k}(\mathbf{Z}_k) W_{j_2k} (\mathbf{Z}_k) (Z_{j_1k}- (1-Z_{j_1 k})) (Z_{j_2k}- (1-Z_{j_2 k}) \bar{Y}^{out}_{j_1k} \widetilde{Y}^{out}_{j_2k}\right) \\
       & - \frac{1}{N^{out}N^{in}}  \sum_{k=1}^K  \sum_{j_1k \in \mathcal{N}^{out}_{k}} \sum_{j_2k \in \mathcal{N}^{out}_{k}} (\bar{\mu}_{j_1k}(1)-\bar{\mu}_{j_1k}(0)))(\widetilde{\mu}_{j_2k}(1)-\widetilde{\mu}_{j_2k}(0))
    \end{split}
\end{equation}
\end{small}
where, for $z\in\{0,1\}$, $\bar{\mu}_{jk}(z)
:=
\sum_{ik\in \mathcal{N}^{out}_{jk}}
{|\mathcal{N}^{out}_{jk}|^{-1}}
Y_{ik}(z,\alpha)$ and $\widetilde{\mu}_{jk}(z)
:=
\sum_{ik\in \mathcal{N}^{out}_{jk}}
{|\mathcal{N}^{in}_{ik}|^{-1}}
Y_{ik}(z,\alpha)$. Equality $(1)$ follows from Assumption~\ref{part_intf}. The last term in $(2)$ need not be positive. Moreover, the second term in $(2)$ need not be bounded by ${(N^{out}N^{in})^{-1}}
\sum_{k=1}^K
\sum_{jk\in\mathcal{N}^{out}_{k}}
\{|\mathcal{N}^{out}_{k}|-1\}
\mathbb{E}
\left[
W^2_{jk}(\mathbf{Z}_{k})
\{Z_{jk}+(1-Z_{jk})\}
\bar{Y}^{out}_{jk}
\widetilde{Y}^{out}_{jk}
\right]$, which is a component of
$C(\hat{\tau}^{out}(\alpha),\hat{\tau}^{in}(\alpha))$.
Therefore, $C(\hat{\tau}^{out}(\alpha),\hat{\tau}^{in}(\alpha))$ is not necessarily an upper bound of $\operatorname{cov}(\hat{\tau}^{out}(\alpha),\hat{\tau}^{in}(\alpha))$. This observation does not contradict the conservative variance bound $\operatorname{var}(\hat{\tau}_{\lambda}(\alpha))
\leq
V^c(\hat{\tau}_{\lambda}(\alpha))$. Equivalently,
\[
\begin{split}
&\lambda^2
\operatorname{var}(\hat{\tau}^{out}(\alpha))
+
(1-\lambda)^2
\operatorname{var}(\hat{\tau}^{in}(\alpha))
+
2\lambda(1-\lambda)
\operatorname{cov}(\hat{\tau}^{out}(\alpha),\hat{\tau}^{in}(\alpha))
\\
&\leq
\lambda^2
V^c(\hat{\tau}^{out}(\alpha))
+
(1-\lambda)^2
V^c(\hat{\tau}^{in}(\alpha))
+
2\lambda(1-\lambda)
C(\hat{\tau}^{out}(\alpha),\hat{\tau}^{in}(\alpha)).
\end{split}
\]
Finally, the last term in $(2)$ of \eqref{cov_hat_tau_out_tau_in}, which contributes to $\operatorname{cov}(\hat{\tau}^{out}(\alpha),\hat{\tau}^{in}(\alpha))$, is not unbiasedly estimable in general. In particular, unbiased estimation cannot be obtained by simply substituting inverse probability weighted estimators for
$\bar{\mu}_{jk}(z)$ and $\widetilde{\mu}_{jk}(z)$.
\end{proof}

\begin{proof}[Proof of Corollary \ref{V_lambda_homo_cond_3}]
Under Condition~\ref{ave_equivalence_cond_3}, $(N^{out})^{-1}\bar{Y}^{out}_{jk}
=
(N^{in})^{-1}\widetilde{Y}^{out}_{jk}$. Therefore,
\[
\begin{split}
&C(\hat{\tau}^{out}(\alpha),\hat{\tau}^{in}(\alpha)) =
\sum_{k=1}^K
\sum_{jk\in\mathcal{N}^{out}_k}
|\mathcal{N}^{out}_{jk}|
\,
\mathbb{E}
\left[
W^2_{jk}(\mathbf{Z}_k)
\{Z_{jk}+(1-Z_{jk})\}
(N^{out}N^{in})^{-1}
\bar{Y}^{out}_{jk}
\widetilde{Y}^{out}_{jk}
\right]  \\
&\quad =
\sum_{k=1}^K
\sum_{jk\in\mathcal{N}^{out}_k}
|\mathcal{N}^{out}_{jk}|
\,
\mathbb{E}
\left[
W^2_{jk}(\mathbf{Z}_k)
\{Z_{jk}+(1-Z_{jk})\}
(N^{out})^{-2}
\{\bar{Y}^{out}_{jk}\}^2
\right]   =
V^c(\hat{\tau}^{out}(\alpha)).
\end{split}
\]
Similarly,
\[
\begin{split}
C(\hat{\tau}^{out}(\alpha),\hat{\tau}^{in}(\alpha))
&=
\sum_{k=1}^K
\sum_{jk\in\mathcal{N}^{out}_k}
|\mathcal{N}^{out}_{jk}|
\,
\mathbb{E}
\left[
W^2_{jk}(\mathbf{Z}_k)
\{Z_{jk}+(1-Z_{jk})\}
(N^{in})^{-2}
\{\widetilde{Y}^{out}_{jk}\}^2
\right]= V^c(\hat{\tau}^{in}(\alpha)).
\end{split}
\]
Hence
\[
C(\hat{\tau}^{out}(\alpha),\hat{\tau}^{in}(\alpha))
=
V^c(\hat{\tau}^{out}(\alpha))
=
V^c(\hat{\tau}^{in}(\alpha)).
\]
Substituting this identity into \eqref{V_c_hat_tau_lambda_int1} gives
\[
\begin{split}
V^c(\hat{\tau}_{\lambda}(\alpha))
&=
\lambda^2 V^c(\hat{\tau}^{out}(\alpha))
+
(1-\lambda)^2 V^c(\hat{\tau}^{in}(\alpha)) +
2\lambda(1-\lambda)
C(\hat{\tau}^{out}(\alpha),\hat{\tau}^{in}(\alpha)) \\
&=
V^c(\hat{\tau}^{out}(\alpha))
=
V^c(\hat{\tau}^{in}(\alpha)),
\end{split}
\]
for every $\lambda\in[0,1]$. This proves the result.
\end{proof}

\begin{proof}[Proof of Lemma \ref{suf_cond_lambda_non_trivial}]
Under the setting of Lemma~\ref{suf_cond_lambda_non_trivial}, we have $N^{out}=(K-t)+tm$ and $N^{in}=(K-t)m+t$. Without loss of generality, suppose that the first $K-t$ clusters are out-star graphs and that clusters $K-t+1,\ldots,K$ are in-star graphs. From the expression for $V^c(\hat{\tau}^{out}(\alpha))$ in Proposition~\ref{V_out_V_in}, consider first an out-star cluster $k\in\{1,\ldots,K-t\}$.
\begin{small}
\begin{equation*}
    \begin{split}
        & \frac{1}{N^{out \ 2 }} \  \mathbb{E}[ W^2_{jk}(\mathbf{Z}_k)   \bar{Y}^{out \ 2}_{jk}]=\mathbb{E}[ W^2_{jk}(\mathbf{Z}_k) (\sum_{ik \in \mathcal{N}^{out}_{jk}}  \frac{1}{N^{out } |\mathcal{N}^{out}_{jk}|} Y_{ik} )^2] = \frac{1}{[(K-t)+tm]^2m^2 } \mathbb{E}\left[ W^2_{jk}(\mathbf{Z}_k) (\sum_{ik \in \mathcal{N}^{out}_{jk}}   Y_{ik})^2\right] 
    \end{split}
\end{equation*}
\end{small}
For clusters with $k \in \{K-t+1, \ldots, K\}$, we have
\begin{small}
\begin{equation*}
    \begin{split}
        & \frac{1}{N^{out \ 2 }} \  \mathbb{E}[ W^2_{jk}(\mathbf{Z}_k)   \bar{Y}^{out \ 2}_{jk}]=\mathbb{E}[ W^2_{jk}(\mathbf{Z}_k) (\sum_{ik \in \mathcal{N}^{out}_{jk}}  \frac{1}{N^{out } |\mathcal{N}^{out}_{jk}|} Y_{ik} )^2] = \frac{1}{[(K-t)+tm]^2 } \mathbb{E}\left[ W^2_{jk}(\mathbf{Z}_k) (\sum_{ik \in \mathcal{N}^{out}_{jk}}   Y_{ik})^2\right].
    \end{split}
\end{equation*}
\end{small}
Based on $V^c(\hat{\tau}^{in}(\alpha))$ in Proposition \ref{V_out_V_in}, for each cluster $k \in \{1, \dots, K-t\}$, we have
\begin{small}
\begin{equation*}
    \begin{split}
        & \frac{1}{N^{in \ 2 }} \  \mathbb{E}[ W^2_{jk}(\mathbf{Z}_k)   \widetilde{Y}^{out \ 2}_{jk}]=\mathbb{E}[ W^2_{jk}(\mathbf{Z}_k) (\sum_{ik \in \mathcal{N}^{out}_{jk}}  \frac{1}{N^{in } |\mathcal{N}^{in}_{ik}|} Y_{ik} )^2] = \frac{1}{[(K-t)m+t]^2 } \mathbb{E}\left[ W^2_{jk}(\mathbf{Z}_k) (\sum_{ik \in \mathcal{N}^{out}_{jk}}   Y_{ik})^2\right]. 
    \end{split}
\end{equation*}
\end{small}
For clusters $k \in \{K - t + 1, \ldots, K\}$, we have
\begin{small}
\begin{equation*}
    \begin{split}
        & \frac{1}{N^{in \ 2 }} \  \mathbb{E}[ W^2_{jk}(\mathbf{Z}_k)   \widetilde{Y}^{out \ 2}_{jk}]=\mathbb{E}[ W^2_{jk}(\mathbf{Z}_k) (\sum_{ik \in \mathcal{N}^{out}_{jk}}  \frac{1}{N^{in } |\mathcal{N}^{in}_{ik}|} Y_{ik} )^2] = \frac{1}{[(K-t)m+t]^2m^2 } \mathbb{E}\left[ W^2_{jk}(\mathbf{Z}_k) (\sum_{ik \in \mathcal{N}^{out}_{jk}}   Y_{ik})^2\right].
    \end{split}
\end{equation*}
\end{small}

For $C(\hat{\tau}^{out}(\alpha),\hat{\tau}^{in}(\alpha))$ in \eqref{covc_in_formula}, for each cluster $k \in \{1,\ldots,K-t\}$,
\begin{equation*}
    \begin{split}
        \mathbb{E} \left[ W^2_{jk}(\mathbf{Z}_k) (N^{out}N^{in})^{-1} \bar{Y}^{out}_{jk} \widetilde{Y}^{out}_{jk} \right] = (K-t+tm)^{-1} m^{-1} [(K-t)m+t]^{-1}  \ \mathbb{E}\left[ W^2_{jk}(\mathbf{Z}_k) (\sum_{ik \in \mathcal{N}^{out}_{jk}}   Y_{ik})^2\right].
    \end{split}
\end{equation*}
For each cluster $k$ with $k \in \{K - t + 1, \dots, K\}$, we have
\begin{equation*}
    \begin{split}
       \mathbb{E} \left[ W^2_{jk}(\mathbf{Z}_k) (N^{out}N^{in})^{-1} \bar{Y}^{out}_{jk} \widetilde{Y}^{out}_{jk} \right] = (K-t+tm)^{-1} [(K-t)m+t]^{-1} m^{-1} \ \mathbb{E}\left[ W^2_{jk}(\mathbf{Z}_k) (\sum_{ik \in \mathcal{N}^{out}_{jk}}   Y_{ik})^2\right].
    \end{split}
\end{equation*}
Therefore, 
\begin{equation*}
    \begin{split}
        & V^{c}(\hat{\tau}^{out}(\alpha))-C(\hat{\tau}^{out}(\alpha),\hat{\tau}^{in}(\alpha)) \\
        & =\sum_{k=1}^{K-t} \sum_{jk \in \mathcal{N}^{out}_k} \left[ \frac{1}{[(K-t)+tm]^2m^2 }-  \frac{1}{(K-t+tm) [(K-t)m+t] m } \right] \mathbb{E}\left[ W^2_{jk}(\mathbf{Z}_k) (\sum_{ik \in \mathcal{N}^{out}_{jk}}   Y_{ik})^2\right] \\
        & + \sum_{k=K-t+1}^{K} \sum_{jk \in \mathcal{N}^{out}_k} \left[ \frac{1}{[(K-t)+tm]^2 }-  \frac{1}{(K-t+tm) [(K-t)m+t] m } \right] \mathbb{E}\left[ W^2_{jk}(\mathbf{Z}_k) (\sum_{ik \in \mathcal{N}^{out}_{jk}}   Y_{ik})^2\right] \\
        & = \sum_{k=1}^{K-t} \sum_{jk \in \mathcal{N}^{out}_k} \left[ \frac{K-t+tm-(K-t+tm)m }{(K-t+tm)^2m^2 ((K-t)m+t) }  \right]\mathbb{E}\left[ W^2_{jk}(\mathbf{Z}_k) (\sum_{ik \in \mathcal{N}^{out}_{jk}}   Y_{ik})^2\right]\\
        & + \sum_{k=K-t+1}^{K} \sum_{jk \in \mathcal{N}^{out}_k} \left[ \frac{[(K-t)m+t] m-(K-t+tm ) }{[(K-t)+tm]^2 [(K-t)m+t] m}\right] \mathbb{E}\left[ W^2_{jk}(\mathbf{Z}_k) (\sum_{ik \in \mathcal{N}^{out}_{jk}}   Y_{ik})^2\right]\\
        & = \frac{t(1-m^2)}{(K-t+tm)^2m^2 ((K-t)m+t)} \sum_{k=1}^{K-t} \sum_{jk \in \mathcal{N}^{out}_k} \mathbb{E}\left[ W^2_{jk}(\mathbf{Z}_k) (\sum_{ik \in \mathcal{N}^{out}_{jk}}   Y_{ik})^2\right]\\
        & + \frac{(K-t)(m^2-1)  }{[(K-t)+tm]^2 [(K-t)m+t] m} \sum_{k=K-t+1}^{K} \sum_{jk \in \mathcal{N}^{out}_k}  \mathbb{E}\left[ W^2_{jk}(\mathbf{Z}_k) (\sum_{ik \in \mathcal{N}^{out}_{jk}}   Y_{ik})^2\right]>_{(1)}0\\
    \end{split}
\end{equation*}
(1) follows from Conditions (a) and (b) in Lemma \ref{suf_cond_lambda_non_trivial}.Meanwhile, 
\begin{equation*}
    \begin{split}
        & V^{c}(\hat{\tau}^{in}(\alpha))-C(\hat{\tau}^{out}(\alpha),\hat{\tau}^{in}(\alpha)) \\
        & =\sum_{k=1}^{K-t} \sum_{jk \in \mathcal{N}^{out}_k} \left[ \frac{1}{[(K-t)m+t]^2 }-  \frac{1}{(K-t+tm) [(K-t)m+t] m } \right] \mathbb{E}\left[ W^2_{jk}(\mathbf{Z}_k) (\sum_{ik \in \mathcal{N}^{out}_{jk}}   Y_{ik})^2\right] \\
        & + \sum_{k=K-t+1}^{K} \sum_{jk \in \mathcal{N}^{out}_k} \left[\frac{1}{[(K-t)m+t]^2m^2 }-  \frac{1}{(K-t+tm) [(K-t)m+t] m } \right] \mathbb{E}\left[ W^2_{jk}(\mathbf{Z}_k) (\sum_{ik \in \mathcal{N}^{out}_{jk}}   Y_{ik})^2\right] \\
        & = \sum_{k=1}^{K-t} \sum_{jk \in \mathcal{N}^{out}_k} \left[ \frac{t(m^2-1)}{[(K-t)m+t]^2m (K-t+tm)  }  \right]\mathbb{E}\left[ W^2_{jk}(\mathbf{Z}_k) (\sum_{ik \in \mathcal{N}^{out}_{jk}}   Y_{ik})^2\right]\\
        & + \sum_{k=K-t+1}^{K} \sum_{jk \in \mathcal{N}^{out}_k} \left[ \frac{(K-t)(1-m^2)}{[(K-t)m+t]^2m^2 (K-t+tm) }\right] \mathbb{E}\left[ W^2_{jk}(\mathbf{Z}_k) (\sum_{ik \in \mathcal{N}^{out}_{jk}}   Y_{ik})^2\right]>_{(1)}0\\
    \end{split}
\end{equation*}
(1) follows from Conditions (a) and (b) in Lemma \ref{suf_cond_lambda_non_trivial}. Therefore,
$$V^c(\hat{\tau}^{in}(\alpha))>C (\hat{\tau}^{out}(\alpha),\hat{\tau}^{in}(\alpha)), \quad V^c(\hat{\tau}^{out}(\alpha))>C (\hat{\tau}^{out}(\alpha), \hat{\tau}^{in}(\alpha)),$$
which implies that
\begin{equation}
\label{non_trivial_solution_int_1}
    V^c(\hat{\tau}^{in}(\alpha))-C (\hat{\tau}^{out}(\alpha) , \hat{\tau}^{in}(\alpha))>0  
\end{equation}
\begin{equation}
\label{non_trivial_solution_int_2}
V^c(\hat{\tau}^{in}(\alpha))-C (\hat{\tau}^{out}(\alpha), \hat{\tau}^{in}(\alpha))< V^c(\hat{\tau}^{in}(\alpha))+ V^c(\hat{\tau}^{out}(\alpha))-2C (\hat{\tau}^{out}(\alpha), \hat{\tau}^{in}(\alpha))
\end{equation} and  
  \begin{equation}
\label{non_trivial_solution_int_3}  V^c(\hat{\tau}^{in}(\alpha))+ V^c(\hat{\tau}^{out}(\alpha))-2C (\hat{\tau}^{out}(\alpha), \hat{\tau}^{in}(\alpha))>0.
\end{equation}
    From \eqref{non_trivial_solution_int_1}--\eqref{non_trivial_solution_int_3} and the expressions for $\lambda_0$ in \eqref{opt_lambda} and $\lambda^*$ in \eqref{lambda_star_formular}, we conclude that $0<\lambda^*<1$.
\end{proof}

\subsection{Technical details of Section \ref{estimand_weight_efficiency}}
\label{append_proof_sec_estimand_weight_efficiency}
\begin{proof}[Proof of Lemma~\ref{suffi_cond3_example}]
In each in-star graph, the non-central units are senders, denoted by $jk$. Since there are $tK$
in-star graphs and $(1-t)K$ out-star graphs,
\[
N^{out}=tK m_{in}+(1-t)K
=K(tm_{in}+1-t),
\]
and
\[
N^{in}=tK+(1-t)K m_{out}
=K(t+(1-t)m_{out}).
\]
For a receiver--sender pair in an in-star graph, we have
$|\mathcal{N}^{out}_{jk}|=1$ and
$|\mathcal{N}^{in}_{ik}|=m_{in}$. Therefore,
\[
N^{in}|\mathcal{N}^{in}_{ik}|
=
K(t+(1-t)m_{out})m_{in},
\]
whereas
\[
N^{out}|\mathcal{N}^{out}_{jk}|
=
K(tm_{in}+1-t).
\]
Since $m_{in},m_{out}\geq 2$, $N^{in}|\mathcal{N}^{in}_{ik}|
>
N^{out}|\mathcal{N}^{out}_{jk}|$. Thus, for each receiver--sender pair $(ik,jk)$ in an in-star graph,
$(ik,jk)\in H_1(t)$.

For each out-star graph, we have
$|\mathcal{N}^{in}_{ik}|=1$ and
$|\mathcal{N}^{out}_{jk}|=m_{out}$. Hence
\[
N^{in}|\mathcal{N}^{in}_{ik}|
=
K(t+(1-t)m_{out}),
\]
and
\[
N^{out}|\mathcal{N}^{out}_{jk}|
=
K(tm_{in}+1-t)m_{out}.
\]
Since $m_{in},m_{out}\geq 2$,
\[
N^{in}|\mathcal{N}^{in}_{ik}|
<
N^{out}|\mathcal{N}^{out}_{jk}|.
\]
Therefore, for each receiver--sender pair $(ik,jk)$ in an out-star graph,
$(ik,jk)\in H_2(t)$.

We next analyse the three quantities in the statement. Since each in-star
graph contributes $m_{in}$ receiver--sender pairs to $H_1(t)$, whereas each
out-star graph contributes $m_{out}$ receiver--sender pairs to $H_2(t)$,
\[
\frac{|H_1(t)|}{|H_2(t)|}
=
\frac{tK m_{in}}{(1-t)K m_{out}}
=
\frac{m_{in}}{m_{out}}\frac{t}{1-t},
\]
which is increasing in $t$. Furthermore,
\[
\Delta_{H_1}(t)
=
\frac{1}{|H_1(t)|}
tK m_{in}
\left[
(N^{out})^{-2}
-
(N^{in}m_{in})^{-2}
\right],
\]
and
\[
\Delta_{H_2}(t)
=
\frac{1}{|H_2(t)|}
(1-t)K m_{out}
\left[
(N^{in})^{-2}
-
(N^{out}m_{out})^{-2}
\right].
\]
Thus
\[
R_s(t)
=
\frac{
tK m_{in}
\left[
(N^{out})^{-2}
-
(N^{in}m_{in})^{-2}
\right]
}{
(1-t)K m_{out}
\left[
(N^{in})^{-2}
-
(N^{out}m_{out})^{-2}
\right]
}.
\]
Using the identity $a^{-2}-b^{-2}
=
{(b^2-a^2)}/{(a^2b^2)}$, we obtain
\[
R_s(t)
=
\frac{t m_{in}}{(1-t)m_{out}}
\cdot
\frac{
(N^{in}m_{in})^2-(N^{out})^2
}{
(N^{in}m_{in})^2(N^{out})^2
}
\cdot
\frac{
(N^{out}m_{out})^2(N^{in})^2
}{
(N^{out}m_{out})^2-(N^{in})^2
}.
\]
After cancellation,
\[
R_s(t)
=
\frac{t m_{out}}{(1-t)m_{in}}
\cdot
\frac{
(N^{in}m_{in})^2-(N^{out})^2
}{
(N^{out}m_{out})^2-(N^{in})^2
}.
\]
Next, factor the numerator and denominator:
\[
(N^{in}m_{in})^2-(N^{out})^2
=
(N^{in}m_{in}-N^{out})
(N^{in}m_{in}+N^{out}),
\]
and
\[
(N^{out}m_{out})^2-(N^{in})^2
=
(N^{out}m_{out}-N^{in})
(N^{out}m_{out}+N^{in}).
\]
A direct calculation gives $N^{in}m_{in}-N^{out}
=
(1-t)K(m_{in}m_{out}-1)$, and $N^{out}m_{out}-N^{in}
=
tK(m_{in}m_{out}-1)$. Substituting these expressions into $R_s(t)$ yields
\[
R_s(t)
=
\frac{t m_{out}}{(1-t)m_{in}}
\cdot
\frac{
(1-t)K(m_{in}m_{out}-1)
(N^{in}m_{in}+N^{out})
}{
tK(m_{in}m_{out}-1)
(N^{out}m_{out}+N^{in})
}.
\]
Canceling common factors gives
\[
R_s(t)
=
\frac{m_{out}}{m_{in}}
\cdot
\frac{
N^{in}m_{in}+N^{out}
}{
N^{out}m_{out}+N^{in}
}.
\]
Substituting the formulas of $N^{out}$ and $N^{in}$, and canceling
the common factor $K$, we obtain
\[
R_s(t)
=
\frac{m_{out}}{m_{in}}
\cdot
\frac{
m_{in}(t+(1-t)m_{out})+(tm_{in}+1-t)
}{
m_{out}(tm_{in}+1-t)+(t+(1-t)m_{out})
}.
\]
Expanding the numerator gives
\[
m_{in}(t+(1-t)m_{out})+(tm_{in}+1-t)
=
m_{in}m_{out}+1
+
t(2m_{in}-m_{in}m_{out}-1).
\]
Similarly, the denominator becomes
\[
m_{out}(tm_{in}+1-t)+(t+(1-t)m_{out})
=
2m_{out}
+
t(m_{in}m_{out}-2m_{out}+1).
\]
Hence
\[
R_s(t)
=
\frac{m_{out}}{m_{in}}
\cdot
\frac{
m_{in}m_{out}+1
+
t(2m_{in}-m_{in}m_{out}-1)
}{
2m_{out}
+
t(m_{in}m_{out}-2m_{out}+1)
}.
\]
Thus $R_s(t)$ is a linear fractional of $t$. Define $a:=m_{in}m_{out}+1$, \  $b:=2m_{in}-m_{in}m_{out}-1$, \  $c:=2m_{out}$, and \ 
$d:=m_{in}m_{out}-2m_{out}+1$. Then
\[
R_s(t)
=
\frac{m_{out}}{m_{in}}
\cdot
\frac{a+bt}{c+dt}.
\]
Differentiating with respect to $t$ gives
\[
R_s'(t)
=
\frac{m_{out}}{m_{in}}
\cdot
\frac{b(c+dt)-d(a+bt)}{(c+dt)^2}
=
\frac{m_{out}}{m_{in}}
\cdot
\frac{bc-ad}{(c+dt)^2}.
\]
A direct calculation yields $bc-ad=-(m_{in}m_{out}-1)^2$. Therefore,
\[
R_s'(t)
=
-
\frac{m_{out}}{m_{in}}
\cdot
\frac{(m_{in}m_{out}-1)^2}
{
[2m_{out}+t(m_{in}m_{out}-2m_{out}+1]^2
}.
\]
Since $m_{in}\geq 2$ and $m_{out}\geq 2$, then $(m_{in}m_{out}-1)^2>0$, and hence $R_s'(t)<0$. Thus $R_s(t)$ is strictly decreasing in $t$. Finally, from the expressions above, ${|H_1(t)|}/{|H_2(t)|}
=
({m_{in}}/{m_{out}})({t}/{(1-t)})$ is increasing in $t$. From $R_{s}(t)={(|H_1(t)|\Delta_{H_1}(t))}/{(|H_2(t)|\Delta_{H_2}(t))}$ and the monotonic relationship between ${|H_1(t)|}/{|H_2(t)|}$ and $R_{s}(t)$, it follows that $\Delta_{H_1}(t)/\Delta_{H_2}(t)$ decreases with $t$.
\end{proof}
}

\subsection{Technical details for conditional spillover effects}
In this section, we provide the proof of Proposition \ref{difference_cond_out_in} and demonstrate that Conditions \ref{cond_equivalence_cond_1}, \ref{cond_equivalence_cond_2}, and \ref{cond_equivalence_cond_3} serve as sufficient conditions for the equivalence of the estimands defined in Definitions \ref{cond_out_spillover} and \ref{cond_in_spillover}. Let $[K]:= \{1,\cdots, K\}$, $[n]:= \{1,\cdots, n\}$ and $[n_k]:=\{1,\cdots,n_k\}$ for $k=1,\cdots, K$.

\begin{proof}[Proof of Proposition \ref{difference_cond_out_in}]
Let $A$ represent the adjacency matrix associated with the graph \( \mathcal{G} \). Based on Definition \ref{cond_in_spillover}, we have 
    \begin{small}    
    \begin{equation*}
        \begin{split}
           & \tau^{in}(x,\alpha) = \frac{1}{N^{in}(x)} \sum_{k=1}^K \sum_{ik \in \mathcal{N}^{in}_k (x)  }  \left[ \mu^{in}_{ik}\left(x,1 ,\alpha \right)- \mu^{in}_{ik}\left(x,0 ,\alpha \right) \right] \\
           & =  \frac{1}{N^{in}(x)} \sum_{k=1}^K \sum_{ik \in \mathcal{N}^{in}_k(x)} \left[ \frac{ 1 }{|\mathcal{N}^{in}_{ik}(x)| } \sum_{jk \in \mathcal{N}^{in}_{ik}(x) } \left( \bar{Y}_{ik}( Z_{jk}=1,\alpha)- \bar{Y}_{ik}( Z_{jk}=0,\alpha)  \right) \right]  \\
           & = \frac{1}{N^{in}(x)} \sum_{k=1}^K \sum_{ik \in \mathcal{N}^{in}_{k}(x)} \left[ \frac{ 1  }{|\mathcal{N}^{in}_{ik}(x)| } \sum_{jk =1 }^{n_k} A_{ji} \left( \bar{Y}_{ik}( Z_{jk}=1,\alpha)-\bar{Y}_{ik}( Z_{jk}=0,\alpha)  \right) \cdot 1\{ X_{jk}=x \}  \right]   \\
           & = \sum_{k=1}^K  \sum_{jk=1}^{n_k}  \sum_{ik \in \mathcal{N}^{in}_k(x)} A_{ji} \frac{1 }{N^{in} (x) \cdot |\mathcal{N}^{in}_{ik}(x)|} \cdot \left( \bar{Y}_{ik}( Z_{jk}=1,\alpha)- \bar{Y}_{ik}( Z_{jk}=0,\alpha)  \right) \cdot 1\{ X_{jk}=x \} \\
           & = \sum_{k=1}^K \sum_{jk=1}^{n_k} 1\{X_{jk}=x\}  \sum_{ik \in \mathcal{N}^{in}_k(x)} A_{ji} \cdot \frac{1 }{N^{in} (x) \cdot |\mathcal{N}^{in}_{ik}(x)|}  \cdot \left( \bar{Y}_{ik}( Z_{jk}=1,\alpha)- \bar{Y}_{ik}(Z_{jk}=0,\alpha)  \right)  \\
           & =_{(1)}  \sum_{k=1}^K   \sum_{jk=1}^{n_k} 1\{X_{jk}=x \}  \sum_{ik \in \mathcal{N}^{out}_{jk}} \frac{1}{N^{in} (x) \cdot |\mathcal{N}^{in}_{ik}(x)|} \cdot \left( \bar{Y}_{ik}( Z_{jk}=1,\alpha)- \bar{Y}_{ik}(Z_{jk}=0,\alpha)  \right)  \\
           & = \sum_{k=1}^K   \sum_{jk \in \mathcal{N}^{out}_{k}(x)}  \sum_{ik \in \mathcal{N}^{out}_{jk}} \frac{1}{N^{in} (x) \cdot |\mathcal{N}^{in}_{ik}(x)|} \cdot \left( \bar{Y}_{ik}( Z_{jk}=1,\alpha)- \bar{Y}_{ik}( Z_{jk}=0,\alpha)  \right)  \\
        \end{split}
    \end{equation*}
       \end{small}
       $(1)$ is by the fact that if $ik \in \mathcal{N}^{in}_{k}(x)$, $ik$ is connected with $jk$ and $X_{jk}=x$, then $\mathcal{N}^{out}_{jk} \in \mathcal{N}^{in}_{k}(x)$. 
Combining Definition \ref{cond_in_spillover}, we obtain the condition for ensuring $\tau^{out}(x,\alpha)\neq\tau^{in}(x,\alpha)$. 
\end{proof}

\begin{proof}[Proof based on Condition \ref{cond_equivalence_cond_1}]
    Given Condition \ref{cond_equivalence_cond_1} is satisfied, we have, Based on Definition \ref{cond_out_spillover} 
\begin{equation}
\label{proof_cond_equivalence_cond_1_int1}
        \begin{split}
        & \sum_{k=1}^K \sum_{jk \in \mathcal{N}^{out}_{k}(x)} \sum_{ik \in \mathcal{N}^{out}_{jk}} \frac{1}{N^{out}(x) \    |\mathcal{N}^{out}_{jk}|} \left(  \bar{Y}_{ik}( Z_{jk}=1,\alpha)- \bar{Y}_{ik}(Z_{jk}=0,\alpha)  \right) \\
        & = \sum_{k=1}^K c_{k,x} \sum_{jk \in \mathcal{N}^{out}_{k}(x)}  \frac{1}{N^{out}(x)  \   |\mathcal{N}^{out}_{jk}|} \  |\mathcal{N}^{out}_{jk}| = \sum_{k=1}^K c_{k,x}  \frac{  |\mathcal{N}^{out}_k(x)|}{N^{out}(x)}. \\
        \end{split}
    \end{equation} 
Meanwhile, based on Definition \ref{cond_in_spillover}, we have 
\begin{equation}
\label{proof_cond_equivalence_cond_1_int2}
        \begin{split}
        & \frac{1}{N^{in}(x)} \sum_{k=1}^K \sum_{ik \in \mathcal{N}^{in}_k (x) } \left\lbrace \frac{ 1 }{|\mathcal{N}^{in}_{ik}(x)| } \sum_{jk \in \mathcal{N}^{in}_{ik}(x) } \left( \bar{Y}_{ik}( Z_{jk}=1,\alpha)- \bar{Y}_{ik}( Z_{jk}=0,\alpha)  \right)  \right\rbrace  \\
        & = \frac{1}{N^{in}(x)} \sum_{k=1}^K c_{k,x} \sum_{ik \in \mathcal{N}^{in}_k(x) } \left(  \frac{1  }{|\mathcal{N}^{in}_{ik}(x)| }  \cdot |\mathcal{N}^{in}_{ik}(x) |  \right) =  \sum_{k=1}^K c_{k,x} \sum_{ik  \in \mathcal{N}^{in}_k(x)} \frac{1}{N^{in}(x)}  \\
        & = \sum_{k=1}^K c_{k,x} \frac{ |\mathcal{N}^{in}_k (x)| }{N^{in}(x)}  =_{(1)}  \sum_{k=1}^K c_{k,x} \frac{ |\mathcal{N}^{out}_{k}(x)|   }{N^{out}(x)}
        \end{split}
    \end{equation}
where $(1)$ is based on Condition \ref{cond_equivalence_cond_2}. Then combining last equalities of equations \eqref{proof_cond_equivalence_cond_1_int1} and \eqref{proof_cond_equivalence_cond_1_int2}, $\tau^{out}(x, \alpha )=\tau^{in}(x, \alpha)$. 
\end{proof}

\begin{proof}[Proof based on Condition \ref{cond_equivalence_cond_2}]
    Based on the homogeneous pairwise spillover effects conditioning on $x$, the conditional outward spillover effect can be written as
    \begin{equation*}
        \begin{split}
        & \sum_{k=1}^K \sum_{jk \in \mathcal{N}^{out}_{k}(x) } \sum_{ik \in \mathcal{N}^{out}_{jk}} \frac{1}{N^{out}(x) \cdot  |\mathcal{N}^{out}_{jk}|} \left(  \bar{Y}_{ik}( Z_{jk}=1,\alpha)- \bar{Y}_{ik}( Z_{jk}=0,\alpha)  \right) \\
        & = \frac{c_x}{N^{out}(x)} \sum_{k=1}^K \sum_{jk \in \mathcal{N}^{out}_k (x)} \frac{1 }{ |\mathcal{N}^{out}_{jk}|  }      \sum_{ik \in \mathcal{N}^{out}_{jk}} 1=c_{x}. \\
        \end{split}
    \end{equation*}
    The conditional inward spillover effect can be written as 
    \begin{equation*}
        \begin{split}
        & \frac{1}{N^{in}(x)} \sum_{k=1}^K \sum_{ik \in \mathcal{N}^{in}_{k}(x)} \left[ \frac{ 1 }{|\mathcal{N}^{in}_{ik}(x)| } \sum_{jk \in \mathcal{N}^{in}_{ik}(x)} \left( \bar{Y}_{ik}(Z_{jk}=1,\alpha)- \bar{Y}_{ik}(Z_{jk}=0,\alpha)  \right)  \right]  \\
        & = \frac{c_x}{N^{in}(x)} \sum_{k=1}^K \sum_{ik \in \mathcal{N}^{in}_{k} (x)} \left[  \frac{1 }{|\mathcal{N}^{in}_{ik}(x)| }  \cdot |\mathcal{N}^{in}_{ik}(x) |  \right] = \frac{c_x}{N^{in}(x)} \sum_{k=1}^K \sum_{ik \in \mathcal{N}^{in}_k(x) } 1  = c_x 
        \end{split}
    \end{equation*}
\end{proof}

\begin{proof}[Proof based on Condition \ref{cond_equivalence_cond_3}]
Based on the notations and the conclusion from Proposition \ref{difference_cond_out_in}, we have 
$ \tau^{out}(x, \alpha)- \tau^{in}(x,\alpha)= \sum_{k=1}^K \sum_{jk \in \mathcal{N}^{out}_{k}(x) } \sum_{ik \in \mathcal{N}^{out}_{jk} } A_{ik,jk} \cdot F_{ik, jk}$. Under condition \ref{cond_equivalence_cond_3}, we have $A_{ik,jk}=0$ for $ik \in \mathcal{N}^{out}_{jk}$, $jk \in \mathcal{N}^{out}_{k}(x)$ and $k \in [K]$, which leads to $ \tau^{out}(x, \alpha)- \tau^{in}(x,\alpha)= \sum_{k=1}^K \sum_{jk \in \mathcal{N}^{out}_{k}(x) } \sum_{ik \in \mathcal{N}^{out}_{jk} } 0 \cdot F_{ik, jk}=0$.
\end{proof}

\bibliographystyle{plainnat}
\bibliography{paper-ref}

\end{document}